\documentclass[aps,twocolumn,prd,floats,amsmath,amssymb,nofootinbib,floatfix,showpacs]{revtex4-1}
\usepackage[bookmarks,breaklinks, colorlinks, urlcolor=magenta, citecolor=blue, link color=green]{hyperref}

\usepackage{epsfig}
\usepackage{graphicx}
\usepackage{color}
\usepackage{mathrsfs}
\usepackage{amssymb}
\usepackage{amsmath}
\usepackage{url}
\usepackage{nicefrac}
\RequirePackage{xspace}

\def\jpsi{\ensuremath{{J\mskip -3mu/\mskip -2mu\psi\mskip 2mu}}}
\def\dsp{\displaystyle}
\def\be {\begin{equation}}
\def\ee {\end{equation}}
\def\bea {\begin{eqnarray}}
\def\eea {\end{eqnarray}}
\def\bc {\begin{center}}
\def\ec {\end{center}}
\def\nn {\nonumber}

\def\gev{\ensuremath{\mathrm{Ge\kern -0.1em V}}}
\def \thl {{\theta_\ell}}
\def \thK {{\theta_{K}}}
\def \azeL{{{\cal A}_0^L}}
\def \azeR{{{\cal A}_0^R}}
\def \apaL{{{\cal A}_\parallel^L}}
\def \apaR{{{\cal A}_\parallel^R}}
\def \apeL{{{\cal A}_\perp^L}}
\def \apeR{{{\cal A}_\perp^R}}
\def \Re{\text{Re}}
\def \Im{\text{Im}}
\def \kstar{{K^{\!*}}}

\def \eff{{\text{eff }}}
\def \braket#1#2#3{\langle #1|#2| #3\rangle}
\def\AFB{A_{\text{FB}}}

\def\eff{\text{eff}}
\def\AFB{A_{\text{FB}}}
\def\Gf{\Gamma_{\!\! f}}
\def\s{\ensuremath{s}\xspace}
\def\invfb   {\ensuremath{\mbox{\,fb}^{-1}}\xspace}

\def\ket|#1>{\left|#1 \right>}

\def\bra<#1|{\left< #1 \right|}

\def\bracket<#1|#2>{\setbox0=\vbox{\hbox{$#1$$#2$}}\left<#1\kern1pt
    \vrule  height\ht0\kern2pt #2\right>} 

\def\dirmat<#1|#2|#3>{\setbox0=\vbox{\hbox{$#1$$#2$$#3$}}\left<#1\kern1pt
    \vrule height\ht0\kern1pt#2\kern1pt \vrule height\ht0\kern1pt
    #3\right>} 

\newcommand{\lqcd}{\ensuremath{\Lambda_{\mathrm{QCD}}}\xspace}

\allowdisplaybreaks
\begin{document}

\title{New Physics Effects and  Hadronic Form Factor Uncertainties in $\mathbf{B\to \kstar\ell^+\ell^-}$}

\author{Diganta Das}
\affiliation{The Institute of Mathematical Sciences, Taramani, Chennai
  600113, India}
\author{Rahul Sinha}
\affiliation{The Institute of Mathematical Sciences, Taramani, Chennai
  600113, India}

\date{\today}

\begin{abstract}
  It is well known that New Physics can contribute to weak decays of
  heavy mesons via virtual processes during its decays.  The discovery
  of New Physics, using such decays is made difficult due to
  intractable strong interaction effects needed to describe it. Modes
  such as $B\to \kstar \ell^+\ell^-$ offer an advantage as they
  provide a multitude of observables via angular analysis.  We show
  how the multitude of ``related observables'' obtained from $B\to
  \kstar \ell^+\ell^-$, can provide many new ``clean tests'' of the
  Standard Model.  The hallmark of these tests is that several of them
  are independent of the unknown universal form factors that describe
  the decay in heavy quark effective theory.  We derive a relation
  between observables that is free of form factors and Wilson
  coefficients, the violation of which will be an unambiguous signal
  of New Physics.  We also derive relations between observables and
  form factors that are independent of Wilson coefficients and enable
  verification of hadronic estimates. We show how form factor ratios
  can be measured directly from helicity fraction with out any
  assumptions what so ever.  We find that the allowed parameter space
  for observables is very tightly constrained in Standard Model,
  thereby providing clean signals of New Physics.  We examine in
  detail both the large-recoil and low-recoil regions of the $\kstar$
  meson and point out special features and derive relations between
  observables valid in the two limits. In the large-recoil regions
  several of the relations are unaffected by corrections to all orders
  in $\alpha_s$. We present yet another new relation involving only
  observables that would verify the validity of the relations between
  form-factors assumed in the low-recoil region.  The several
  relations and constraints derived will provide unambiguous signals
  of New Physics if it contributes to these decays.
\end{abstract}

\pacs{11.30.Er,13.25.Hw, 12.60.-i}

\maketitle

\section{Introduction}
\label{sec:Intro}

It is well known that physics beyond the Standard Model referred to as
New Physics can either be discovered by direct production of new
particles at high energies or by indirect searches at high luminosity
facilities where new physics can contribute virtually to loop
processes.  The most well known example of the latter kind is the muon
magnetic moment. Unfortunately, even though muon is a lepton, hadronic
contributions have to be estimated and turn out to be the limiting
factor in the search for New Physics.  Indirect searches for New
Physics often involve precision measurement of a single quantity as in
the case of muon magnetic moment. The single measurement is compared
to a theoretical estimate that needs to be accurately
calculated. There are however, certain decays which involve
measurement of several related observables.  Well known examples of
such decays are $B\to V_1V_2$ where $B$ decays to two vector mesons
$V_1$ and $V_2$ and the semi-leptonic penguin decay $B\to
\kstar\ell^+\ell^-$. The heavy meson decays to such modes occur in
multiple partial waves and allow a measurement of a multitude of
related observables. In this paper we will show how the observables
obtained from an angular analysis of $B\to \kstar\ell^+\ell^-$ allow
for a cleaner signals of New Physics if it exists.

It is hoped that flavor changing neutral current transitions in $b\to
s$ and $b\to d$ will be altered by physics beyond the Standard Model
(SM) and their study would reveal possible signal of new physics (NP)
if it exists. However, understanding the hadronic flavor changing
neutral current decays requires estimating hadronic effects which
cannot be completely accurately done. Experimental data collected by
the Belle and Babar collaborations at the B-factories, CLEO, Tevatron
and now LHCb seems to indicate that new physics does not show up as a
large and unambiguous effects in flavor physics. This has bought into
focus the need for theoretically cleaner observables, i.e. observables
that are relatively free from hadronic uncertainties.  In the search
for new physics it is therefore crucial to effectively separate the
effect of new physics from hadronic uncertainties that can contribute
to the decay.

One of the modes that is regarded as significant in this attempt is
$B\to \kstar \ell^+\ell^-$ an angular analysis of which is known to
result in a multitude of observables \cite{Sinha:1996sv,
 Kruger:1999xa,Altmannshofer:2008dz}, each of which are
function of invariant dilepton mass $q^2$.  
Throughout, our discussions we will neglect the lepton and $s$-quark
masses, ignore the very small $CP$ violation arising within the
Standard Model~\cite{Sinha:1996sv, Kruger:1999xa} and exclude studying
the resonant region in $q^2$.  In this limit, any observable that is
chosen may eventually be expressed in terms of six real transversity
amplitudes that correspond to the three states of polarizations of
$K^*$ and the left or right chirality of the lepton $\ell^-$. We can
hence have at best six independent observables.  Several different
experiments BaBar, Belle, CDF and LHCb have studied the mode $B\to
\kstar \ell^+\ell^-$~\cite{:2009zv,:2008ju,Aubert:2006vb,CDF,
  Aaltonen:2011qs,Aaltonen:2011ja,Aaij:2011aa}, providing valuable
data as a function of the dilepton invariant mass $q^2$ by studying
uni-angular distributions. The partial branching fraction is measured
in chosen $q^2$ bins by preforming a complete angular integration. A
study of the angular distribution of the direction of the lepton in an
appropriately chosen frame (see Section~\ref{sec:III}) has also
already been done by all the four experiments to measure the
forward-backward asymmetry $\AFB$ and the longitudinal polarization
fraction $F_L$, in terms of integrated dilepton invariant mass regions
of $q^2$.  CDF and LHCb have in addition performed an angular study of
the azimuthal angle defined as the angle between the planes formed by
the leptons and the decay products of $\kstar$ i.e. $K$, $\pi$. We
will show that the $F_\perp$ helicity fraction can be obtained from a
uni-angular distribution of azimuthal angle.  Future experimental
studies at LHC-B, Belle II and Super-B will enable the study this mode
with significantly larger statistics and make possible the analysis
with multi-angular distributions and the measurement of all the
observables.

In a recent paper~\cite{Das:2012qe} it was shown that the multitude of
related observables obtained via an angular analysis in $B\to \kstar
\ell^+\ell^-$, can provide many ``clean tests'' of the Standard
Model. The hallmark of these tests is that several of them are
independent of the universal form factors $\xi_\|$ and $\xi_\perp$
required to describe the decay using heavy quark effective
theory. Indeed, in the large recoil region considered in
Ref.~\cite{Das:2012qe}, these relations are even more interesting as
they are unaffected by corrections to all orders in $\alpha_s$. {\em We
will refer to such relations that are independent of universal form
factors and are unaffected by corrections to all orders in
$\alpha_s$ as ``clean relations.''} A variety of relations were derived
which included relations between observables and form factors that are
independent of Wilson coefficients. Such relations are inherently
clean and important as they enable verification of hadronic
estimates. We show how form factor ratios can be measured directly
from ratios of helicity amplitudes measured at the zero crossings of
asymmetries with out any assumptions what so ever. Another achievement
is the derivation of a relation between observables alone, based entirely on
the assumption that the amplitudes have form given by the Standard
Model, but which is never-the-less independent of form factors and
Wilson coefficients. This relation would provide an unambiguous test
of the standard model relying purely on observables. We also presented
a clean expression for the ``effective photon vertex'' involving the
same operator that also contributes to the process $B\to\kstar
\gamma$. We emphasize that the amplitude for $B\to \kstar \gamma$
involves the universal form factor $\xi_\|$ and is inherently not
clean. It is hence some what surprising that the same vertex can be
expressed independent of the universal form-factors
in heavy quark effective theory in a way that is valid at order
$1/m_b$ to all orders in $\alpha_s$. While $C_9$ and $C_{10}$
individually depend on form factors, we find that the expression for
the ratio $C_9/C_{10}$ is clean. Based purely on the signs of the form
factors and the fact that zero crossing of the forward backward
asymmetry has been observed, we convincingly concluded that the signs
of the Wilson coefficients are in agreement with Standard Model. We
found that there exist three sets of equivalent solutions to each of the
three Wilson coefficients involving different observables. However,
only two of the sets are independent.  It was shown that the allowed
parameter space for observables is very tightly constrained in
Standard Model, thereby providing clean signals of New Physics.

In this paper we not only derive all the expressions presented in
Ref.~\cite{Das:2012qe} in detail but also derive several new
expression and constraints.  We extend the analysis to examine in
detail both the large-recoil and low-recoil regions of the $\kstar$
meson and probe special features and relations valid in the two
limits.  We present yet another new relation involving only
observables that would verify the validity of the relations between
form-factors assumed in the low-recoil region.  Under this
approximation mentioned earlier we have the six real presumably
non-zero amplitudes that are described in terms of eight combinations
of form factors and Wilson coefficients.  We elaborate in this paper
how the six observables can be used to verify Standard Model and to
distinguish possible new physics contributions from hadronic effects
which in the usual approach hinder the discovery of new physics. This
is made possible by fortunate advances in our understanding of these
form-factors that permit us to make two reliable inputs in terms of
ratios of form-factors which are well predicted at order $1/m_b$ to
all orders in $\alpha_s$ and are free from universal form factors
$\xi_\|$ and $\xi_\perp$ in heavy quark effective theory.

In this paper we briefly review the theoretical framework of $B\to
\kstar\ell^+\ell^-$ in the Standard Model in
Section~\ref{sec:framework}.  In Section~~\ref{sec:III} we express the
differential decay distribution in terms of angular variables and
helicity amplitudes. We also define observables that are directly
measurable by angular analysis. The helicity amplitudes are expressed
in terms of form factors ~\ref{sec:IV}, where we also setup essential
notations used throughout the paper.  While most of our analysis is
independent of the values of form factors we do use the various
symmetries possible in the heavy quark limit to emphasize the variety
of interesting results possible with $B\to \kstar \ell^+\ell^-$. In
Section~\ref{sec:large-recoil} we discuss the symmetry relations among
form factors arising in the large recoil limit of the $\kstar$
meson. A similar discussion for the low recoil limit is presented in
Section~\ref{sec:low-recoil}. Our model independent analysis is
described in details in Section~\ref{sec:wilson-solns}, where we also
derive the bulk of new relations. The results presented in this
section are in general valid for all $q^2$. The large recoil limit is
obtained simply by assigning the form factors the expressions or
values valid in this limit. The low recoil limit requires special
attention and is discussed in Section~\ref{sec:low-recoil-detail},
where we consider special features and derive more new interesting new
results. In Section~\ref{sec:conclusions} we summarize the main
results of the paper. The derivation involved in the solutions of
Wilson coefficients are given in Appendix~\ref{sec:appendix-1} and the
numerical values of form factors and inputs used by us are presented
in Appendix~\ref{sec:appendix-inputs}.

\section{Theoretical Framework}
\label{sec:framework}

The decay $B (p)\to \kstar(k)\ell^-(q_1)\ell^-(q_2)$ is described
within the standard model by an effective Hamiltonian
$\mathcal{H}_\eff$ that involves separation of long-distance QCD
effects from the short distance QCD and weak interaction effects. The
effective short distance Hamiltonian for $b\to s \ell^+\ell^-$
transition is well understood and given by:
\cite{Kruger:1999xa,Altmannshofer:2008dz,Buchalla:1995vs}
\begin{eqnarray}
  \label{eq:2}
  \mathcal{H}_\eff&=&\frac{G_F\alpha}{\sqrt{2}\pi} V_{tb}V_{ts}^* 
  [
  C_9^{\eff}(\bar{s}\gamma_\mu P_L b)\bar{\ell}\gamma^\mu \ell \nonumber\\
  &+& C_{10}(\bar{s}\gamma_\mu P_L b)
  \bar{\ell}\gamma^\mu\gamma_5\ell \nonumber\\ 
  &-& \frac{2C_7^{\eff}}{q^2}(\bar{s} i \sigma_{\mu\nu} q^\nu m_b P_R
  b) \bar{\ell}\gamma^\mu \ell     ]
\end{eqnarray}
where $q_\nu=q_{1\nu}+q_{2\nu}=p_\nu-k_\nu$ and we have defined
\begin{equation*}
  \label{eq:3}
  P_{L,R}=\frac{(1\mp \gamma_5)}{2}
\end{equation*}
and $q^2$ is the dilepton invariant mass squared.
In the above we have ignored the $\s$ quark mass and throughout this
paper we will ignore the lepton mass.  The Wilson coefficients
$C_{7,9,10}^{\eff}$ are evaluated at the scale $\mu=m_b=4.8\gev$ at
NNLL accuracy~\cite{Altmannshofer:2008dz}:
\begin{equation}
C_7^{\text{eff}}=-0.304,~~C_9^{\text{eff}}=4.211+Y(q^2),~~C_{10}=-4.103\nn
\end{equation}
where,
\be
C_7^{\text{eff}}=C_7-\frac{1}{3}C_3-\frac{4}{9}C_4-\frac{20}{3}C_5-\frac{80}{9}C_6\nn
\ee
and the function $Y(q^2)$
is given by~\cite{Altmannshofer:2008dz,Buras:1993xp,Buras:1994dj,Beneke:2001at}
\begin{eqnarray}
Y(q^2)&=&h(q^2,m_c)\Bigg(\frac{4}{3}C_1+C_2+6C_3+60C_5\Bigg)\nn\\
&&-\frac{1}{2} h(q^2,m_b)\Bigg(7 C_3+\frac{4}{3}C_4+76C_5+\frac{64}{3}C_6\Bigg)\nn\\
&&-\frac{1}{2}h(q^2,0)\Bigg(C_3+\frac{4}{3}C_4+16C_5+\frac{64}{3}C_6\Bigg)\nn\\
&&+\frac{4}{3}C_3+\frac{64}{9}C_5+\frac{64}{27}C_6\nn
\end{eqnarray}
The function $h(q^2,m_q)$ reads as:
\begin{eqnarray}
 h(q^2,m_q)&=&-\frac{4}{9}\Bigg(\ln\frac{m_q^2}{\mu^2}-\frac{2}{3}-y\Bigg)
-\frac{4}{9}(2+y)\sqrt{|y-1|}\nn\\&&
\times\Bigg[\Theta(1-y)\Bigg(\ln\frac{1+\sqrt{1-y}}{\sqrt{y}×}-i\frac{\pi}{2×}\Bigg)\nn\\&&
+\Theta(y-1)\text{arctan}\frac{1}{\sqrt{y-1}×}\Bigg]\nn
\end{eqnarray}
where we have defined $y=4m_q^2/q^2$, and we have neglected the small
weak phase.

The $B\rightarrow \kstar$ hadronic matrix elements of the local quark
bilinear operators $\bar{s}\gamma_\mu P_L b$ and $\bar{s} i
\sigma_{\mu\nu} q^\nu m_b P_R b$ can be parametrized in terms of the
$q^2$-dependent QCD form factors $V, A_{1,2}, T_{1,2,3}$ as
\begin{widetext}
  \begin{eqnarray}
    \label{eq:5}
 \dirmat<\bar \kstar(k)|\bar s \gamma_\mu(1-\gamma_5)b|\bar B(p)> 
  &=& -i\epsilon_\mu(m_B+m_{\kstar}) A_1(q^2) 
   + p_\mu(\epsilon^*.q)\frac{2\,A_2(q^2)}{m_B+m_{\kstar}} \nonumber\\
  &&\qquad  
    +i\epsilon_{\mu \nu \rho \sigma}\epsilon^{*\nu}p^\rho k^\sigma
   \frac{2\,V(q^2)}{m_B+m_{\kstar}}\\ 
\dirmat<\bar \kstar(k)|\bar s \sigma_{\mu\nu}q^\nu(1+\gamma_5)b|\bar B(p)> 
  &=&i\epsilon_{\mu\nu\rho\sigma}\epsilon^{*\nu}p^\rho k^\sigma 2\,T_1(q^2)
   + T_2(q^2)[\epsilon^*_\mu(m_B^2-m_{\kstar}^2)-2\,(\epsilon^*.q) p_\mu]
   \nonumber\\ 
   &&\qquad  -(\epsilon^*.q)\,q^2\frac{2\,T_3(q^2)}{m_B^2-m_{\kstar}^2}p_\mu~,
  \end{eqnarray}
\end{widetext}
where, $q_\nu=p_\nu-k_\nu$. We have dropped terms proportional to
$q_\mu$ since the terms $q_\mu\bar{\ell}\gamma^\mu\gamma_5\ell $ and
$q_\mu\bar{\ell}\gamma^\mu\ell $ do not contribute in the limit of
vanishing lepton mass.  The form factors have been studied using QCD
sum rules on the light cone, QCD factorization in the heavy quark
limit, soft-collinear theory and using operator product expansion that
is valid for large dilepton mass $\sqrt{q^2}$.  The decay $B\to \kstar
\ell^+\ell^-$ has the advantage that it can be studied as a function
of the dilepton mass or $q^2$. If one excludes the resonant region and
the very small CP violation arising within SM, all the Wilson
coefficients and form factors contributing to the decay are real. In
this paper as mentioned above we will make both these assumptions.

The complete angular distribution requires the polarization of
$\kstar$ or a study of the angular distribution of the $\kstar$ decay
into $K\pi$. This is readily done in the narrow width approximation
for the $\kstar$ since the decay of $\kstar\to K\pi$ is itself well
understood in terms of an effective Lagrangian. The resulting matrix
elements is described in a model independent approach in terms of
three reliably calculable effective Wilson coefficients that represent
short distance contributions and six (in the limit of vanishing lepton
mass) $B\to \kstar$ form factors.  The $B\to K\pi\ell^+\ell^-$ matrix
element can hence, be written as

\begin{widetext}
\begin{eqnarray}
  \mathcal{M} = \frac{G_F\alpha}{\sqrt{2}\pi}V_{tb}V_{ts}^*
  \bigg\{\bigg[
  C_9^\text{eff}\braket{K \pi}{\bar{s}\gamma^{\mu}P_Lb}{\bar
    B}{\bar{l}\gamma_\mu l} 
  + C_{10}\braket{K \pi}{\bar{s}\gamma^{\mu}P_Lb}{\bar B}
  \bar{l}\gamma_\mu\gamma_5l 
  -\frac{2C_7m_b}{q^2}\braket{K\pi}{\bar{s}i\sigma_{\mu\nu}q^{\nu}P_Rb}
  {\bar  B}\bar{l}\gamma_\mu l 
  \bigg\}\bigg].
\end{eqnarray}
\end{widetext}

\section{Angular Distribution and Observables}
\label{sec:III}

The decay $\bar{B}(p)\to K^{*}(k)\ell^+(q_1)\ell^-(q_1)$ with
$\kstar(k)\to K(k_1)\pi(k_2)$ on the mass shell, is completely described
by four independent kinematic variables. These are the lepton-pair
invariant mass squared $q^2=(q_1+q_2)^2$, the angle $\phi$ between the
decay planes formed by $\ell^+\ell^-$ and $K\pi$ respectively and the
angles $\theta_\ell$ and $\theta_K$ defined as follows: assuming that
the $\kstar$ has a momentum along the positive $z$ direction in $B$ rest
frame, $\theta_K$ is the angle between the $K$ and the $+z$ axis and
$\theta_\ell$ is the angle of the $\ell^-$ with the $+z$ axis.  The
differential decay distribution of $B\to \kstar\ell^+\ell^-$ can be
written as
\begin{widetext}
\begin{align}
 \label{eq:helicity}
  \frac{d^4\Gamma(B\to \kstar\ell^+\ell^-)}{dq^2\, d\cos\thl\,
    d\cos\thK\, d\phi}
& =
  I(q^2,\thl,\thK,\phi)= \frac{9}{32\pi}\Big[ I_1^s \sin^2\thK + I_1^c
  \cos^2\thK + (I_2^s \sin^2\thK + I_2^c \cos^2\thK) \cos 2\thl
  \nonumber \\ &+ I_3 \sin^2\thK \sin^2\thl \cos 2\phi + I_4 \sin
  2\thK \sin 2\thl \cos\phi + I_5 \sin 2\thK \sin\thl \cos\phi + I_6^s
  \sin^2\thK \cos\thl\nonumber \\ &+ I_7 \sin 2\thK \sin\thl \sin\phi
  + I_8 \sin 2\thK \sin 2\thl \sin\phi + I_9 \sin^2\thK \sin^2\thl
  \sin 2\phi\Big]\,.
\end{align}
\end{widetext}
We note that $I$'s are $q^2$ dependent but we have chosen to suppress
the explicit $q^2$ dependence for simplicity. Throughout the paper we
will not explicitly state the $q^2$ dependence of observables and
variables, however, the dependence is implicit.  A study of the
angular distribution of the decay will allow us to measures all the
$I$'s.  Since the $\kstar$ in ${B\to \kstar \ell^+\ell^-}$ decay, is
created on shell it has three polarization states.  Hence, we can
express $I$'s explicitly in terms of the six transversity amplitudes
${\cal A}_{\perp,\parallel,0}^{L,R}$, where $\perp$, $\|$ and $0$
represent the polarizations and $L$, $R$ denote the chirality of the
lepton $\ell^{-}$. We can write the nine observables explicitly in
terms of the six transversity amplitudes ${\cal
  A}_{\perp,\parallel,0}^{L,R}$ as:
\begin{align}
  I_1^s & = \frac{3}{4} \Big[|\apeL|^2 + |\apaL|^2 +
    (L\to R) \Big], \nn\\ 
  I_1^c & =  \Big[|\azeL|^2 + (L\to R)\Big],\nn\\
  I_2^s & = \frac{1}{4}\Big[ |\apeL|^2+ |\apaL|^2 + (L\to
    R)\Big],\nn\\ 
  I_2^c & = - \Big[|\azeL|^2 + (L\to R)\Big],\nn\\
  I_3 & = \frac{1}{2}\Big[ |\apeL|^2 - |\apaL|^2  + (L\to
    R)\Big],\nn\\  
  I_4 & = \frac{1}{\sqrt{2}}\Big[\Re (\azeL^{}\apaL^*) +
    (L\to R)\Big],\nn\\
  I_5 & = \sqrt{2}\Big[\Re(\azeL^{}\apeL^*) - (L\to R)
\Big]\nn\\
  I_6^s  & = 2\Big[\Re (\apaL^{}\apeL^*) - (L\to R) \Big],\nn\\
  I_7 & = \sqrt{2}\Big[\Im (\azeL^{}\apaL^*) - (L\to R)
  \Big],\nn\\
  I_8 & = \frac{1}{\sqrt{2}}\Big[\Im(\azeL^{}\apeL^*) +
    (L\to R)\Big],\nn\\
  I_9 & =\Big[\Im (\apaL^{*}\apeL) + (L\to R)\Big],\nn
\end{align}
In the above we have ignored the lepton mass.  As mentioned above we
will assume that the resonant region is excluded in the analysis and that
$CP$ violation arising within the standard model is negligible.  In
the absence of $CP$ violation the conjugate mode
$\bar{B}\to\bar{K}^*\ell^+\ell^-$ has an identical decay distribution
except that $I_{5,6,8,9}$ switch signs to become $-I_{5,6,8,9}$ in the
differential decay distribution~\cite{Sinha:1996sv,
  Kruger:1999xa}. The helicity amplitudes ${\cal
  A}_{\perp,\parallel,0}^{L,R}$, are then all real and only six of the
$I$'s can be non-zero and independent. In fact, it is easy to see that
$I_7$, $I_8$ and $I_9$ must vanish in the limit of vanishing $CP$
violation.

The explicit form of these transversity amplitudes are
\begin{widetext}
\begin{subequations}
\begin{align}
\label{trans-amp1}
{\cal A}_{\perp}^{L,R}  =&  N \sqrt{2} \sqrt{\lambda(m_B^2, m_\kstar^2,q^2)} \bigg[ 
\left[ (C_9^\eff \mp C_{10}^{\eff}) \right] \frac{ V(q^2) }{ m_B + m_\kstar} 
 + \frac{2m_b}{q^2} C_7^\eff T_1(q^2)
\bigg],\\
{\cal A}_{\parallel}^{L,R}  =& - N \sqrt{2}(m_B^2 - m_\kstar^2) \bigg[ \left[ (C_9^\eff \mp C_{10}^\eff ) \right] 
\frac{A_1(q^2)}{m_B-m_\kstar}+\frac{2 m_b}{q^2}C_7^\eff  T_2(q^2)
\bigg],\\
\label{trans-amp3}
{\cal A}_{0}^{L,R}  =&  - \frac{N}{2 m_\kstar \sqrt{q^2}}  \bigg(
 \left[ (C_9^\eff  \mp C_{10}^\eff) \right]
 \times 
\Big[ (m_B^2 - m_\kstar^2 - q^2) ( m_B + m_\kstar) A_1(q^2) 
 -\lambda(m_B^2, m_\kstar^2,q^2) \frac{A_2(q^2)}{m_B + m_\kstar}
\Big] 
\nonumber\\
& \qquad + {2 m_b} C_7^\eff  \Big[
 (m_B^2 + 3 m_\kstar^2 - q^2) T_2(q^2)
-\frac{\lambda(m_B^2, m_\kstar^2,q^2)}{m_B^2 - m_\kstar^2} T_3(q^2) \Big]
\bigg)
\end{align}
\end{subequations}
where,
\begin{equation}
N= V_{tb}^{\vphantom{*}}V_{ts}^* \left[\frac{G_F^2 \alpha^2}{3\cdot 2^{10}\pi^5 m_B^3}
 q^2 \sqrt{\lambda(m_B^2, m_\kstar^2,q^2)}\right]^{1/2},
\end{equation}
with ${\lambda}(m_B^2, m_\kstar^2,q^2)= m_B^4 + m_{\kstar}^4 + q^4 - 2 (m_B^2 m_{\kstar}^2+ m_{\kstar}^2
q^2 + m_B^2 q^2)$. We note that the helicity amplitudes ${\cal
  A}_{\perp,\parallel,0}^{L,R}$ are functions of $q^2$, for simplicity
we have suppressed the explicit dependence on $q^2$.
\begin{align}
  \label{eq:13}
  I(q^2,\thl,\thK,\phi)&= \dsp \frac{9}{16\pi} \Bigg[
  \frac{\big(|\apeL|^2 +|\apeR|^2+|\apaL|^2 +|\apaR|^2\big)}{4}
  \sin^2\thK (1+\cos^2\thl) +\big(|\azeL|^2 +|\azeR|^2\big) \cos^2\thK
  \sin^2\thl \nn\\&+ \frac{ \big(|\apeL|^2 +|\apeR|^2-|\apaL|^2
    -|\apaR|^2\big)}{4}\cos2\phi \sin^2\thK \sin^2\thl+ \Re
  (\apaL\apeL^*-\apaR\apeR^*)\cos\thl\sin^2\thK \nn \\ & + \frac{\Re
    (\azeL\apeL^*-\azeR\apeR^*)}{\sqrt{2}}\cos\phi\sin\thl\sin(2\thK)
  +\frac{\Re(\azeL\apaL^*+\azeR\apaR^*)}{2\sqrt{2}}
  \cos\phi\sin(2\thl)\sin(2\thK) \Bigg].
\end{align}
It is easy to see that integration over $\cos\thK$, $\cos\thl$ and
$\phi$ results in the differential decay rate with respect to the
invariant lepton mass, which is given by the sum of the modulus
squared of all the transversity amplitudes at the same invariant lepton
mass: 
\end{widetext}
\begin{equation}
  \label{eq:1}
  \frac{d\Gamma}{dq^2}=\sum_{\lambda=0,\|,\perp}(|{\cal
    A}_\lambda^L|^2+|{\cal  A}_\lambda^R|^2) 
\end{equation}
%
It is obvious from Eq.~(\ref{eq:13}) that a complete study of the
angular distribution will allow us to measure six observables. We
define the relevant observables to be the three helicity fractions
defined as follows:
\begin{subequations}
\begin{align}
  \label{F0}
  F_L&= \dsp \frac{|\azeL|^2 +|\azeR|^2}{\dsp \Gamma_{\!f}}~,\\
  \label{Fparallel}
  F_\|&= \dsp\frac{|\apaL|^2 +|\apaR|^2}{\dsp\Gamma_{\!f}}~,\\
  \label{Fperp}
  F_\perp&= \frac{|\apeL|^2 +|\apeR|^2}{\dsp\Gamma_{\!f}}~,
\end{align}
\end{subequations}
where $ \Gamma_{\!\!
  f}\equiv\sum_\lambda(|{\cal A}_\lambda^L|^2+|{\cal
  A}_\lambda^R|^2)$ and, $F_L+F_\|+F_\perp=1$. The well known forward--backward asymmetry $\AFB$,
\begin{equation}
  \label{eq:AFB}
  \AFB=\dsp\frac{\Big[\dsp\int_0^1-\dsp \int_{-1}^0\Big]\dsp
    d\cos\thl \frac{d^2 
      (\Gamma+\bar{\Gamma})}{d q^2d\cos\thl}}{\dsp\int_{-1}^1 \dsp d\cos\thl
    \frac{d^2 (\Gamma+\bar{\Gamma})}{d q^2d\cos\thl}}~,
\end{equation}
and two new angular asymmetries,
\begin{widetext}
\begin{align}
  \label{eq:A4}
  A_{4}=&\frac{\Big[\dsp\int_{\pi/2}^{3\pi/2}d\phi-\dsp\int_{-\pi/2}^{\pi/2}d\phi \Big]
    \Big[\dsp\int_0^1 d\cos\thK-\dsp\int_{-1}^0 d\cos\thK \Big]
    \Big[\int_0^1 d\cos\thl-\int_{-1}^0 d\cos\thl \Big]\dsp
    \frac{d^4(\Gamma-\bar{\Gamma})}{dq^2d\cos\thl d\cos\thK d\phi}} 
  {\dsp\int_0^{2\pi}d\phi\int_{-1}^1d\cos\thK \int_{-1}^1d\cos\thl\,
    \frac{d^4(\Gamma+\bar{\Gamma})}{dq^2d\cos\thl d\cos\thK d\phi}}~,\\
  \label{eq:A5}
  A_{5}=&\frac{\int_{-1}^1
    d\cos\thl\Big[\int_{\pi/2}^{3\pi/2}d\phi-\int_{-\pi/2}^{\pi/2}d\phi
    \Big] \Big[\int_0^1 d\cos\thK-\int_{-1}^0 d\cos\thK \Big]
    \dsp\frac{d^4(\Gamma+\bar{\Gamma})}{dq^2d\cos\thl d\cos\thK d\phi} }
  {\int_{-1}^1d\cos\thl\int_0^{2\pi}d\phi\int_{-1}^1d\cos\thK \,
   \dsp \frac{d^4(\Gamma+\bar{\Gamma})}{dq^2d\cos\thl d\cos\thK d\phi}}~.
\end{align}
\end{widetext}
$\AFB$, $A_4$ and $A_5$ can be written directly in terms of the
transversity amplitudes as follows:
\begin{eqnarray}
A_4&=&\frac{\sqrt{2}}{\pi}\frac{\Re(\mathcal{A}_0^L\mathcal{A}_\|^{L*})+\Re(\mathcal{A}_0^R\mathcal{A}_\|^{R*})}{\Gamma_{\!f}},\\
A_5&=&\frac{3}{2\sqrt{2}}\frac{\Re(\mathcal{A}_0^L\mathcal{A}_\perp^L-\mathcal{A}_0^R\mathcal{A}_\perp^R)}{\Gamma_{\!f}},\\
A_{\text FB}&=&\frac{3}{2}\frac{\Re({\cal A}_\|^L{\cal A}_\perp^L-{\cal
    A}_\|^R{\cal A}_\perp^R)}{\Gamma_{\!f}}.
\end{eqnarray}

A complete angular analysis requires much larger data set than are
currently analyzed, hence angular distributions in terms of only one
angular variable have been studied.  The angular distribution as a
function of $q^2$ and $\cos\theta_\ell$ with $\phi$ and $\cos\theta_K$
integrated out is given by:
\begin{align}
  \label{eq:12}
  \frac{ d^2\Gamma}{d q^2 d\cos\theta_\ell} =&\Gamma
  \Big[\AFB \cos\theta_\ell+\frac{3}{8}(1-F_L)\,
  (1+\cos^2\theta_\ell) \nn\\&\qquad+\frac{3}{4}F_L (1-\cos^2\theta_\ell)\Big]~.
\end{align}
Angular analysis in terms of $\cos\thl$ enables the measurement of
both $F_L$ the longitudinal helicity fraction and the
forward--backward asymmetry $\AFB$.  The other helicity fractions
$F_\perp$ or $F_\|$ can be measured from the angular distributions as
well but it has been believed that one need to perform a full angular
analysis. It is, however, easy to see that a combination of $F_L$ and
$F_\perp$ can be measured if the angular distribution in terms of
$\phi$ is studied. The angular distribution in
$\phi$ is given by:
\begin{equation}
  \label{eq:phi_dist}
 \frac{ d^2\Gamma}{d q^2 d\phi} =\frac{\Gamma}{2\pi}
 \Big[1-\frac{1-F_L-2F_\perp}{2}\cos 2\phi+I_9 \sin 2\phi \Big].~~ 
\end{equation}
The distribution in $\phi$ allows us to measure $1-F_L-2F_\perp$. If
$F_L$ is measured independently one can obtain $F_\perp$. The
distribution also allows us to measure $I_9$, which is immeasurably
small in SM~\cite{Sinha:1996sv}, and assumed to be zero in our
study. Recently the angular analysis in $\phi$ has been
studied~\cite{Aaltonen:2011ja,Moriond-talk} by CDF and LHCb
collaborations. In the next section we will show that $1-F_L-2F_\perp$ is also
small in the SM as a consequence of heavy quark effective theory. We
will conclude in Sec.~\ref{sec:framework} that the angular
distribution will be almost constant for $q^2\approx 0$, with small
variation in $\cos\phi$ at large $q^2$.

There is yet another technique to measure $F_\perp$ which involves
studying angular distributions in terms of only one angular
variable. However, this approach requires independent analysis in the
transversity frame defined with $\jpsi$ at rest. In this frame the
lepton makes an angle $\theta_{\rm{tr}}$ with the $z$-axis. The
expression for the differential decay rate as a function of
$\cos\theta_{\rm{tr}}$ is given by:
\begin{align}
  \label{eq:14}
  \frac{d\Gamma}{d q^2d\cos\theta_{\rm{tr}}}=&\Gamma\Big[\frac{3}{8}
  (1-F_\perp) (1+\cos^2\theta_{\rm{tr}})\nn\\&\qquad +\frac{3}{4} F_\perp
  (1-\cos^2\theta_{\rm{tr}})\Big]   
\end{align}
Clearly, $F_\perp$ the perpendicular polarization fraction can be
measured from a fit to $\cos\theta_{\rm{tr}}$ in the transversity frame. The
errors in $F_L$ and $F_\perp$ measured in this fashion will be
correlated and the correlation will have to be taken care of.

\section{Notation: Observables in terms of Form-Factors}
\label{sec:IV}

The six transversity amplitudes in Eq.~(\ref{trans-amp1}) to
(\ref{trans-amp3}) are written in terms of the Wilson coefficients and
the form factors in most general form as,
\begin{subequations}
\begin{align}
\label{amp-def1}
 \mathcal{A}_\perp^{ L,R}&=C_{L,R}\mathcal{F}_\perp-\widetilde{\mathcal{G}}_\perp\\
\label{amp-def2}
 \mathcal{A}_\parallel^{ L,R}&=C_{L,R}\mathcal{F}_\parallel-\widetilde{\mathcal{G}}_\parallel\\
\label{amp-def3}
 \mathcal{A}_0^{ L,R}&=C_{L,R}\mathcal{F}_0-\widetilde{\mathcal{G}}_0
\end{align}
\end{subequations}
where to leading order, $C_{L,R}=C_9^\eff\mp C_{10}$ and
$\widetilde{\mathcal{G}}_\lambda= C_7^\eff
\mathcal{G}_\lambda$. $C_7^\eff$, $C_9^\eff$ and $C_{10}$ are the
Wilson coefficients that represent short distance
corrections. $\mathcal{F}_\lambda$ and
$\widetilde{\mathcal{G}}_\lambda$ are defined below in terms of
$q^2$-dependent QCD form factors that parameterize the $B\to \kstar$
matrix element~\cite{Altmannshofer:2008dz} and are suitably defined to
include both factorizable and non-factorizable contributions at any
given order. The treatment of the form factors depends largely on the
recoil energy of the $\kstar$ or equivalently $q^2$ and will have to
be treated differently in the limit of heavy quark effective theory.
In the large recoil limit (see Section~\ref{sec:large-recoil}) the
next to leading order effects including factorizable and
non-factorizable corrections  can be parametrically included by
replacements $C_9^\eff\to C_9$ and defining
$\widetilde{\mathcal{G}}_\lambda = C_7^\eff \mathcal{G}_\lambda+
\cdots$, with the dots representing the next to leading and higher
order terms.  Hence the Wilson coefficient and form factor can be
lumped together into a single factor $\widetilde{\cal G}_\lambda$.  We
note that even at leading order it is impossible to determine
$C_7^\eff$ with the value of $\mathcal{G}_\lambda$ being determined.
The treatment of form factors in the low recoil limit (see
Section~\ref{sec:low-recoil} for details) differs significantly from
the large recoil. In the low recoil limit the leading corrections are
the non-perturbative effects up to and including terms suppressed by
$\lqcd/Q$ (where $Q=\{m_b,\sqrt{q^2}\}$)and and include the
next-to-leading order corrections from the charm quark mass $m_c$ and
the strong coupling at $\mathcal{O}(m^2_c/Q^2,\alpha_s)$.

The form factors $\mathcal{F}_\lambda$ and
$\widetilde{\mathcal{G}}_\lambda$ can be related to the form factors
$V$, $A_{1,2}$ and $T_{1,2,3}$ introduced in Eqs.~(\ref{trans-amp1})
-- (\ref{trans-amp3}) ~\cite{Altmannshofer:2008dz} by comparing these
expressions for ${\cal A}_\lambda^{L,R}$ with those in
Eqs.~{(\ref{amp-def1}) -- (\ref{amp-def3})}. Including higher order QCD
correction and non-factorizable corrections $\mathcal{F}_\lambda$ and
$\widetilde{\mathcal{G}}_\lambda$ can be written as:
\begin{subequations}
\label{eq:form-factors}
\begin{align}
  \widetilde{\mathcal{G}}_\perp&=-\!N\sqrt{2\lambda(m_B^2,
    m_\kstar^2,q^2)} \frac{2m_b}{q^2}C_7^\eff T_1(q^2)+\cdots\\    
  \widetilde{\mathcal{G}}_\|&=N\sqrt{2}(m_B^2-m_\kstar^2)
  \frac{2m_b}{q^2}C_7^\eff T_2(q^2)+\cdots\\ 
  \mathcal{F}_\perp&=N\sqrt{2\lambda(m_B^2, m_\kstar^2,q^2)}\,
  \frac{V(q^2)}{m_B+m_\kstar}\\  
  \mathcal{F}_\|&=-\!N\sqrt{2}(m_B+m_\kstar)A_1(q^2)\\
  \mathcal{F}_0&=\frac{-\!N}{2m_\kstar\sqrt{q^2}}
  \Big[(m_B^2-m_\kstar^2-q^2)(m_B +  m_\kstar)A_1(q^2) \nn \\ 
  &\quad-\lambda(m_B^2, m_\kstar^2,q^2)\frac{A_2(q^2)}{m_B+m_\kstar}\Big]\\ 
  \widetilde{\mathcal{G}}_0&=\frac{N}{2m_\kstar\sqrt{q^2}}
  2m_b\Big[(m_B^2+3m_\kstar^2-q^2) C_7^\eff T_2(q^2) \nn \\ 
  &\quad-\lambda(m_B^2, m_\kstar^2,q^2)\frac{C_7^\eff T_3(q^2)}
  {m_B^2-m_\kstar^2}\Big]+\cdots.
\end{align}  
\end{subequations}

With the help of Eq.~(\ref{amp-def1}) to (\ref{amp-def3}) the
observables $F_L$, $F_\|$, $F_\perp$, $\AFB$, $A_4$ and $A_5$ can be
written in terms of the Wilson coefficients and from factors as:
\begin{subequations}
\begin{align}
\label{eq1:FL}
&F_L\Gf= 2(C_9^2+C_{10}^2)\mathcal{F}_0^2+2\widetilde{\mathcal{G}}_0^2-
4C_9\mathcal{F}_0\widetilde{\mathcal{G}}_0\\
\label{eq1:Fparallel}
&F_\|\Gf= 2(C_9^2+C_{10}^2)\mathcal{F}_\|^2+2\widetilde{\mathcal{G}}_\|^2-
4C_9\mathcal{F}_\|\widetilde{\mathcal{G}}_\| \\
\label{eq1:Fperp}
&F_\perp\Gf=2(C_9^2+C_{10}^2)\mathcal{F}_\perp^2+2\widetilde{\mathcal{G}}_\perp^2-
4C_9\mathcal{F}_\perp\widetilde{\mathcal{G}}_\perp \\
\label{eq1:A4}
&\frac{\pi A_4\Gf}{2\sqrt{2}}= \widetilde{\mathcal{G}}_\|\widetilde{\mathcal{G}}_0 + (C_9^2 + C_{10}^2)\mathcal{F}_0\mathcal{F}_\|\nonumber\\
&\qquad\qquad -C_9 (\mathcal{F}_\|\widetilde{\mathcal{G}}_0+\widetilde{\mathcal{G}}_\|\mathcal{F}_0)\\
\label{eq1:A5}
&\frac{\sqrt{2}A_5\Gf}{3}=C_{10} (\mathcal{F}_\perp\widetilde{\mathcal{G}}_0+\widetilde{\mathcal{G}}_\perp\mathcal{F}_0)- 
2C_9C_{10}\mathcal{F}_0\mathcal{F}_\perp\\
\label{eq1:AFB}
&\frac{A_{\text FB}\Gf}{3}=C_{10}(\mathcal{F}_\|\widetilde{\mathcal{G}}_\perp
+\mathcal{F}_\perp\widetilde{\mathcal{G}}_\|)-2C_9C_{10} \mathcal{F}_\|\mathcal{F}_\perp
\end{align}
\end{subequations}
We use Eq.~(\ref{eq1:FL})--(\ref{eq1:AFB}) to solve the Wilson
coefficients in terms of the observables and the form factors. This
solutions are achieved by defining new variables
\begin{subequations}
  \label{eq:uvwz}
  \begin{align}
    r_\|=&\frac{\widetilde{\mathcal{G}}_\|}{\mathcal{F}_\|}-C_9,\\
    r_\perp=&\frac{\widetilde{\mathcal{G}}_\perp}{\mathcal{F}_\perp}-C_9,\\
    r_0=&\frac{\widetilde{\mathcal{G}}_0}{\mathcal{F}_0}-C_9,\\
    r_{\!\wedge}=&\frac{\widetilde{\mathcal{G}}_\|+\widetilde{\mathcal{G}}_0}
    {\mathcal{F}_\|+\mathcal{F}_0}-C_9. 
 \end{align}
\end{subequations}
 In terms of these new variables $r_\|$, $r_\perp$,
$r_0$ and $r_{\!\wedge}$ the observables in Eq.~(\ref{eq1:FL})--(\ref{eq1:AFB}))
can be written conveniently as:
\begin{subequations}
\begin{align}
\label{eq2:Fparallel}
&F_\|\Gf =2\mathcal{F}_\|^2\big(r_\|^2+C_{10}^2\big)\\[1.5ex]
\label{eq2:Fperp}
&F_\perp\Gf = 2\mathcal{F}_\perp^2 \big(r_\perp^2+C_{10}^2\big)\\[1.5ex]
\label{eq2:FL}
&F_L\Gf =2\mathcal{F}_0^2 \big(r_0^2+C_{10}^2\big)\\[1.5ex]
\label{eq2:A4}
&(F_L+F_\|+\sqrt{2}\pi A_4)\Gf = 2(\mathcal{F}_0+
  \mathcal{F}_\|)^2 \big(r_{\!\wedge}^2+C_{10}^2\big)\\[1.5ex]
\label{eq2:A5}
&\sqrt{2}A_5\Gf = 3\mathcal{F}_\perp\mathcal{F}_0 C_{10}\big(r_0+r_\perp\big)\\[1.5ex]
\label{eq2:AFB}
&\AFB\Gf= 3\mathcal{F}_\perp\mathcal{F}_\| C_{10}\big(r_\|+r_\perp\big)\\[1.5ex]
\label{eq2:A5AFB}
&\big(\AFB+\!\sqrt{2}A_5\big)\Gf= 3 \mathcal{F}_\perp(\mathcal{F}_0+
\mathcal{F}_\|)C_{10}\big(r_{\!\wedge}+r_\perp\big)
\end{align}
\end{subequations}
It is easy to see that only six of the seven equations above are
independent; the last Eq.~(\ref{eq2:A5AFB}) is easily obtained from
Eqs.~(\ref{eq2:A5}) and (\ref{eq2:AFB}).  Considerable notational
simplification is achieved by defining the following six ratios of
form factors:
\begin{eqnarray}
  \label{eq:P_1}
  \mathsf{P_1}&=&\frac{\mathcal{F}_\perp}{\mathcal{F}_\|},~~
  \label{eq:P_2}
  \mathsf{P_2}=\dsp \frac{\mathcal{F}_\perp}{\mathcal{F}_0},~
  \label{eq:P_3}
  \mathsf{P_3}=\frac{\mathcal{F}_\perp}{\mathcal{F}_0+\mathcal{F}_\|}=
  \frac{\mathsf{P_1}\mathsf{P_2}}{\mathsf{P_1}+\mathsf{P_2}}\\ 
  \label{eq:P_1p}
  \mathsf{P_1'}&=&\dsp\frac{\widetilde{\mathcal{G}}_\perp}{\widetilde{\mathcal{G}}_\|},~~
  \label{eq:P_2p}
  \mathsf{P_2'}=\frac{\widetilde{\mathcal{G}}_\perp}{\widetilde{\mathcal{G}}_0},~~
  \label{eq:P_3p}
  \mathsf{P_3'}=\frac{\widetilde{\mathcal{G}}_\perp}
  {\widetilde{\mathcal{G}}_\|+\widetilde{\mathcal{G}}_0}=
  \frac{\mathsf{P_1'}\mathsf{P_2'}}{\mathsf{P_1'}+\mathsf{P_2'}}.
\end{eqnarray}
Clearly, $r_{\!\wedge}$ introduced in Eq.~\eqref{eq:uvwz} is not
independent and is easily obtained from a combination of $r_\|$ and
$r_0$. The expression for $r_\wedge$ in terms of $r_\|$ and $r_0$ and
form factors ratios $\mathsf{P_1}$ and $\mathsf{P_2}$ is easily
derived to be,
\begin{equation}
  \label{eq:3e}
  r_{\!\wedge}=\frac{r_\|\mathsf{P_2}+r_0 \mathsf{P_1}}{\mathsf{P_2}+\mathsf{P_1}}
\end{equation}

Naively we have nine theoretical parameters, the three Wilson
coefficients $C_7$, $C_9$ and $C_{10}$ and the six form factors
$\mathcal{F}_0$, $\mathcal{F}_\|$, $\mathcal{F}_\perp$,
$\mathcal{G}_0$, $\mathcal{G}_\|$ and $\mathcal{G}_\perp$, describing
the six observables $\Gf$, $F_L$, $F_\perp$, $A_4$, $A_5$ and
$\AFB$. As mentioned earlier $C_7^\eff$ and $\mathcal{G}_\lambda$
cannot be distinguished and they are lumped together beyond leading
order, so that we have only eight independent theoretical parameters,
the two Wilson coefficients $C_9$ and $C_{10}$ and six form factors
$\mathcal{F}_0$, $\mathcal{F}_\|$, $\mathcal{F}_\perp$,
$\widetilde{\mathcal{G}}_0$, $\widetilde{\mathcal{G}}_\|$ and
$\widetilde{\mathcal{G}}_\perp$.  It is obvious that with two
theoretical inputs in addition to the observables we should in
principle be able to solve for the remaining six theoretical
parameters purely in terms of these two reliable inputs and
observables.  Fortunately, advances in our understanding of these
form-factors permit us a judicious choice of the two reliable inputs
which depends on the energy of recoiling $K^*$ (or equivalent
$q^2$). At large recoil the two inputs are the ratios of form-factors
$\mathsf{P_1}$ and $\mathsf{P'_1}$ which are well predicted at next to
leading order in QCD corrections and free from form factors $\xi_\|$
and $\xi_\perp$ in heavy quark effective theory.  While the choice of
$\mathsf{P_1}$ and $\mathsf{P'_1}$ works well at low $q^2$, at low
recoil another condition equating the three ratios
$\widetilde{\mathcal G}_\lambda/ \mathcal{F}_\lambda$ for
$\lambda=\{0,\|,\perp\}$ is needed.

The decay mode $B\to \kstar\ell^+\ell^-$ have been studied with form
factors calculated in different models. For example, in
Refs.~\cite{exclusiveB} the mode have been studied using light-cone
hadron distribution amplitudes~\cite{exclusive} combined with QCD sum
rules on the light cone~\cite{LCSR}. In Refs.~\cite{Ali:1999mm} the 
mode was studied using naive factorization and QCD sum rules
on the light cone. In in Refs.~\cite{Beneke:2001at,
  Bobeth:2008ij,Egede:2008uy} it has been studied in the heavy quark
limit using QCD
factorization\cite{Beneke:1999br,Beneke:2000ry}. Soft-collinear
effective
theory~\cite{Bauer:2000ew,Bauer:2000yr,Bauer:2001ct,Beneke:2002ph,Hill:2002vw}
that is valid for small $q^2$ (large recoil of $\kstar$) has been used
to study the decay in Ref.~\cite{Ali06}, while operator product
expansion that is valid for large $q^2$ (low recoil) has been studied
in~\cite{hep-ph/0404250}.

In the next two subsections that follow, we will digress to consider
the $B\to \kstar\ell^+\ell^-$ form factors and their relations in the two
limits of the $K^*$ meson recoil energy. We will present our model
independent analysis in the next section
(Sec.~\ref{sec:wilson-solns}). We will assume $\mathsf{P_1}$ and
$\mathsf{P_1'}$ as inputs for most of the paper as the results are
valid throughout the $q^2$ domain, except when
$\mathsf{P_1}=\mathsf{P_1'}$. We will show that the validity of the
large recoil limit approximation can be verified by a direct
measurement of $\mathsf{P_1}$ in terms of helicity fractions, at the
zero-crossing point of $\AFB$, i.e., at $\AFB=0$.  The low recoil
limit is considered at the end in Sec.~\ref{sec:low-recoil-detail},
where we will also examine the special case
$\mathsf{P_1}=\mathsf{P_1'}$. The validity of the low recoil limit can
also be tested through a relation derived purely between observables
which is valid only in the low recoil limit.  In both the recoil
regions we derive several important relations between observables,
Wilson coefficients and form factors.  We find that the six
observables are not independent as there exists one constraint
relation that involves observables alone and hence free from the
details of recoil energy approximation as well. As a consequence we
find that $\mathcal{F}_\|$ cannot be solved for and must be taken as
an additional input as well.

\subsection{Form factor in the large recoil limit}
\label{sec:large-recoil}

In $B\to K^*$ transition at low $q^2$, the light meson
$K^*$ carries a large energy $E_{\kstar}$. Since the initial
$B$ meson contains the heavy $b$ quark, in this limit 
the form factors can be expanded in small
ratios of $\lqcd/m_b$ and $\lqcd/E_\kstar$ 
\cite{hep-ph/9812358}. This reduces the independent
$B\to K^*$ form factors from seven to two universal 
form-factors $\xi_\perp$ and $\xi_\|$. In terms of these 
two form-factors, the seven form factors can be written 
up to $1/m_b$ and $\alpha_s$ corrections as \cite{hep-ph/9812358,BF00}:
\begin{subequations}
\begin{align}
\label{Beneke-Rln1}
A_1(q^2)&=\frac{2E_{K^*}}{m_B+m_{K^*}}\xi_\perp(E_{K^*})\\
A_2(q^2)&=\frac{m_B}{m_B-m_{K^*}}[\xi_\perp(E_{K^*})-\xi_\|(E_{K^*})]\\
A_0(q^2)&=\frac{E_{K^*}}{m_{K^*}}\xi_\|(E_{K^*})\\
V(q^2)&=\frac{m_B+m_{K^*}}{m_B}\xi_\perp(E_{K^*})\\
T_1(q^2)&=\xi_\perp(E_{K^*})\\
T_2(q^2)&=\frac{2E_{K^*}}{m_B}\xi_\perp(E_{K^*})\\
\label{Beneke-Rln7}
T_3(q^2)&=\xi_\perp(E_{K^*})-\xi_\|(E_{K^*}),
\end{align}
\end{subequations}
where, $E_{K^*}$ is the energy of the $K^*$ meson,
\begin{equation*}
E_{K^*}=\frac{m_B^2+m_{K^*}^2-q^2}{2m_B}\nn.
\end{equation*}
We note that the form factor $A_0$ does not appear in our expressions
in the massless lepton limit.  In the large recoil limit $T_2/T_1$ and
$V/A_1$ are well predicted and reduce to the simple form
\begin{eqnarray}
\label{eq:f1}
 \frac{T_2}{T_1}&=&\frac{2E_\kstar}{m_B},\\
\label{eq:f2}
\frac{V}{A_1}&=&\frac{(m_B+m_\kstar)^2}{2E_\kstar m_B}.
\end{eqnarray}
Note that these ratios are independent of the universal form factors
$\xi_\|$ and $\xi_\perp$ and are valid to all orders in the strong
coupling constant~\cite{Beneke:2001at}.

In addition to the  order $\alpha_s$ corrections to the hadronic form
factors, there also exist ``non-factorizable'' corrections, which can be
significant in the heavy quark and large recoil limit. 
Following Ref.~\cite{Beneke:2001at}, these non-factorizable
corrections can be incorporated in next to leading order in QCD by the
following transformations ~\cite{Kruger:2005ep}
\begin{subequations}
\begin{align}
C_7^{\text{eff}}T_i&\rightarrow\mathcal{T}_i\\
C_9^{\text{eff}}&\rightarrow C_9
\end{align}  
\end{subequations}
where the Wilson Coefficients are taken at the next-to-next-to leading order,
and the $\mathcal{T}_i$ are defined as 
\begin{equation}
\label{NNLO}
\mathcal{T}_1=\mathcal{T}_\perp,~~
\mathcal{T}_2=\frac{2E_{K^*}}{m_B}\mathcal{T}_\perp,~~
\mathcal{T}_3=\mathcal{T}_\perp+\mathcal{T}_\|
\end{equation}
The complete expressions of $\mathcal{T}_{\perp,\|}$ are given in 
Ref.~\cite{Beneke:2001at}.

The form factor ratios $\mathsf{P_{1,2,3}}$ and $\mathsf{P'_{1,2,3}}$
can be written with the help of the
Eqs.~(\ref{Beneke-Rln1})-~(\ref{Beneke-Rln7}). The expressions for the
ratio's $\mathsf{P_1}$ and $\mathsf{P_1'}$ are of particular interest,
since these form factor ratio do not receive any QCD correction in the
heavy quark effective theory and are independent of the both form
factor $\xi_\|$ and $\xi_\perp$ to all orders in $\alpha_s$ and to
leading order in the $1/m_b$ expansion.  We will take expressions for
$\mathsf{P_1}$ and $\mathsf{P_1'}$ as input and find that they are
given by the simple form,
\begin{subequations}
\begin{align}  
\label{eq:P_1-simple}
  \mathsf{P_1}=\frac{\mathcal{F}_\perp}{\mathcal{F}_\|}
  &=\frac{\sqrt{\lambda(m_B^2, m_\kstar^2,q^2)}}{(m_B+m_\kstar)^2}\,
  \frac{V(q^2)}{A_1(q^2)}\nn\\&\equiv\Bigg[\frac{-\sqrt{\lambda(m_B^2,
      m_\kstar^2,q^2)}}{2E_\kstar m_B}\Bigg],\\~~~~ 
\label{eq:P_1prime-simple}
  \mathsf{P_1'}=\dsp\frac{\widetilde{\mathcal{G}}_\perp}
  {\widetilde{\mathcal{G}}_\|}&=\frac{-\sqrt{\lambda(m_B^2, 
      m_\kstar^2,q^2)}}{m_B^2-m_{K^*}^2}\frac{\mathcal{T}_1}{\mathcal{T}_2}\nn\\ 
&=\Bigg[\frac{-\sqrt{\lambda(m_B^2,
    m_\kstar^2,q^2)}\,m_B}{2E_{K^*}(m_B^2-m_{K^*}^2)}\Bigg] 
\end{align}  
\end{subequations}

It may be noted that the form factor ratios $\mathsf{P_1}$ and
$\mathsf{P_1'}$ do not depend on the universal form factors $\xi_\|$
and $\xi_\perp$ and are unaltered by the inclusion of non-factorizable
corrections and higher order corrections in QCD. $\mathsf{P_1}$ and
$\mathsf{P_1'}$ are hence used by us as reliable theoretical inputs.
On the other hand it is easy to see that $\mathsf{P_{2,3}}$ and
$\mathsf{P_{2,3}'}$ depend on universal form factors and hence receive
corrections from higher order and non-factorizable contributions that
results in a more complicated expression. In our approach
$\mathsf{P_{2,3}}$ and $\mathsf{P_{2,3}'}$ will be obtained in terms
of observables and $\mathsf{P_1}$ and $\mathsf{P_1'}$ in
Eqs. \eqref{eq:P2}, \eqref{eq:P2p}, \eqref{eq:P3} and \eqref{eq:P3p}.

The expressions \eqref{eq:P_1-simple} and \eqref{eq:P_1prime-simple}
are valid for large recoil region where $q^2$ is small and are usually
considered extremely accurate for $q^2$ between $1~\gev$ and $6~\gev$
~\cite{BF00}. The region $q^2 < 1 ~\gev$ is ignored to eliminate
resonance contributions which might not only introduce uncertainties
but also introduce complex contributions which we have assumed are
absent.  Unless otherwise stated, large recoil region would mean
$0.10\gev^2\le q^2\le 12.86\gev^2$.  We stress that once the
non-factorizable corrections are taken into account, the Wilson
coefficient $C_7$ can no longer be separated from the hadronic form
factor. The $C_7$ and the the hadronic form factors lump together into
effective photon vertex $\widetilde{\cal{G}}_\lambda$, which as we
will show, can be expressed in terms of observables and the form
factors $\mathsf{P_1}$ and $\mathsf{P_1'}$.

\subsection{Form factor in the low recoil limit}
\label{sec:low-recoil}
A model independent description for the case of low recoil energy of
the $\kstar$ in $B\to \kstar\ell^+\ell^-$ decay was put forward by
Grinstein and Pirjol \cite{hep-ph/0404250} in the modified Heavy Quark
Effective Theory framework. In this approach \cite{hep-ph/0404250},
``near the zero point $q^2\approx (m_B-m_\kstar)^2$, the long distance
contributions to $B\to \kstar \ell^+\ell^-$ can be computed as short
distance effect using simultaneous heavy quark and operator product
expansion in $1/Q$ with $Q=\{m_b,\sqrt{q^2}\}$.'' In view of this the
subleading $m_{K^*}/m_B$ terms are neglected and non-factorizable
corrections are ignored.
An elaborate study of the predictions for $B\to K^*\ell^+\ell^-$ was
undertaken in Ref. \cite{arXiv:1006.5013} where the next-to-leading
order corrections from the charm quark mass $m_c$ and strong coupling
at ${\cal O}(m_c/Q^2,\alpha_s)$ were included.  The result is a
relations between the $B\to \kstar \ell^+\ell^-$ form factors that reduces
the number of
independent hadronic form factors reduces to only three, i.e., $V,
A_1$ and $A_2$ can be expressed in terms of the form factors $T_1,
T_2, T_3$ as:
\begin{subequations}
\label{eq:pirjol}
\begin{align}
T_1(q^2)&=\kappa V(q^2)\\
T_2(q^2)&=\kappa A_1(q^2)\\
T_3(q^2)&=\kappa A_2(q^2)\frac{m_B^2}{q^2}
\end{align}
\end{subequations}
where, the expression of $\kappa$ is given in \cite{arXiv:1006.5013}.

In the low recoil limit the non-factorizable corrections and higher
order corrections in $\alpha_s$ are ignorable hence we have
$\widetilde{\mathcal{G}}_\lambda= C_7^\eff\mathcal{G}_\lambda$ for all
$\lambda=\{0, \|,\perp \}$. The condition in Eq.~\eqref{eq:pirjol}
together with Eq.~\eqref{eq:form-factors} on ignoring $m_{K^*}/m_B$
terms can be recast as
\begin{eqnarray}
\label{eq:form-factors-equal}
\frac{\mathcal{G}_\|}{\mathcal{F}_\|}
&=&\frac{\mathcal{G}_\perp}{\mathcal{F}_\perp}
=\frac{\mathcal{G}_0}{\mathcal{F}_0}\equiv\hat{\kappa}=-\kappa\frac{2
  m_B m_b}{q^2}.
\end{eqnarray}
This can easily be seen to imply that,
\begin{equation}
\label{eq:relation-P}
  \mathsf{P_1}=\mathsf{P_1'},\qquad \mathsf{P_2}=\mathsf{P_2'},\qquad
  \mathsf{P_3}=\mathsf{P_3'} 
\end{equation}
and hence
\begin{equation}
\label{eq:r-equal}
 r_\|=r_\perp=r_0=r_{\!\wedge}\equiv r.
\end{equation}

In the low recoil limit the form factor ratios $\mathsf{P_1}$ and
$\mathsf{P_1'}$ are easily derived to be 
\begin{eqnarray}
\label{P1-low}
\mathsf{P_1}&=& \mathsf{P_1'}=\frac{-\sqrt{\lambda(m_B^2, m_\kstar^2,q^2)}}
{(m_B+m_{K^*})^2}\frac{V(q^2)}{A_1(q^2)}.
\end{eqnarray}
Note that in this limit as well the two ratios $\mathsf{P_1}$ and $\mathsf{P_1'}$ are
independent of the universal form factors $\xi_\|$ and
$\xi_\perp$. The low-recoil approximation is expected to work well in
region $14.18\gev^2\le q^2\le 19\gev^2$. Conventionally the low-recoil
region is meant to imply this range of $q^2$.  In
Sec.~\ref{sec:low-recoil-detail} we will reconsider the low-recoil
region to study the special feature that emerge in the low-recoil
region. In the low recoil limit we need to take special care of the
fact that $\mathsf{P_1}=\mathsf{P_1'}$.

\section{Model independent analysis}
\label{sec:wilson-solns}

In this section we present a new model independent approach that
offers a possibility of isolating hadronic effects from genuine new
physics contributions.  We begin by deriving the solutions for the
Wilson coefficients $C_{9}$,  $C_{10}$ and the effective photon
vertex $\widetilde{\mathcal G}_\|$, in terms of observables and the
minimum number of required form factor ratios, some of which are more
or less independent of hadronic uncertainties.

The first set of solutions are obtained using three independent
Eqs.~(\ref{eq2:Fparallel}), (\ref{eq2:Fperp}) and (\ref{eq2:AFB}), and one
easily solves (see Appendix~(\ref{sec:appendix-1})) for
$r_\|+r_\perp$ to be,
\begin{equation*}
r_\|+r_\perp= \pm\frac{\sqrt{\Gf}}{\sqrt{2}\mathcal{F}_\perp}\Big(\mathsf{P_1^2}
F_\|+F_\perp\pm \mathsf{P_1} \sqrt{4 F_\|F_\perp-\tfrac{16}{9} A_{\text
    FB}^2}\Big)^{\nicefrac{1}{2}} 
\end{equation*}
However, $r_\|+r_\perp=0$ when $\AFB=0$ from Eq.~(\ref{eq2:AFB}). The term inside the round bracket of the above equation becomes a whole
square if $\AFB=0$, hence,
\begin{equation}
  \label{eq:u+v}
r_\|+r_\perp\Big|_{\AFB=0}=\pm\frac{\sqrt{\Gf}}{\sqrt{2}\mathcal{F}_\perp}
\big(\sqrt{F}_\perp\pm \mathsf{P_1}\sqrt{F}_\|\big)=0  
\end{equation}
 Since, the
expression for $r_\|+r_\perp$ should be valid for all values of the
observables, the right hand side could go to zero only if positive
sign ambiguity is chosen, taking into account that $\mathsf{P_1}$ is
negative. This fixes the sign ambiguity inside the round
bracket. The condition $r_\|+r_\perp=0$ gives us the familiar relation for the zero
crossing of $\AFB$. The definitions of $r_\|$ and $r_\perp$ straight forwardly
imply that $\AFB=0$ at:
\begin{align}
  \label{eq:AFB-zero-1}
  2\,C_9&=C_7\big(\frac{{\cal G}_\perp}{{\cal F}_\perp}+\frac{{\cal
      G}_\|}{{\cal F}_\|}\big ),\nn\\
  &=-\frac{4 m_b}{q^2}C_7\frac{T_1(q^2)}{V(q^2)}(m_B+m_{\kstar})
  \Big(1-\frac{m_{\kstar}^2}{2m_B^2}\Big),\nn\\
&=-\frac{4 m_b m_B}{q^2}C_7
  \Big(1-\frac{m_{\kstar}^2}{2m_B^2}\Big)+{\cal O}(\alpha_s)~.
\end{align}
where we have used Eqns.~\eqref{eq:f1} and \eqref{eq:f2}. The ${\cal
  O}(\alpha_s)$ dependence arises from the ratio $T_1(q^2)/V(q^2)$,
which also depends on $\xi_\perp
     (q^2)$~\cite{BF00}.

Notice that, Eq.~(\ref{eq:u+v}) implies that when $\AFB=0$ we
must have a exact equality
\begin{equation}
  \label{eq:P_1exp}
\mathsf{P_1}=-\frac{\sqrt{F_\perp}}{\sqrt{F_\|}}\Bigg|_{\AFB=0}
\end{equation}
enabling a measurement of $\mathsf{P_1}$ in terms of the ratio of helicity
fractions. If zero crossing were to occur it would provide an
interesting test of our understanding of form factors. 
Very recently LHCb has confirmed~\cite{Moriond-talk} zero crossing of
$\AFB$ for the first time. The zero crossing is observed at
$q^2=4.9^{+1.1}_{-1.3}\gev^2$, which is consistent
with the predictions of the Standard Model and lies in the large recoil
region. Eq.~\eqref{eq:P_1exp} can hence be used to measure
$\mathsf{P_1}$ at the zero crossing of $\AFB$.  A
confirmation of the estimate of  $\mathsf{P_1}$ with  direct
helicity measurements would leave no doubt
on the reliable predictability HQET in the large recoil region.

The solution of $C_{10}$ in terms of the observables and hadronic form factors
now reads as:
\begin{equation}  
\label{eq:C10}
  C_{10}=\dsp\frac{\sqrt{\Gf}}{\sqrt{2}\mathcal{F}_\|}\frac{2}{3}\frac{
    \AFB}{\Big[\pm\sqrt{\mathsf{P_1^2} F_\|+F_\perp+ \mathsf{P_1} Z_1}\,\Big]}~.
\end{equation}
where $Z_1$ is defined as,
\begin{equation}
 \label{eq:Z_1}
Z_1=\sqrt{4F_\|F_\perp-\frac{16}{9}A_{\text{FB}}^2}
\end{equation}
This solution allows us to measure $C_{10}$ directly in terms of
observables, ``clean'' form factors $\mathsf{P_1}$, $\mathsf{P_1'}$
and on $\mathcal{F}_\|$.   In Tables~\ref{table:C10-FP-value}
and~\ref{table:C10-FP-value-II} we have present the predicted values
of $F_\perp$ and $C_{10}$ using $F_L$ and $\AFB$ values
from~\cite{Aaij:2011aa} and~\cite{Moriond-talk} respectively. In
Table~\ref{table:C10-FP-value-II} we also estimate $F_\perp$ which is
computed directly from data using Eq.~\eqref{eq:phi_dist} and the
value of $S_3$ quoted in Ref.~\cite{Moriond-talk}.

\begin{table*}[ht]
 \centering
\begin{tabular}{|c|c|c|c|c|c|c|c|c|}
\hline
\hline
$q^2 (\gev^2)$&0.10-2.00&2.00-4.30&4.30-8.68&10.09-12.86&14.18-16.00&16.00-19.00&1-6\\
\hline
$F_\perp$~(T)&0.44$\pm$0.01&0.14$\pm$0.06&0.19$\pm$0.03&0.25$\pm$0.04&-&0.14$\pm$0.016&0.21$\pm$0.05\\
\hline
$C_{10}$~(T) &14.36$\pm$1.68&2.81$\pm$0.78&3.00$\pm$0.384&2.34$\pm$0.372&-&3.11$\pm$0.39&3.81$\pm$0.58\\
\hline
\hline
\end{tabular}
\caption{The predictions for $F_\perp$ (Eq.~(\ref{eq:AFB-FP})) and
  $C_{10}$ (Eq.~(\ref{eq:C10})) using $0.37\invfb$ LHCb~\cite{Aaij:2011aa} data for $F_L$,
  $\AFB$ and $d\Gamma/dq^2$. ``(T)'' in the first column indicates the
  values quoted are theoretical estimates. The form factor
  $\mathcal{F}_\|$ and the ratios $\mathsf{P_1}$ and $\mathsf{P_1'}$
  are averaged over each $q^2$ bin using heavy-to-light  form factor
  at large recoil (for $0.10~\gev^2\le q^2\le 12.86 \gev^2$)
  and  heavy-to-light form factor at low recoil (for
  $16~\gev^2\le q^2\le 19~\gev^2$). The region 
  $14.18~\gev^2\le q^2\le 16~\gev^2$ is neglected as the
  form factors can not be calculated reliably in this region. The
  unusual large value of $C_{10}$ in the $0.10~\gev^2\le q^2\le
  2~\gev^2$ region could be  due to failure in estimating
  $\mathcal{F}_\|$ or perhaps be a signal new physics. It is
  unlikely~\cite{Korchin:2010uc,Korchin:2011ze}
  that such a large effect can be  due to the  contributions from low lying resonances in the experimental data.  It may be noted that
  estimate of $F_\perp$ does not depend on universal form factors and
  is clean in the low recoil limit. } 
\label{table:C10-FP-value}
\end{table*}

\begin{table*}[ht]
 \centering
\begin{tabular}{|c|c|c|c|c|c|c|c|c|}
\hline
\hline
$q^2 (\gev^2)$&0.10-2.00&2.00-4.30&4.30-8.68&10.09-12.86&14.18-16.00&16.00-19.00&1-6\\
\hline
$F_\perp$~(E)
&$0.36^{+0.14}_{-0.11}$&$0.11^{+0.09}_{-0.15}$&$0.31\pm
0.09$&$0.145^{+0.12}_{-0.13}$&$0.35\pm0.13$&$0.08^{+0.13}_{-0.14}$&$0.22^{+0.10}_{-0.11}$\\ 
\hline
$F_\perp$~(T)&0.31$\pm$0.03&0.15$\pm$0.04&0.20$\pm$0.03&0.22$\pm$0.03&-&0.12$\pm$0.01&0.17$\pm$0.03\\
\hline
$C_{10}$~(T) &12.91$\pm$1.07&2.60$\pm$0.779&2.88$\pm$0.32&2.0$\pm$0.25&-&2.55$\pm$0.29&3.26$\pm$0.45\\
\hline\hline
\end{tabular}
\caption{ The same as Table~\ref{table:C10-FP-value} but with $1.0\invfb$
  LHCb data~\cite{Moriond-talk}. 
  ``(E)'' in the first column indicates the
  values quoted are experimental estimates. $F_\perp$ (E) is computed
  directly from data using  Eq.~\eqref{eq:phi_dist} and the value of
  $S_3$ quoted in Ref.~\cite{Moriond-talk}. The values of $C_{10}$
  seem to decrease  with the larger data set used and are marginally lower than
  theoretical estimates. Unfortunately, the cause of discrepancy in $C_{10}$ can
  not be fixed, it could either be due to failure in estimating
  $\mathcal{F}_\|$ or perhaps be a signal new physics.  Note that in the $0.10~\gev^2\le q^2\le
  2~\gev^2$ region  $C_{10}$ is still large even with improved
  statistics.  We
  emphasize that the two values of $F_\perp$ are in good agreement
  almost  throughout the $q^2$ region.} 
\label{table:C10-FP-value-II}
\end{table*}

\begin{center}
\begin{figure*}[htbp]
 \begin{center}
   \includegraphics*[width=0.4\textwidth]{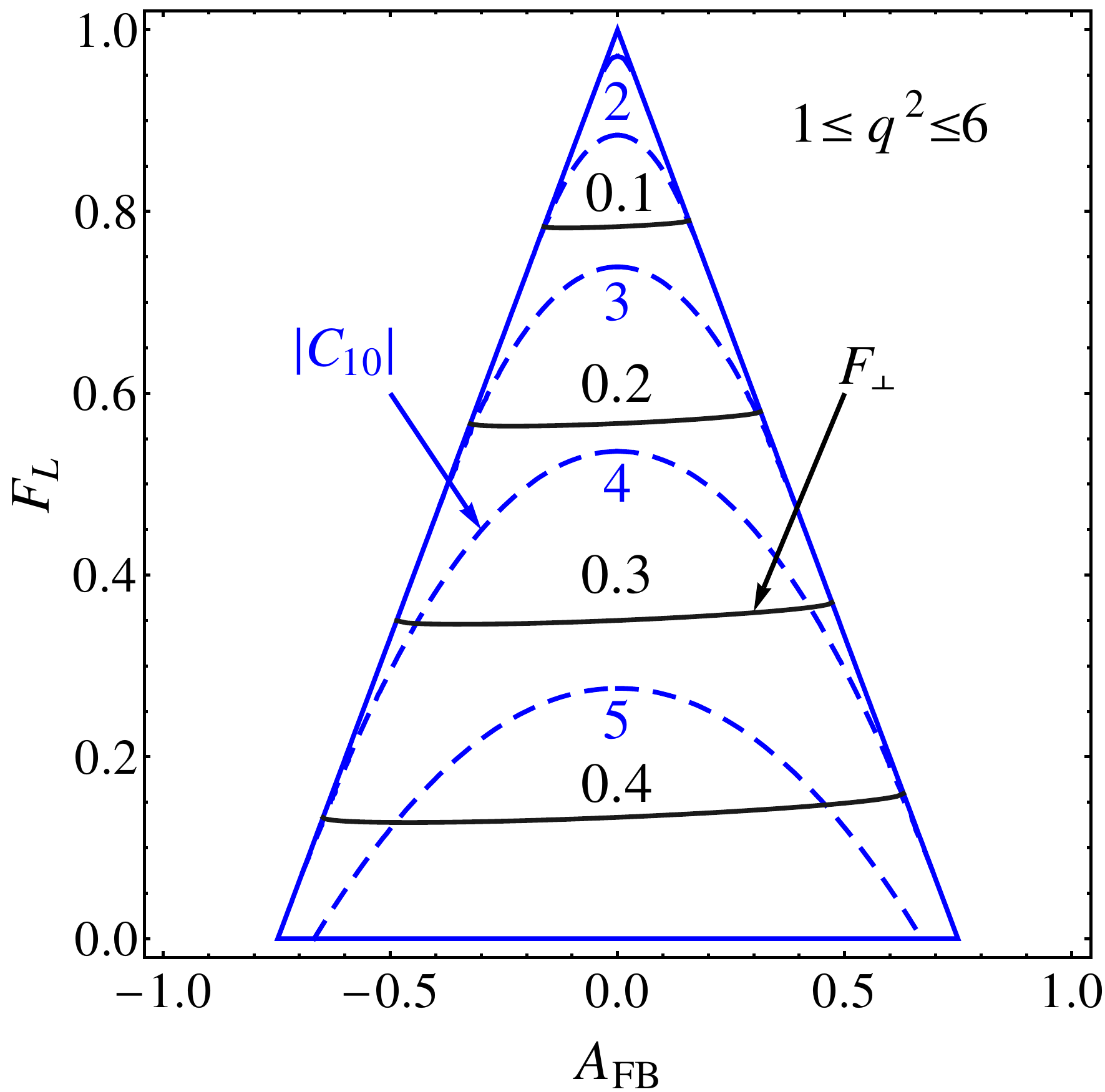}%
    \includegraphics*[width=0.4\textwidth]{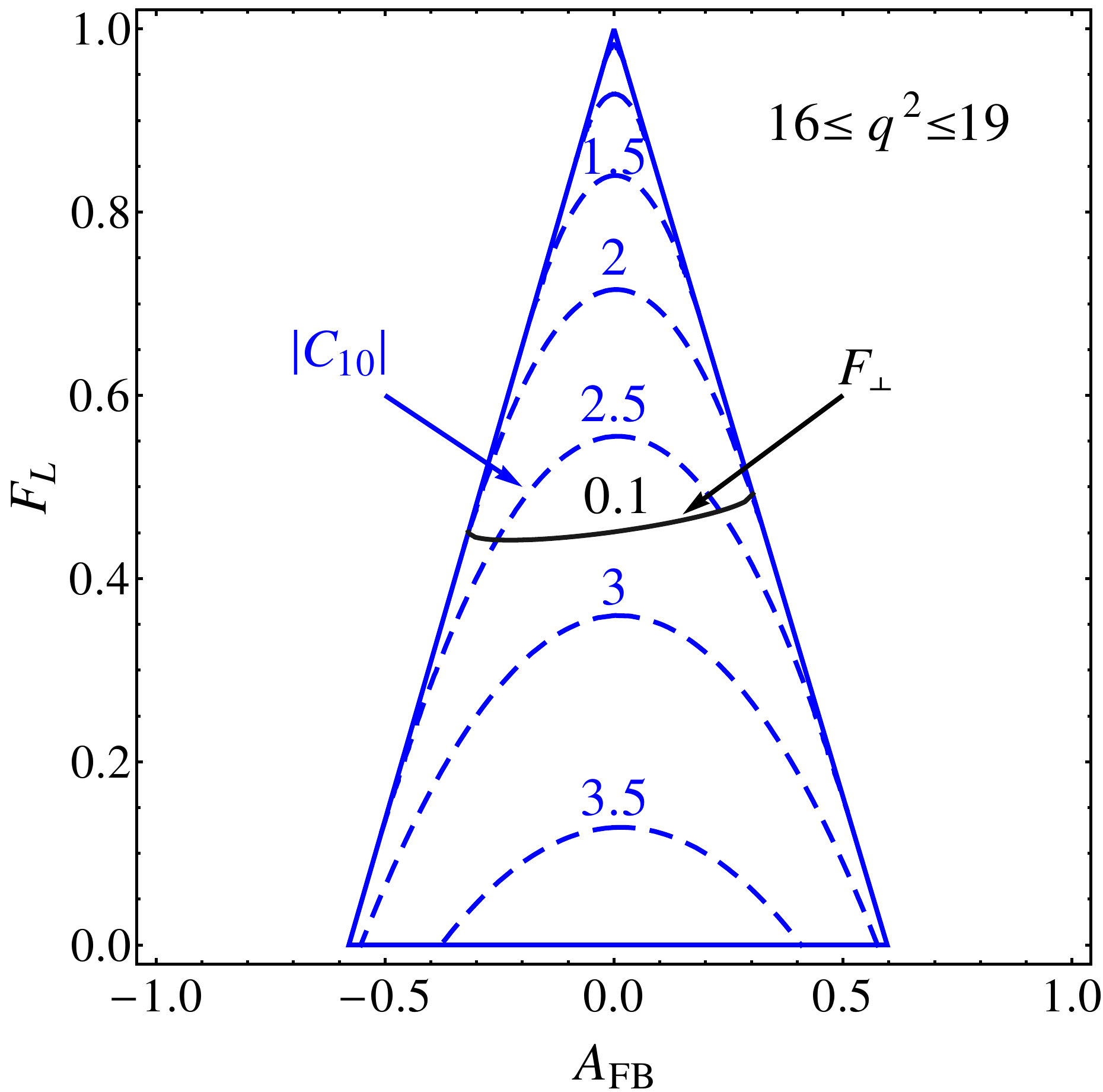}%
    \caption{The allowed $F_L-\AFB$ parameter space depicted by the
      solid (blue) triangle for $R=-1$, is obtained by demanding that
      $F_\perp$ (see Eq.~\ref{eq:AFB-FP}) is real valued. In the
      figure to the left $\mathsf{P_1}$ and $\mathsf{P_1'}$ values are
      averaged over $1~\gev^2$ to $6~\gev^2$ using heavy-to-light
      form factor at large recoil (see Sec.~(\ref{sec:large-recoil})),
      and in the figure to the right we've used $\mathsf{P_1}$ and
      $\mathsf{P_1'}$ values averaged over $16~\gev^2$ to $19~\gev^2$
      using heavy-to-light form factor at low recoil (see
      Sec.~(\ref{sec:low-recoil})). Inside the triangles, the solid
      (black) lines correspond to the $F_\perp$ values the dashed
      (blue) lines correspond to the $C_{10}$ values as function of
      $F_L$ and $\AFB$. }
   \label{fig:C10-FP}
  \end{center}
\end{figure*}
\begin{figure*}[hbtp]
 \begin{center}
   \includegraphics*[width=0.4\textwidth]{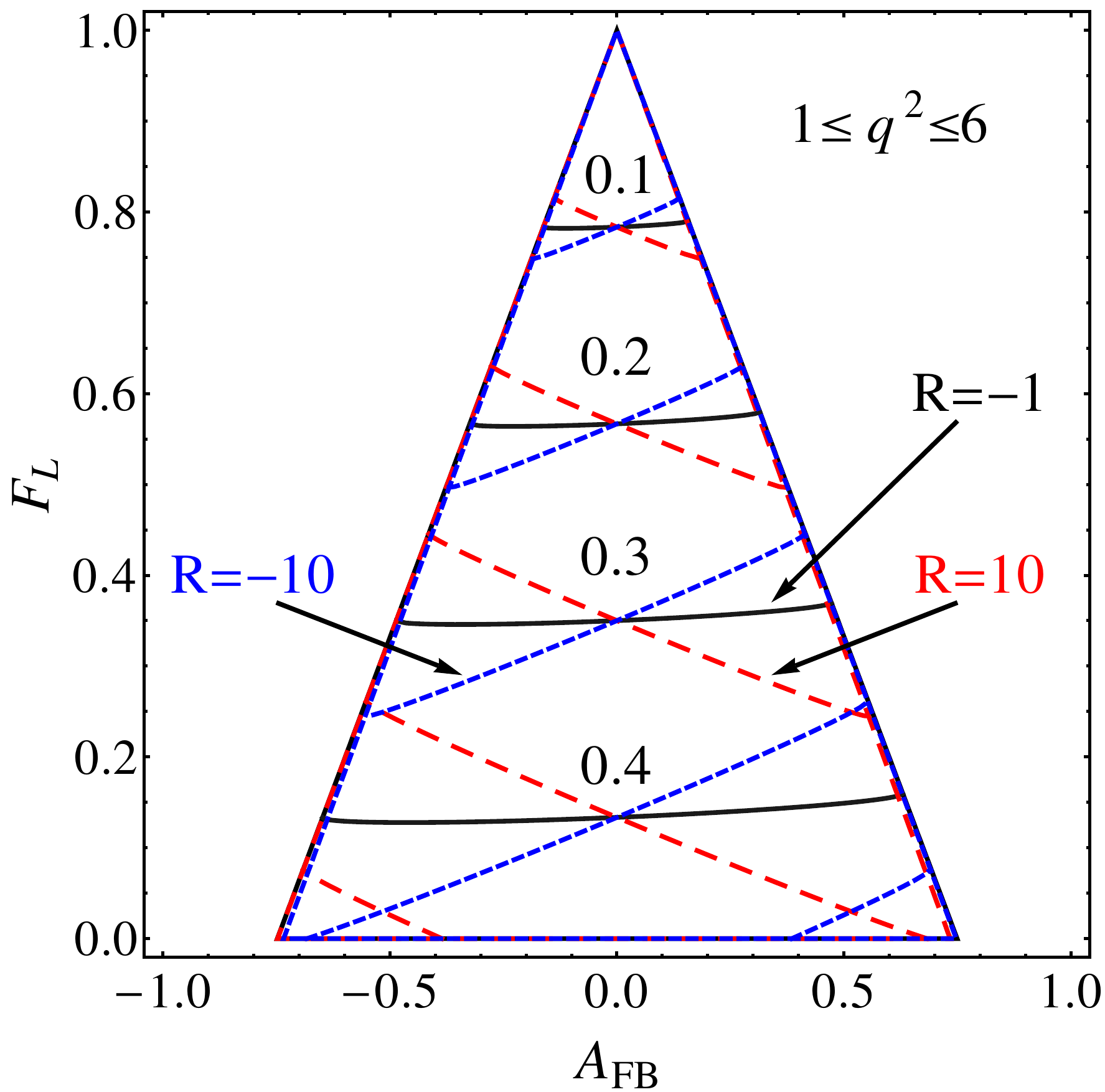}%
   \includegraphics*[width=0.4\textwidth]{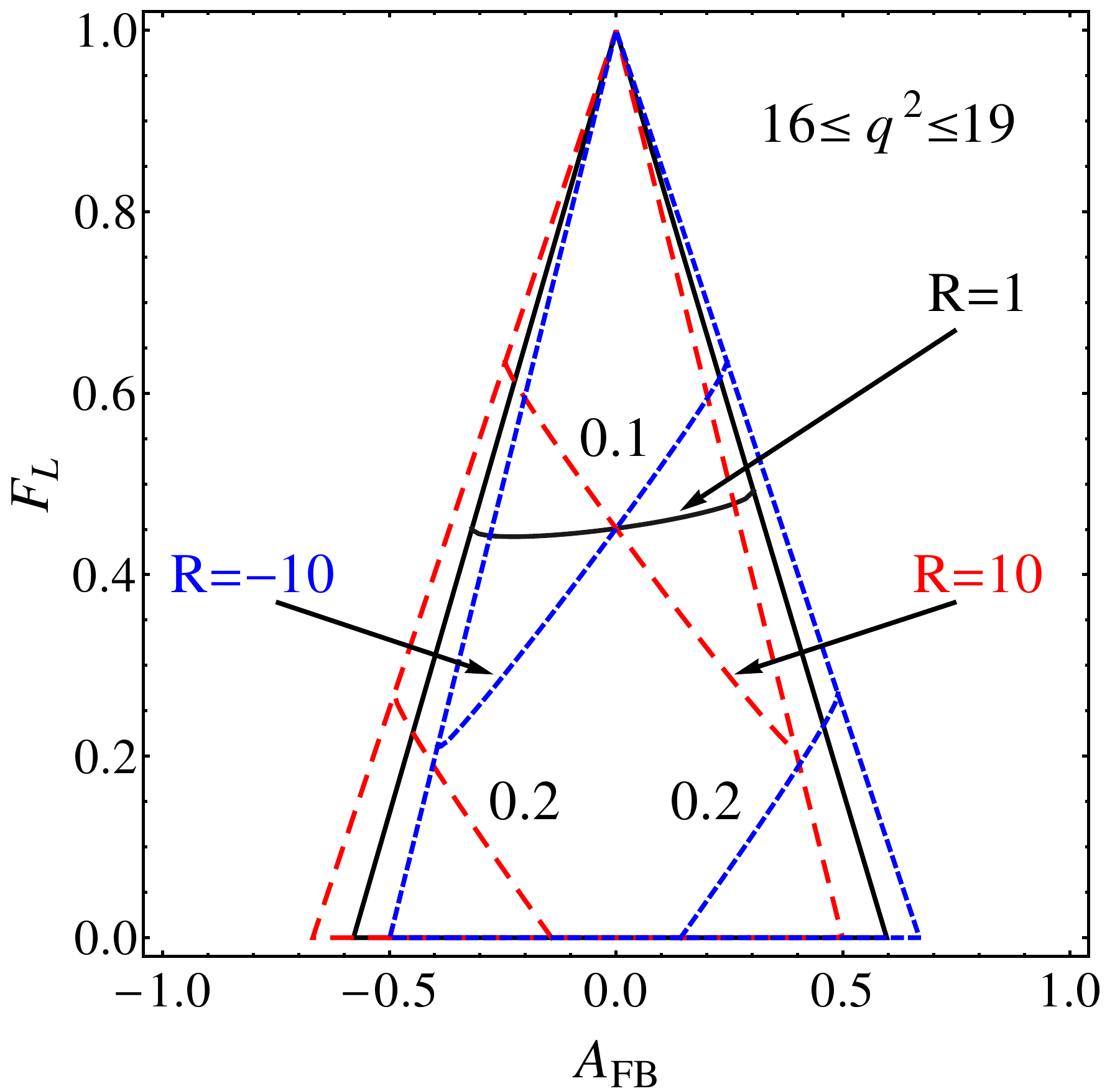}%
   \caption{The same as Fig.~\ref{fig:C10-FP}, except that the
     variation of the parameter space is studied as a function of
     $R$. The large-dashed (red) triangle and the identical lines
     correspond to $R=10$.  The $R=-10$ case is depicted by small
     dashed (blue) line.  $R=-1$ case is shown for reference with
     solid (black) lines.  In the figure to the left $\mathsf{P_1}$
     and $\mathsf{P_1'}$ values are averaged over $1~\gev^2$ to
     $6~\gev^2$ and in the figure to the right we've used
     $\mathsf{P_1}$ and $\mathsf{P_1'}$ values averaged over
     $16~\gev^2$ to $19~\gev^2$.}
   \label{fig:C10-FP-R}
  \end{center}
\end{figure*}
\end{center}

A rather unexpected observation is that as long as $4 F_\|F_\perp\geq 
\frac{16}{9} A_{\text FB}^2~$, the term $\mathsf{P_1^2}
F_\|+F_\perp+\mathsf{P_1} Z_1$ is always positive. This is easily seen by
an (infinite) series expansion in $\AFB$:
\begin{align}
  \label{eq:expansion}
  \mathsf{P_1^2} F_\|+F_\perp+\mathsf{P_1}
  Z_1 &=(\mathsf{P_1}\sqrt{F_\|}+\sqrt{F_\perp})^2-\frac{4
    \AFB^2\mathsf{P_1}}{9\sqrt{F_\|F_\perp}}\nn\\\quad &-\frac{4
    \AFB^4\mathsf{P_1}}{81(F_\|F_\perp)^{\nicefrac{3}{2}}}+{\cal O}(\AFB^6) 
\ge 0,
\end{align}
where every terms is positive since $\mathsf{P_1}$ is negative. Since, the Wilson coefficient $C_{10}$ is
real in the Standard model, $Z_1$ must be real restricting the
observables $F_\|$, $F_\perp$ and $\AFB$ such that:
\begin{equation}
\label{eq:realZ1}
4 F_\|F_\perp\geq 
\frac{16}{9} A_{\text FB}^2.
\end{equation}
The violation of this condition will be a clear signal of new
physics. On the other hand, if the experiments find a real value that
does not agree with the estimates of Standard model value, it could
either be a signal of new physics or of the uncertainties in form
factor estimations.  The Wilson coefficient $C_9$, can also be solved
(see Appendix \ref{sec:appendix-1}) in terms of observables and form
factor ratios,
\begin{eqnarray}
  \label{eq:C9}
  C_9&=&\dsp\frac{\sqrt{\Gf}}{\sqrt{2}\mathcal{F}_\|}\, \frac{(F_\|\mathsf{P_1}
    \mathsf{P_1'}-F_\perp) -\tfrac{1}{2}(\mathsf{P_1}-\mathsf{P_1'})Z_1}{\Big[\pm(\mathsf{P_1}\!-\!\mathsf{P_1'})\sqrt{\mathsf{P_1^2}
    F_\|+F_\perp+ \mathsf{P_1} Z_1}\,\Big]},
\end{eqnarray}
All the discussions following Eq.(\ref{eq:C10}) equally is applicable
to this solutions. The way the matrix element decomposition is defined
in the heavy quark and large energy limit, at next-to-leading logarithmic 
order~\cite{Beneke:2001at}, does not allow us to factor out the Wilson coefficient $C_7^\text{eff}$ from
the hadronic form factors $T_i$.  Hence, the solution of $C_7$ is not
possible. However we can solve for the effective photon vertex
$\widetilde{\mathcal{G}}_\|$, which we can express in terms of the
observables and $\mathsf{P_1}$, $\mathsf{P_1'}$. The solution of
$\widetilde{\mathcal{G}}_\|$ is:
\begin{equation}
  \label{eq:C7}
  \widetilde{\mathcal{G}}_\|=\dsp\frac{\sqrt{\Gf}}{\sqrt{2}\,}\,\frac{(\mathsf{P_1^2}
    F_{\|}-F_\perp)} {\Big[\pm(\mathsf{P_1}\!-\!\mathsf{P_1'})\sqrt{\mathsf{P_1^2}
    F_\|+F_\perp+ \mathsf{P_1} Z_1}\,\Big]} ,
\end{equation}

To obtain the three expressions, Eqs.~(\ref{eq:C10}), (\ref{eq:C9}) and
(\ref{eq:C7}) we removed the sign ambiguities in the solution by
looking at the behavior of the solutions at the $\AFB$ zero crossing
points.  All our solutions for the Wilson coefficients depend
explicitly on the assumption that $\AFB\neq 0$, hence, the Wilson
coefficients can be determined at any $q^2$ except at the zero
crossing of $\AFB$. The denominator of $\widetilde{\cal G}_\|$ and
$C_9$ depend on $\mathsf{P_1}-\mathsf{P_1'}$, so the behavior of the
Wilson coefficients at the point $\mathsf{P_1}\to \mathsf{P_1'}$ needs
careful examination. Unlike the zeros of $\AFB$, which can be
experimentally determined and hence avoided, the crossing point for
$\mathsf{P_1}$ and $\mathsf{P_1'}$, a priori, can only be determined
based on calculations and hence may be uncertain. We note that in this
limit we have $r_\|-r_\perp=0$, where as in the limit $\AFB=0$ we had
$r_\|+r_\perp=0$. Naively, $C_9$ and $\widetilde{\mathcal{G}}_\|$
appear to be divergent in the limit $\mathsf{P_1}\to \mathsf{P_1'}$,
as can be seen from Eqs.~\eqref{eq:C9} and \eqref{eq:C7} and indeed
Eq.~\eqref{eq:21} cannot be used to determine the Wilson coefficients
$C_7$ and $C_9$. However, it is easily seen that the Wilson
coefficients are finite when $\mathsf{P_1}\to \mathsf{P_1'}$. Consider
the combination $\widetilde{\mathcal{G}}_\|-\mathcal{F}_\| C_9$, which
is seen from Eqs.~\eqref{eq:C7} and \eqref{eq:C9} to be manifestly
finite in the limit $\mathsf{P_1'}\to \mathsf{P_1}$:
\begin{equation}
  \label{eq:P12P1plimit}
  \widetilde{\mathcal{G}}_\|-\mathcal{F}_\|
  C_9=\sqrt{\frac{\Gf}{2}}\frac{F_\| \mathsf{P_1}+\tfrac{1}{2} Z_1}{\sqrt{\mathsf{P_1^2}
    F_\|+F_\perp+ \mathsf{P_1} Z_1}}.
\end{equation}
We will show that the combination
$\widetilde{\mathcal{G}}_\|-\mathcal{F}_\| C_9$ can be determined and
indeed if $\mathcal{F}_\|$ is assumed $\widetilde{\mathcal{G}}_\|$ and
$C_9$ can be individually determined and are finite.

In the limit $\mathsf{P_1'}=\mathsf{P_1}$, Eq.~\eqref{eq:uvwz} implies
\begin{equation}
  \label{eq:r}
  r_\|^2+C_{10}^2=r_\perp^2+C_{10}^2=\frac{F_\| \Gf}{2
    \mathcal{F}_\|^2}= \frac{F_\perp \Gf}{2 \mathcal{F}_\perp^2}. 
\end{equation}
We thus have
\begin{equation}
  \label{eq:new-P-relation}
\mathsf{P_1^2}=\mathsf{P_1'}^2=\frac{F_\perp}{F_\|}=
\frac{\mathcal{F}_\perp^2}{\mathcal{F}_\|^2},  
\end{equation}
which enables a measurement of $\mathsf{P_1}$. Indeed if the hadronic
estimate $\mathsf{P_1^2}=F_\perp/F_\|$ is verified by measurement even
when $\AFB\neq 0$, we can conclude with certainty that
$\mathsf{P_1}=\mathsf{P_1'}$. Hadronic estimates can thus be verified
experimentally. Note that a similar condition at $\AFB=0$ also
provided a measurement of $\mathsf{P_1}$ in Eq.~\eqref{eq:P_1exp}.

Many more important results can be derived from the expressions derived 
so far. We can use Eqs.~(\ref{eq:C10}) and (\ref{eq:C9}) to
obtain the ratio $C_9/C_{10}$:
\begin{align}
  \label{eq:C9by10}
  R\equiv \frac{C_9}{C_{10}}=\frac{2 (F_\|\mathsf{P_1}
    \mathsf{P_1'}-F_\perp) -(\mathsf{P_1}-\mathsf{P_1'}) 
    Z_1}{\tfrac{4}{3}A_{\text FB}(\mathsf{P_1}-\mathsf{P_1'})}~.
\end{align}
We emphasize that $C_9/C_{10}$, defined henceforth as $R$, is
expressed as a ``clean parameter'' in terms of observables and the two 
ratio's of form factors which are predicted exactly in heavy quark
effective theory.  Our expressions so far depend on the helicity
fractions $F_\|$ and $F_\perp$, however, $F_L$ has been measured and
since $F_L+F_\|+F_\perp=1$, we can express $F_\|$ in terms of $F_L$
and $F_\perp$. All our conclusion throughout the paper can be
re-expressed in terms of just two helicity fractions $F_L$ and
$F_\perp$.  Eq.~(\ref{eq:C9by10}) can be used to experimentally test
the ratio of $C_9$ and $C_{10}$.  On the other hand if the ratio
$R=C_9/C_{10}$ is known very accurately, $F_\perp$ can be
predicted using Eq.~\eqref{eq:C9by10} in terms of $F_L$ and $A_{\text{FB}}$ as:
\begin{widetext}
\begin{equation}
  \label{eq:AFB-FP}
F_\perp=\frac{-4 R \AFB (\mathsf{P_1}-\mathsf{P_1'}) (1+\mathsf{P_1} \mathsf{P_1'})+3 (1-F_L)
   (\mathsf{P_1}+\mathsf{P_1'})^2- (\mathsf{P_1}-\mathsf{P_1'})\sqrt{T_\perp}
   }{3(1+\mathsf{P_1^2}) (1+\mathsf{P_1'}^2)}
\end{equation}
where,
\begin{equation*}
  T_\perp=9 (1-F_L)^2 (\mathsf{P_1'}+\mathsf{P_1})^2-24 R \AFB (1-F_L) (\mathsf{P_1}-\mathsf{P_1'}) (1-\mathsf{P_1} \mathsf{P_1'})-16
  \AFB^2 \big[R^2 (\mathsf{P_1}-\mathsf{P_1'})^2+(1+\mathsf{P_1^2}) (1+\mathsf{P_1'}^2)\big]\nn
\end{equation*}
The sign of the term containing $\sqrt{T_\perp}$ could either be
positive or negative. Of the two possible solutions for $F_\perp$, in
Eq.~(\ref{eq:AFB-FP}) we have chosen the solution which gives the
correct value of $R$ obtained from Eq.~(\ref{eq:C9by10}).  This
solution corresponds to the one with the negative ambiguity as shown
in Eq.~(\ref{eq:AFB-FP}).  As can be seen from the
Eq.~(\ref{eq:AFB-FP}), the transversity amplitude $F_\perp$ is
expressed in terms of two observables $F_L$ and $\AFB$ which has
already been measured. Using the measured values of $F_L$ and $\AFB$
from Ref.~\cite{Aaij:2011aa} and ~\cite{Moriond-talk} we have
tabulated the perdicted values of $F_\perp$ in
Tables~\ref{table:C10-FP-value} and \ref{table:C10-FP-value-II}
respectively.

In order that $F_\perp$ take real values,
$T_\perp$ must be positive. The positivity of $T_\perp$ imposes
constraints on the possible values for $F_L$ and $\AFB$, which cannot
therefore be arbitrarily chosen.   
The requirement for real solution for $F_\perp$ hence implies a
constraint on $\AFB$ in terms of $\mathsf{P_1}$, $\mathsf{P_1'}$, $R$ and observable
$F_L$:
\begin{align}
\begin{gathered}
  \label{eq:AFB-FL}
 \dsp\frac{-3 (1-F_L)}{4}\, T_{-}\leq A_{FB}\leq\frac{3 (1-F_L)}{4}\,
 T_{+}\\[2.5ex] 
T_{\pm}=\dsp\frac{(\mathsf{P_1}+\mathsf{P_1'})^2}{\dsp\sqrt{(1+\mathsf{P_1^2})(1+\mathsf{P_1'}^2)}
  \sqrt{(\mathsf{P_1}+\mathsf{P_1'})^2+R^2(\mathsf{P_1}-\mathsf{P_1'})^2} \mp(1-\mathsf{P_1} \mathsf{P_1'})\,(\mathsf{P_1}-\mathsf{P_1'})\,R }
\end{gathered}
\end{align}
\end{widetext}
It is easy to see that $T_{\pm}\approx 1$ when $\mathsf{P_1}\approx
\mathsf{P_1}'\approx -1$. Given the values of $\mathsf{P_1}$ and
$\mathsf{P_1}'$ from Table~\ref{table:nonlin-2}, we expect
$T_{\pm}\approx 1$. The allowed domain for $\AFB$ is hence almost free
from $R$ as long as $\mathsf{P_1}\approx \mathsf{P_1}'\approx -1$. In
Fig.~\ref{fig:C10-FP}, we have depicted the permitted $F_L-\AFB$
parameter space by the solid (blue)
triangle for $R=-1$. In the figure to the left $\mathsf{P_1}$ and
$\mathsf{P_1'}$ values are averaged over $1~\gev^2$ to $6~\gev^2$
using heavy-to-light form factor at large recoil (see
Sec.~(\ref{sec:large-recoil})), and in the figure to the right we've
used $\mathsf{P_1}$ and $\mathsf{P_1'}$ values averaged over
$16~\gev^2$ to $19~\gev^2$ using hevay-to-light form factor at low
recoil (see Sec.~(\ref{sec:low-recoil})). Inside the triangles, the
solid (black) lines correspond to the $F_\perp$ values the dashed
(blue) lines correspond to the $C_{10}$ values as function of $F_L$
and $\AFB$.
In Fig.~\ref{fig:C10-FP-R} the varation of the parameter
space is studied as a function of $R$. The large-dashed (red) triangle
and the identical lines correspond to $R=10$.  The $R=-10$ case is
depicted by small dashed (blue) line.  $R=-1$ case is shown for
reference with solid (black) lines.  In the figure to the left
$\mathsf{P_1}$ and $\mathsf{P_1'}$ values are averaged over
$1~\gev^2$ to $6~\gev^2$ and in the figure to the right we've used
$\mathsf{P_1}$ and $\mathsf{P_1'}$ values averaged over $16~\gev^2$ to
$19~\gev^2$.
\begin{center}
\begin{figure*}[!bhtp]
 \begin{center}
   \includegraphics[width=0.4\textwidth]{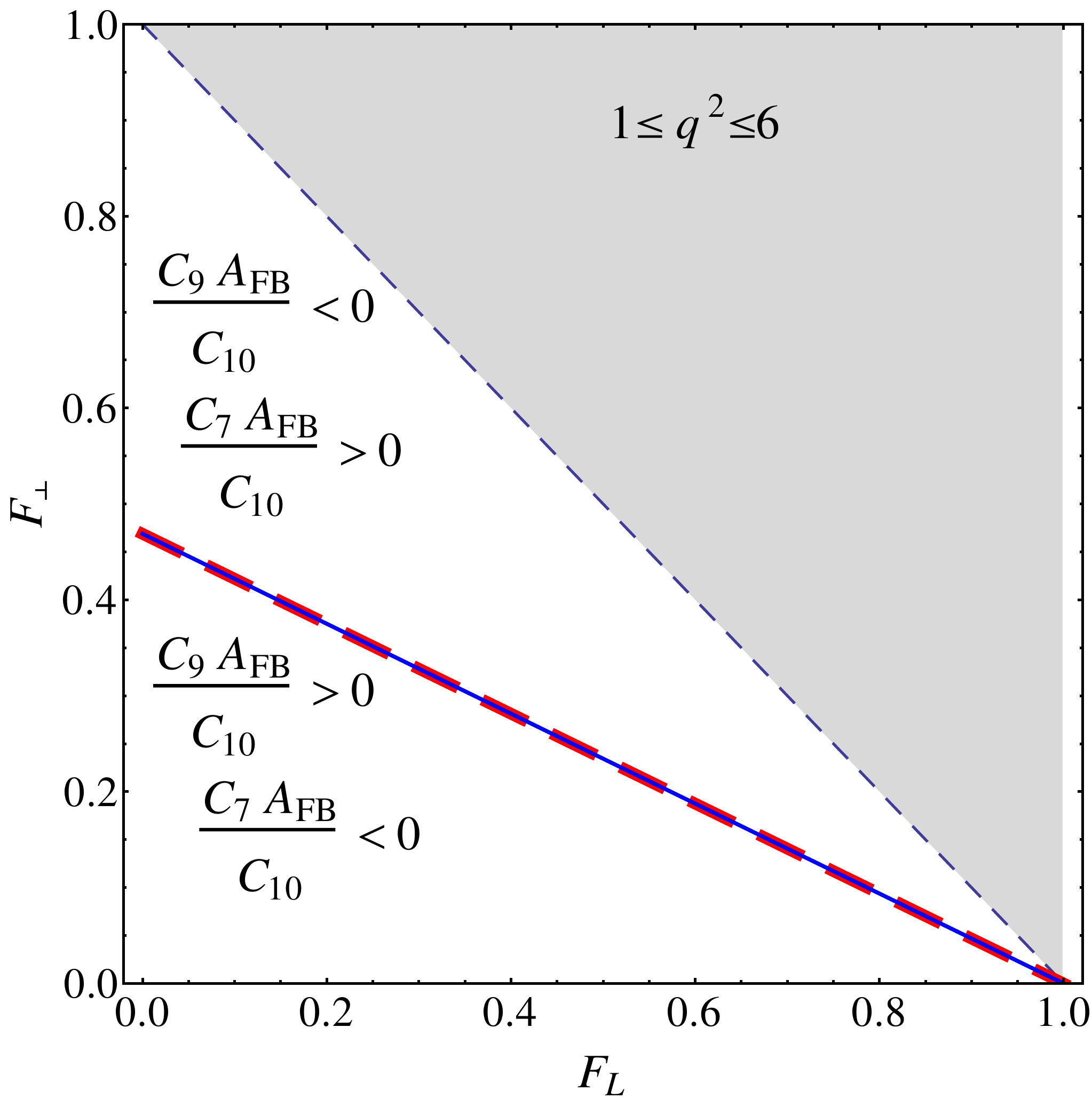}
   \includegraphics[width=0.4\textwidth]{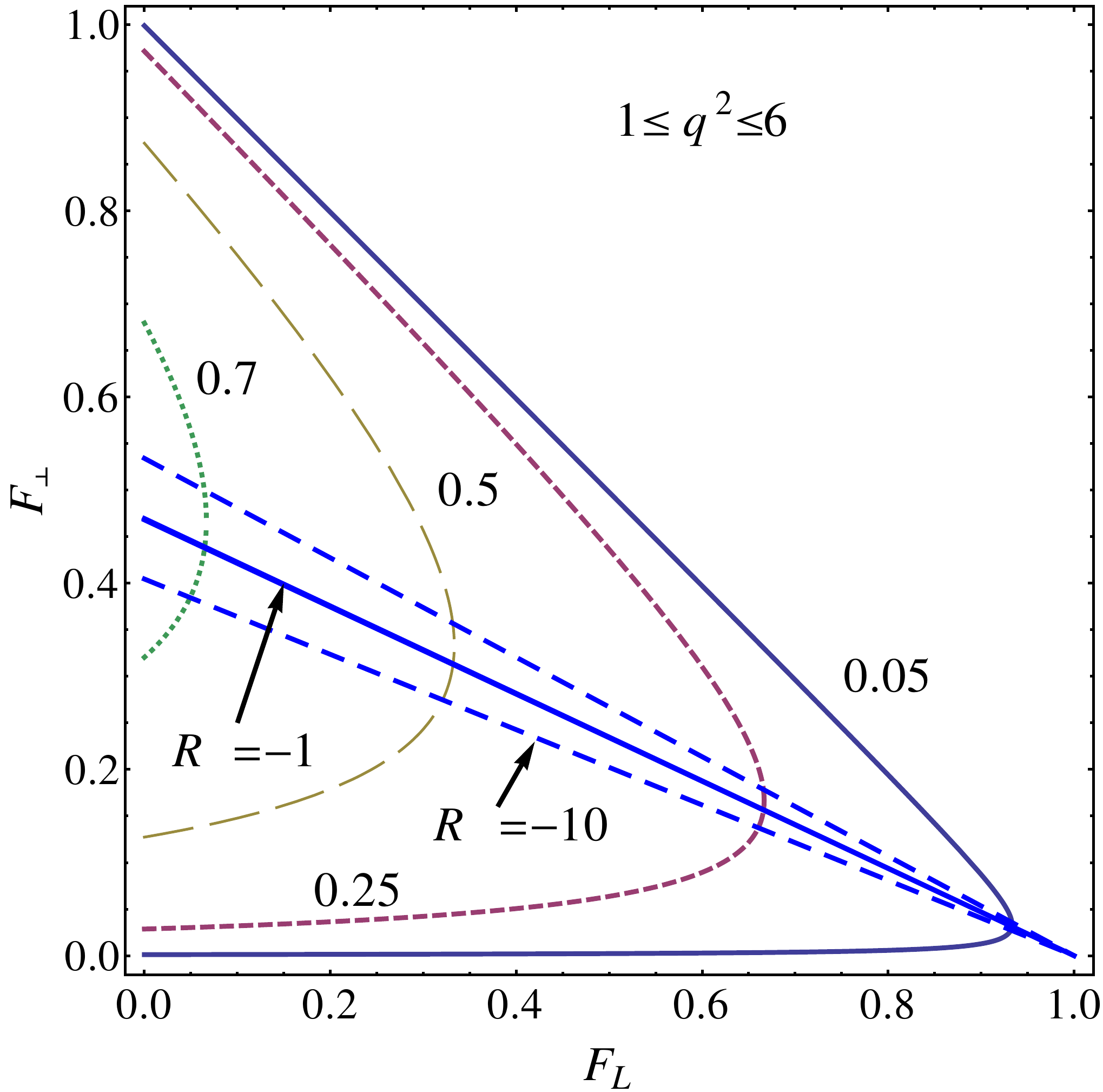}
   \caption{The constraints on $F_L-F_\perp$ parameter space arising
     from Eq.~\eqref{eq:17} with the value of $\mathsf{P_1}$ and
     $\mathsf{P_1'}$ averaged over $1~\gev^2\le q^2\le 6~\gev^2$.
     The allowed region for $R=-1$ is depicted by the diagonal thick
     solid (blue) line that predicts $F_\perp$ to lie in a very narrow
     region, well approximated by a line.  The allowed $F_L$-$F_\perp$
     parameter space for $|R|=10$ are also depicted as a wedge of
     dashed (blue) lines.  The shaded region in the left figure is
     forbidden by $F_L+F_\perp+F_\|=1$.  In the figure on the left the
     thick dashed (red) line correspond to the solution of $F_\perp$
     from Eq.~(\ref{eq:A_FB}) for $\AFB=0$. This line divides the
     allowed domain into two regions fixing the sign of $\AFB$
     relative to $C_9/C_{10}$ and $C_7/C_{10}$ as depicted in the
     figure.  The additional cures in the right figure correspond to
     the constraint on $F_L-F_\perp$ arising from $Z_1^2>0$ for
     different values of $\AFB$: 0.05, 0.25, 0.5, 0.7, where all the
     regions to the left of these curves are allowed.  }
   \label{fig:wedgeAFB-HQET}
  \end{center}
\end{figure*}
\end{center}
\begin{center}
\begin{figure*}[!bhtp]
 \begin{center}
   \includegraphics[width=0.4\textwidth]{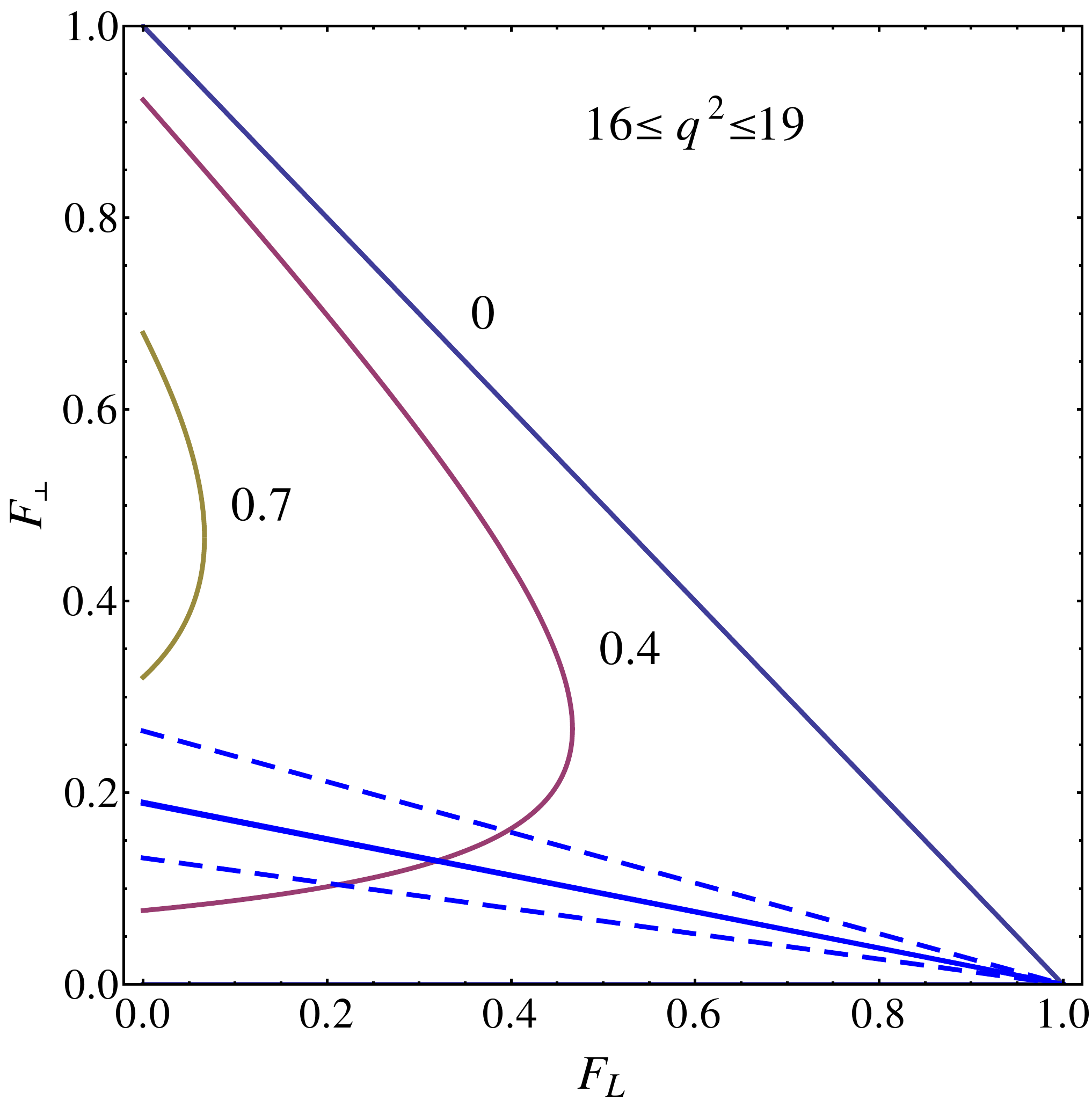}
   \includegraphics[width=0.4\textwidth]{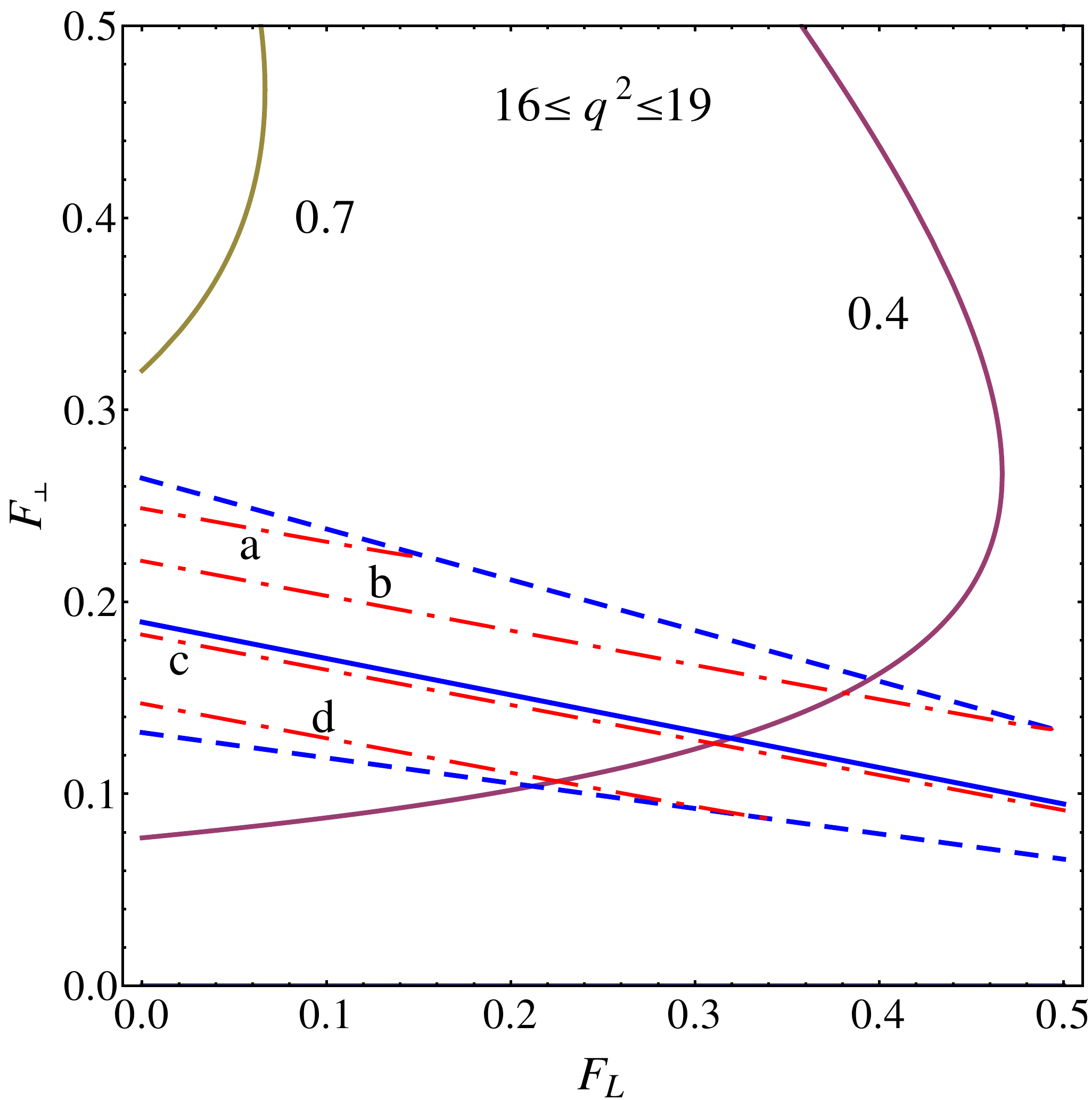}
   \caption{ The same as in Fig.~\ref{fig:wedgeAFB-HQET} except that
     $\mathsf{P_1}$ and $\mathsf{P_1'}$ averaged over $16~\gev^2\le
     q^2\le 19~\gev^2$.  The figure to the right is the inset of the
     figure to the left. In this figure the solid and the dashed
     diagonal (blue) lines are the same as in the figure to the
     right. The dot-dash (red) lines labeled by ``a,b,c,d'' correspond
     to $\AFB=0.5$, $0.3$, $0$, $-0.3$ respectively for $R=-10$. The
     line ``c'' (for $\AFB=0$) divides the domain and 
     corresponds to the thick dashed (red) line in
     Fig.~\ref{fig:wedgeAFB-HQET}. The $\AFB$, $F_L$ and $F_\perp$ must
     be consistent as shown by the dot-dash lines. For $R=-1$ similar
     lines exist for different value of $\AFB$ but overlap with the
     solid blue line. Hence they are not depicted in the figure.}
   \label{fig:wedgeAFB-LCSR}
  \end{center}
\end{figure*}
\end{center}

Interestingly, Eq.~(\ref{eq:C9by10}) can also be inverted to express
$A_{\text FB}$ in terms of $\mathsf{P_1}$, $\mathsf{P_1'}$ and $R$:
\begin{align}
  \label{eq:A_FB}
  \AFB=\frac{3 \left(R X-\sqrt{Y (\mathsf{P_1}-\mathsf{P_1'})^2(1+R^2)-X^2}\right)}{4
    (\mathsf{P_1}-\mathsf{P_1'}) \left(1+R^2\right)}
\end{align}
where, $X=2(F_\| \mathsf{P_1} \mathsf{P_1'}-F_\perp)$ and $Y=4
F_\|F_\perp$.  Note that the Eq.~(\ref{eq:C9by10}) is quadratic in
$\AFB$, and should have resulted in a two-fold ambiguity in the
solution. One easily confirms that only the solution with positive
sign in front of the square root is valid by substituting the
observables in terms form-factors and the Wilson coefficients.  The
usefulness of the result in Eq.~(\ref{eq:A_FB}) is that it constrains
the $F_L$--$F_\perp$ parameter space. This is easily derived by
requiring that $\AFB$ in Eq.~\eqref{eq:A_FB} is real, implying the
positivity of the argument of the radical in the expression for
$\AFB$:
\begin{widetext}
\begin{align}
  \label{eq:17}
  \dsp 1+&\frac{\mathsf{P_1^2}+\mathsf{P_1'}^2+R^2
    (\mathsf{P_1}-\mathsf{P_1'})^2- (\mathsf{P_1}-\mathsf{P_1'})
    \sqrt{R^2+1}
    \sqrt{R^2(\mathsf{P_1}-\mathsf{P_1'})^2+(\mathsf{P_1'}+
      \mathsf{P_1})^2}}{2\mathsf{P_1^2}  
    \mathsf{P_1'}^2}\qquad \qquad \qquad \nn \\
  &\dsp \qquad \qquad \leq \frac{1-F_L}{F_\perp}\leq
  1+\frac{\mathsf{P_1^2}+\mathsf{P_1'}^2+R^2
    (\mathsf{P_1}-\mathsf{P_1'})^2+
    (\mathsf{P_1}-\mathsf{P_1'})\sqrt{R^2+1}
    \sqrt{R^2(\mathsf{P_1}-\mathsf{P_1'})^2+(\mathsf{P_1'}+
      \mathsf{P_1})^2}}{2\mathsf{P_1^2}
    \mathsf{P_1'}^2}~.
\end{align}
\end{widetext}
The constraint implied by this bound is depicted in
Figs.~\ref{fig:wedgeAFB-HQET} and \ref{fig:wedgeAFB-LCSR} where we
have considered two different values corresponding to different bins
of averaged $q^2$ values.  $\mathsf{P_1}$ and $\mathsf{P_1'}$ are
averaged over the $q^2$ region as described in the figure caption.
The reader will note the rigorous constraint within Standard model
i.e. $R=-1$, depicted in the figures by the diagonal thick solid
(blue) line that predicts $F_\perp$ to lie in a very narrow region,
well approximated by a line that is a function of $F_L$ and with the
slope depending on the domain of $q^2$. It is obvious from
Eq.~\eqref{eq:17} that as $R^2$ increases from unity a wider region
around this solid line is allowed. The allowed $F_L$-$F_\perp$
parameter space for $|R|=10$ are also depicted as a wedge of dashed
(blue) lines.  In Fig.~\ref{fig:wedgeAFB-LCSR} on the right we have
shown an enlarged region where for $|R|=10$ we have plotted the values
of $\AFB$ evaluated using Eq.~\eqref{eq:A_FB}.  As the figures shows
$F_L-F_\perp$ correlation is not particularly sensitive to $R$. Also
plotted in these figures are the constraints on the $F_L$-$F_\perp$
parameter space arising from $Z_1^2>0$ for different values of
$\AFB$. The plots also include other details which will be discussed
later.

It is interesting to note that irrespective of the value of $R$, in
the limit $\mathsf{P_1'}\to \mathsf{P_1}$ one obtains
$(1-F_L)/F_\perp=1+1/\mathsf{P_1^2}$. In the limit $m_B\to \infty$ and
the energy of the $\kstar$, $E_\kstar\to \infty$, it is easy to see
that $\mathsf{P_1}=\mathsf{P_1'}\to -1$, and we find that
$F_\|=F_\perp$.  In this limit Eq.~(\ref{eq:phi_dist}) will result in
a constant distribution in $\phi$. Since $\mathsf{P_1}$ and
$\mathsf{P_1'}$ values differ slightly we expect only a very
small coefficient of $\cos\phi$.

The measurements of $F_L$ and $F_\perp$ must be consistent with value
of $\AFB$ and there exists a domain of $R$, $\mathsf{P_1}$ and
$\mathsf{P_1'}$ for which the consistency may hold. These values must
be verified to be consistent with the values of observables. Bounds on
$\mathsf{P_1^2}$ can be obtained from the equations derived so far, in
terms of observables alone.  Extrimizing $\mathsf{P_1^2}$ in terms of
all the non observables in Eq.~(\ref{eq:C10}), we can get following
bounds on $\mathsf{P_1^2}$
\begin{equation}
  \label{eq:7a}
  \mathsf{P_1^2}\lessgtr  \frac{4 F_\|F_\perp-\tfrac{16}{9} A_{\text
    FB}^2}{F_\|^2} \quad\forall\, F_\|F_\perp \lessgtr
 \frac{2}{7}\Big(\frac{4A_{\text{FB}}}{3}\Big)^2 
\end{equation}
For $\AFB=0$, we have already noted the exact equality $\mathsf{P_1^2}=F_\perp/F_\|$. 
Analytical bound on $\mathsf{P_1'}$ are also possible, but are harder to obtain.

We now derive some useful relations that involve $C_7$ and are hence valid
only at the leading order. Eqs.~(\ref{eq:C10}) and (\ref{eq:C7})can
be re-expressed in this limit as: 
\begin{align}
  \label{eq:C7by10}
  & \frac{C_7}{C_{10}}= \frac{3}{2}\frac{{\cal
      F}_\|}{\cal G_\|}\frac{(\mathsf{P_1^2} F_{\|}-F_\perp)}{\AFB (\mathsf{P_1}-\mathsf{P_1'})}~
\end{align}
where we have used the fact that
$\widetilde{\mathcal{G}}_\|=C_7\mathcal{G}_\|$ at leading order.  We
emphasize that $C_7/C_{10}$ is not as clean as $C_9/C_{10}$, which is
expressed in Eq.~(\ref{eq:C9by10}) in terms of observables and ratio's
of two form factors which are predicted exactly in heavy quark
effective theory. $C_7/C_{10}$ on the other-hand depends on
$\mathcal{F}_\|/\mathcal{G}_\|$ which in turn depends on the heavy
quark effective theory form factor $\xi_\perp$. It may nevertheless be
noted that the sign of $\mathcal{F}_\|/\mathcal{G}_\|$ is quite
accurately predicted to be negative, since $A_1(q^2)$ and $T_2(q^2)$
are always positive.  Eq.~(\ref{eq:C7by10}) directly implies a
constraint on the sign of $C_7/C_{10}$. It is easy to conclude that
$(C_7/C_{10}) A_{\text{FB}}\gtrless 0$ only if $\mathsf{P_1^2} \lessgtr F_\perp/F_\|$ when
$\mathsf{P_1}-\mathsf{P_1'}>0$. Eq.~(\ref{eq:7a}) together with Eq.~(\ref{eq:C7by10})
can be used to obtain more useful bounds that are purely in terms of
observables alone, albeit  they are not completely
exhaustive. Eq.~(\ref{eq:7a}) implies:
\begin{equation}
  \label{eq:8}
  \mathsf{P_1^2} F_{\|}-F_\perp\lessgtr \frac{Z_1-F_\|F_\perp}{F_\|}\quad\forall\,
  F_\|F_\perp \lessgtr \frac{2}{7}\Big(\frac{4A_{\text{FB}}}{3}\Big)^2,
\end{equation}
which in turn implies for $(\mathsf{P_1^2} F_\|-F_\perp) <0$ that,
\begin{equation}
  \label{eq:6}
\frac{C_7}{C_{10}} A_{\text{FB}}>0  \qquad \forall \quad F_\|F_\perp <
  \frac{32}{63}A_{\text{FB}}^2~.    
\end{equation}
If, however, $(P^2 F_\|-F_\perp) >0$ we obtain an analogous condition
\begin{equation}
  \label{eq:7l}
\frac{C_7}{C_{10}} A_{\text{FB}}<0  \qquad \forall \quad F_\|F_\perp >
 \frac{16}{27}A_{\text{FB}}^2~.   
\end{equation}
The above bounds have nothing to say on the sign of $C_7/C_{10}$ in
the region,
\begin{equation}
  \label{eq:9}
 \frac{32}{63}\AFB^2\le F_\|F_\perp\le  \frac{16}{27}\AFB^2
\end{equation}
and may not be particularly useful in general. One can
nevertheless draw conclusions on the signs of the Wilson coefficients
by combining Eq.~(\ref{eq:C9by10})
together with Eq.~(\ref{eq:C7by10}) to write:
\begin{eqnarray}
  \label{eq:7}
  \Big(\frac{2}{3} \frac{C_9}{C_{10}}\,\mathsf{P_1^{'\!'}}-\frac{4}{3}
  \frac{C_7}{C_{10}}\,\mathsf{P_1}\Big) \AFB &=&(\mathsf{P_1}^2
  F_{\|}+F_\perp+ \mathsf{P_1}Z_1)\nn\\&>&0,  
\end{eqnarray}
where, $\mathsf{P_1^{'\!'}}=(\mathcal{G}_\|/\mathcal{F}_\|)\,
(\mathsf{P_1}+\mathsf{P_1'})>0$ since each of
$(\mathcal{G}_\|/\mathcal{F}_\|)$, $\mathsf{P_1}$ and $\mathsf{P_1'}$
are always negative.
Defining,
\begin{equation}
  \label{eq:E1E2}
 E_1\equiv \frac{C_9}{C_{10}} \AFB, \qquad
 E_2\equiv \frac{C_7}{C_{10}} \AFB, 
\end{equation}
for convenience, Eq.~\eqref{eq:7} reads
\begin{equation}
\label{eq:E1E2-7}
  \frac{2}{3}\,\mathsf{P_1^{'\!'}}
  E_1 -\frac{4}{3}\,\mathsf{P_1}
E_2 >0
\end{equation}
In SM, $C_7/C_{10} >0$ and $C_9/C_{10}<0$, hence the sign of $E_2$
($E_1$) will be same (opposite) to that observed for $\AFB$. If for
any $q^2$ we find $\AFB>0$, Eq.~(\ref{eq:E1E2-7}) cannot be satisfied
unless the contribution from the $E_2$ term exceeds the $E_1$ term, or
the sign of the $E_2$ term is wrong in SM. In the SM the $E_2$ term
dominates at large recoil i.e. small $q^2$, hence, $\AFB$ must be
positive at small $q^2$ to be consistent with SM.  If $\AFB<0$ is
observed for all $q^2$ i.e. no zero crossing of $\AFB$ is seen, one
can convincingly conclude that $C_7/C_{10}<0$ in contradiction with
SM. However, if zero crossing of $\AFB$ is confirmed with $\AFB>0$ at
small $q^2$ it is possible to conclude that the signs $C_7/C_{10}>0$
and $C_9/C_{10}<0$ are in conformity with SM, as long as other
constraints like $Z_1^2>0$ hold. In Ref.~\cite{Moriond-talk} the zero
crossing is indeed seen. However, in the $2\gev^2 \le q^2\le
4.3\gev^2$ bin, $Z_1^2>0$ is only marginally satisfied.  We emphasize
that these conclusions drawn from Eq.~(\ref{eq:7}) are exact and not
altered by any hadronic uncertainties.

As mentioned in the text earlier, there are three set of solutions
of Wilson coefficients $C_9$ and
$C_{10}$ and the effective photon vertices $\widetilde{\mathcal G}_0$
and $\widetilde{\mathcal G}_0+\widetilde{\mathcal G}_\|$.  We next
discuss the second and the third set of solutions.  The method of
solutions is identical to first set of solutions (see
Eqs.~(\ref{eq:C10}), (\ref{eq:C9}) and (\ref{eq:C7})) and has been
discussed in Appendix~\ref{sec:appendix-1}. Using Eqs.~(\ref{eq2:Fperp}),
(\ref{eq2:FL}) and ~(\ref{eq2:A5}) 
we can easily solve for
$r_0+r_\perp$ as
\begin{equation}
\label{eq:0-perp}
  r_0+r_\perp = \pm\frac{\sqrt{\Gf}}{\sqrt{2}\mathcal{F}_\perp}\Big(\mathsf{P_2^2}
  F_L+F_\perp \pm \mathsf{P_2} Z_2\Big)^{\nicefrac{1}{2}}\\
\end{equation}
where we have defined
\begin{equation}
  \label{eq:Z_2}
Z_2 =\sqrt{4F_L F_\perp-\frac{32}{9}A_5^2},
\end{equation}
and the form factor ratios $\mathsf{P_2}$ has been previously defined in
Eq.~(\ref{eq:P_2}). It is easy to derive that
\begin{equation}
\label{eq3:z+v}
r_0+r_\perp\Big|_{A_5=0}=\pm\frac{\sqrt{\Gf}}{\sqrt{2}\mathcal{F}_\perp}
        \Big(\sqrt{F_\perp}\pm \mathsf{P_2}\sqrt{F_L}\Big)  =0
\end{equation}
since, Eq.~(\ref{eq2:A5}) implies that we have $r_0+r_\perp=0$ at
$A_5=0$.  Once again, repeating the arguments made when $\AFB=0$, the
expression of $r_0+r_\perp=0$ is valid for all values of the
observables. The right hand side of Eq.~(\ref{eq3:z+v}) can be zero
only when positive sign ambiguity is chosen, since $\mathsf{P_2}$ is
negative.  At the zero crossing points of $A_5$ 
we also have the following  exact equality,
\begin{equation}
  \label{eq:10-2}
\mathsf{P_2} =\dsp -\frac{\dsp\sqrt{F_\perp}}{\dsp\sqrt{F_L}}\Bigg|_{A_5=0}
\end{equation}
enabling measurements of form factor ratio $\mathsf{P_2}$ in terms of
observables, as long as the zero crossing of $A_5$ occurs in the large
recoil region (we have verified at leading order that this is indeed true). We now write the second set of solutions of Wilson
coefficients $C_9$ and $C_{10}$ and the effective photon vertex
$\widetilde{\mathcal G}_0$:
\begin{eqnarray}
 \label{eq2:solC10}
C_{10} &=&\frac{\sqrt{\Gf}}{\sqrt{2}\mathcal{F}_0}\frac{2}{3}
\frac{\sqrt{2}A_5} {\Big[\pm\sqrt{\mathsf{P_2^2}F_L+F_\perp
  + \mathsf{P_2}Z_2}\Big]},~\\
\label{eq2:solC9}
C_9&=& \frac{\sqrt{\Gf}}{\sqrt{2}\mathcal{F}_0}\frac{(F_L \mathsf{P_2}
  \mathsf{P_2'}-F_\perp) -\frac{1}{2}(\mathsf{P_2}-\mathsf{P_2'}) Z_2}{\Big[\pm(\mathsf{P_2}-\mathsf{P_2'})\sqrt{\mathsf{P_2^2} F_L+F_\perp+ \mathsf{P_2}
    Z_2}\Big]},~\\
   \label{eq2:solC7}
\widetilde{\mathcal{G}}_0&=&\frac{\sqrt{\Gf}}{\sqrt{2}}\frac{(\mathsf{P_2^2}
    F_{L}-F_\perp)}{\bigg[\pm(\mathsf{P_2}-\mathsf{P_2'})\sqrt{\mathsf{P_2^2}
    F_L+F_\perp+ \mathsf{P_2} Z_2}\bigg]}.~
\end{eqnarray}
It is easy to derive these relations which are identical to the ones
derived in Eqs.~(\ref{eq:C10}), (\ref{eq:C9}) and (\ref{eq:C7}) except
for the replacements: $F_\|\to F_L$, $\AFB \to \sqrt{2}A_5$,
$\mathcal{F}_\| \to \mathcal{F}_0$, $\mathcal{G}_\| \to
\mathcal{G}_0$, which also imply that $r_\|\to r_0$, $\mathsf{P_1}\to
\mathsf{P_2}$ and $\mathsf{P_1'} \to \mathsf{P_2'}$.
Straightforward extrimization with respect to all the non observables
in Eq.~(\ref{eq2:solC10}) gives the
following bounds on the form factor ratios $\mathsf{P_2}$
\begin{equation}
  \label{eq1:P2sq}
  \mathsf{P_2^2}\lessgtr  \frac{4 F_L F_\perp-\tfrac{32}{9}A_5^2}{F_L^2}
  \quad\forall\, F_L F_\perp \lessgtr
  \frac{2}{7}(\frac{4\sqrt{2}A_{5}}{3})^2 \nn\\
\end{equation}
Eq.~(\ref{eq2:solC10})--(\ref{eq2:solC7}) give:
\begin{align}
  \label{eq2:C9by10}
 & \frac{C_9}{C_{10}}=\frac{2 (F_L\mathsf{P_2} \mathsf{P_2'}-F_\perp) -(\mathsf{P_2}-\mathsf{P_2'})
    Z_2}{\tfrac{4}{3}\sqrt{2}A_5 (\mathsf{P_2}-\mathsf{P_2'})}~,\\[3ex]
  \label{eq2:C7by10}
 & \frac{\widetilde{\mathcal G}_0}{C_{10}}= \frac{3}{2}{\mathcal
      F}_0\frac{(\mathsf{P_2^2} F_{L}-F_\perp)}{ \sqrt{2} A_5(\mathsf{P_2}-\mathsf{P_2'})},
\end{align} 
Eq.~(\ref{eq2:C9by10}) can be inverted to obtain expressions for $A_5$
akin to the expression for $\AFB$ obtained in Eq.~(\ref{eq:A_FB}).
One easily derives:
\begin{equation}
  \label{eq:A_5}
\sqrt{2}A_5=\frac{3\big(RX_2\!-\!\sqrt{Y_2(\mathsf{P_2}-\mathsf{P_2'})^2(1+R^2)-X_2^2}\,
  \big)}{4(\mathsf{P_2}-\mathsf{P_2'})(1+R^2)}
\end{equation}
where $X_2=2(F_L\mathsf{P_2}\mathsf{P_2'}-F_\perp)$ and  $Y_2=4F_LF_\perp$.
Eqs.~(\ref{eq2:C9by10}) and (\ref{eq2:C7by10}) can be combined to obtain
\begin{eqnarray}
  \label{eq2:7a}
 (\frac{2}{3}\frac{C_7}{C_{10}}\mathsf{P_2^{'\!'}} -\frac{4}{3}
 \frac{C_9}{C_{10}}\mathsf{P_2}) {A_5}&=&\frac{(\mathsf{P_2^2} F_L+F_\perp+ \mathsf{P_2}Z_2)}{\sqrt{2} }\quad\\&>&0\nn
\end{eqnarray}
where
$\mathsf{P_2^{'\!'}}=(\mathcal{G}_0/\mathcal{F}_0)\,(\mathsf{P_2}+\mathsf{P_2'})>0$,
since $\mathcal{G}_0/\mathcal{F}_0$, $\mathsf{P_2}$ and
$\mathsf{P_2'}$ are all negative. While this is not easily seen as in
the case of $\mathsf{P_1^{'\!'}}$ we have numerically verified at
leading order that this is true for the entire $q^2$ domain.  We have
shown earlier by doing a power expansion in $\AFB$, that
$(\mathsf{P_1^2}F_\|+F_\perp+\mathsf{P_1} Z_1)$ is always positive. It
is easy to see that similar arguments can be made for the positivity
of $(\mathsf{P_2^2}F_L+F_\perp+\mathsf{P_2}Z_2)$ by considering
expansions in $A_5$. Hence if the term in the bracket must be positive
$A_5$ must be positive. At large recoil the term in the bracket is
expected to be positive

\begin{figure*}[thb]
  \centering
\includegraphics[width=0.4\textwidth]{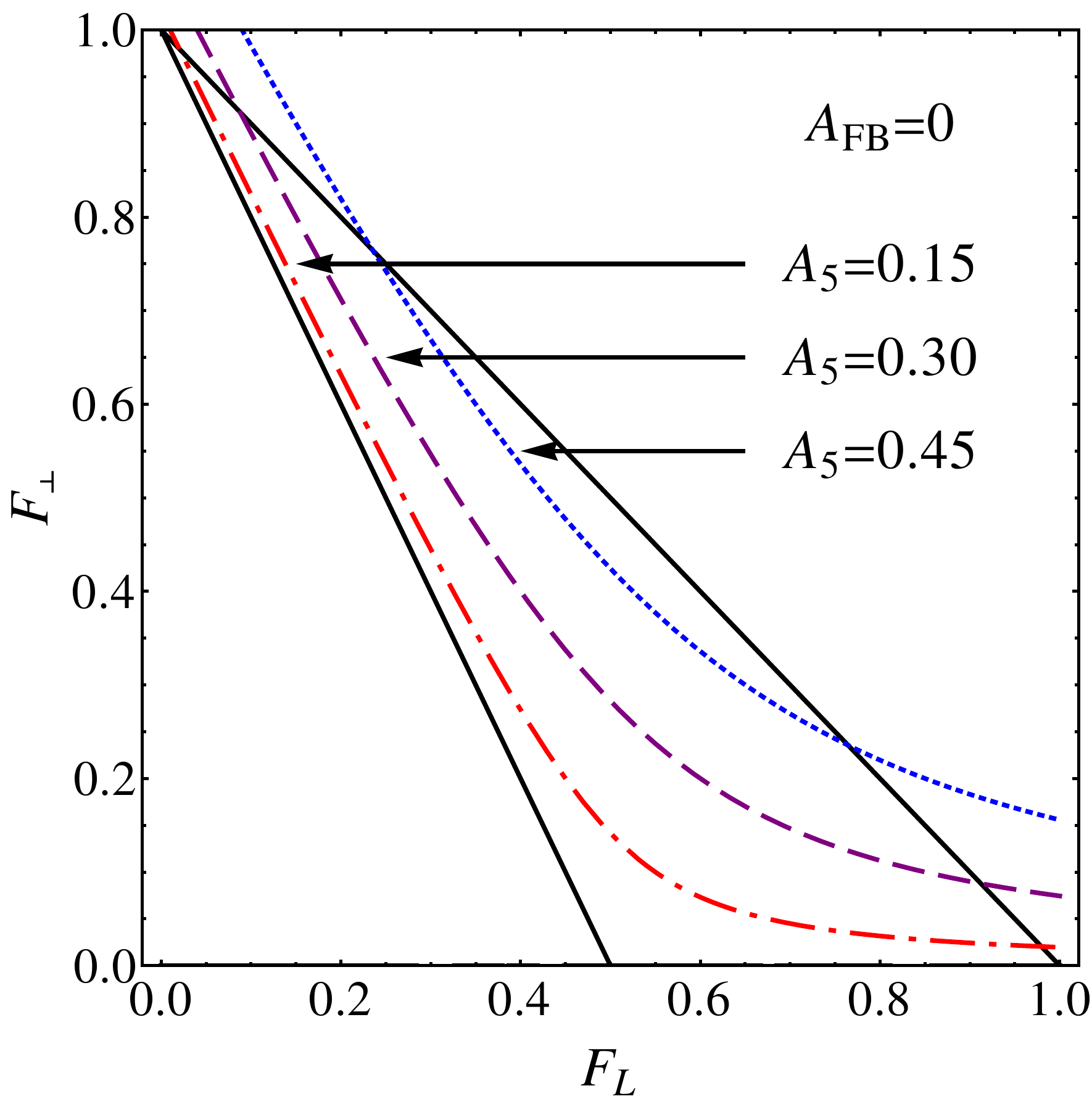}
\includegraphics[width=0.4\textwidth]{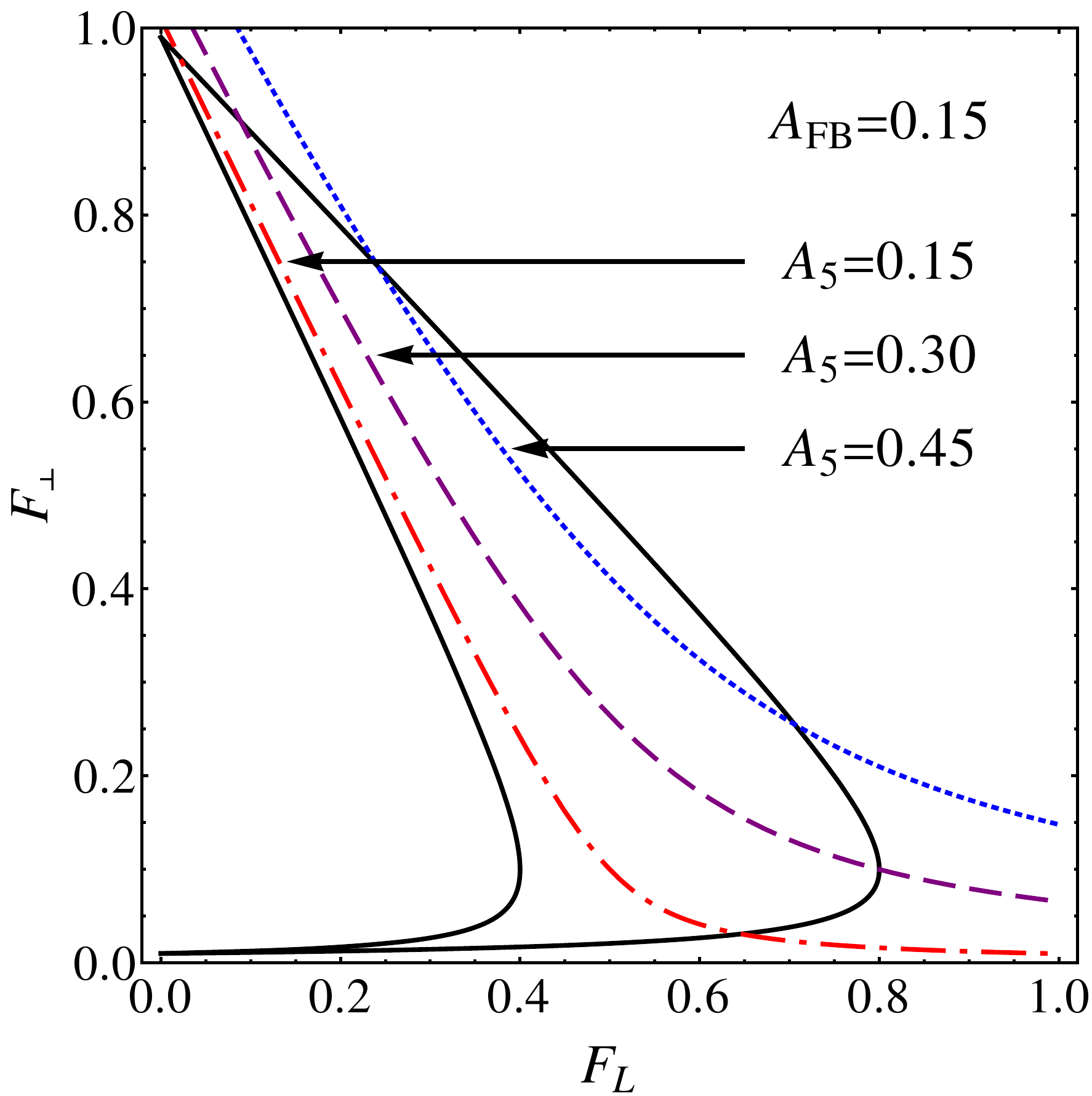}
\includegraphics[width=0.4\textwidth]{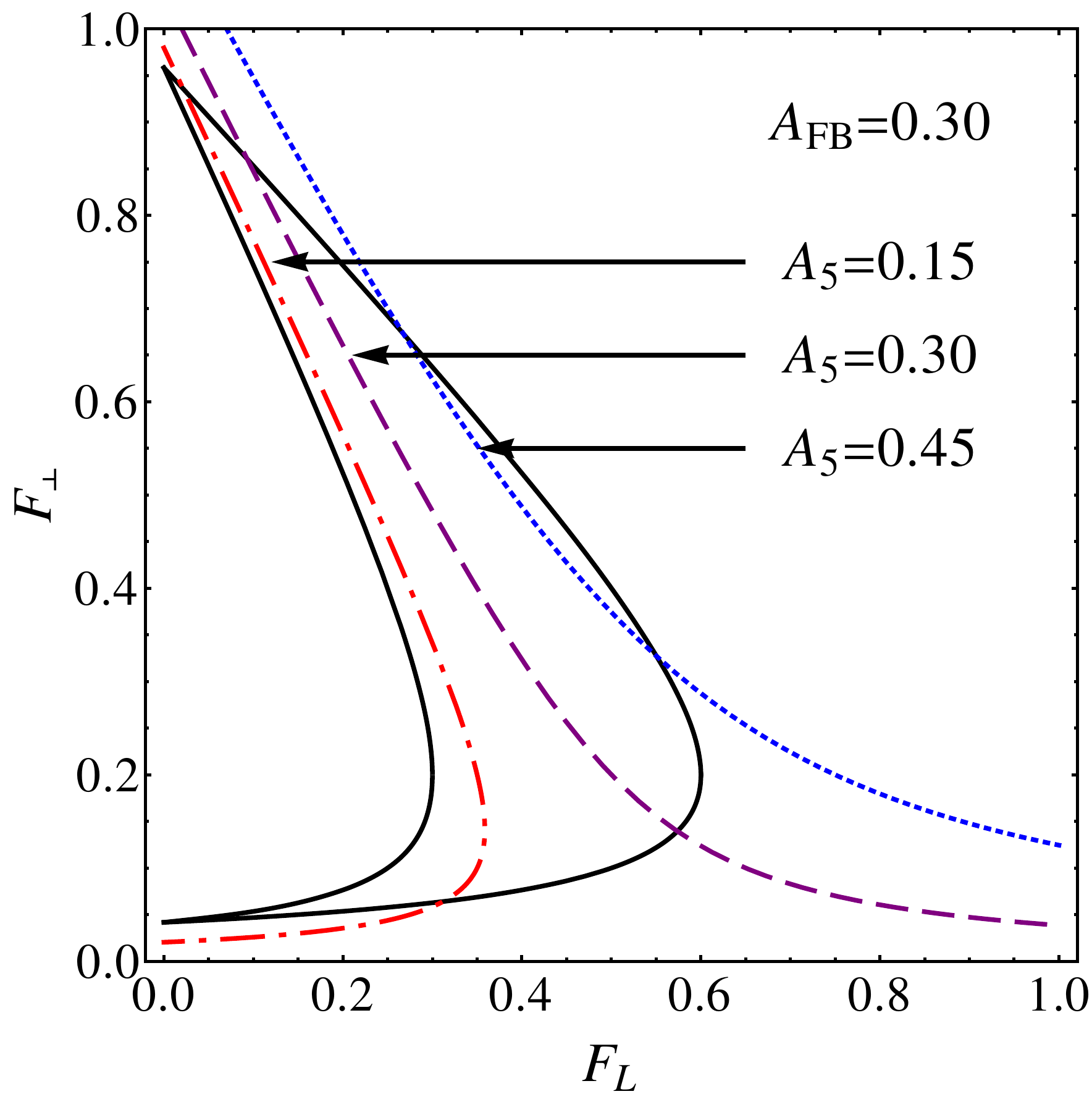}
\includegraphics[width=0.4\textwidth]{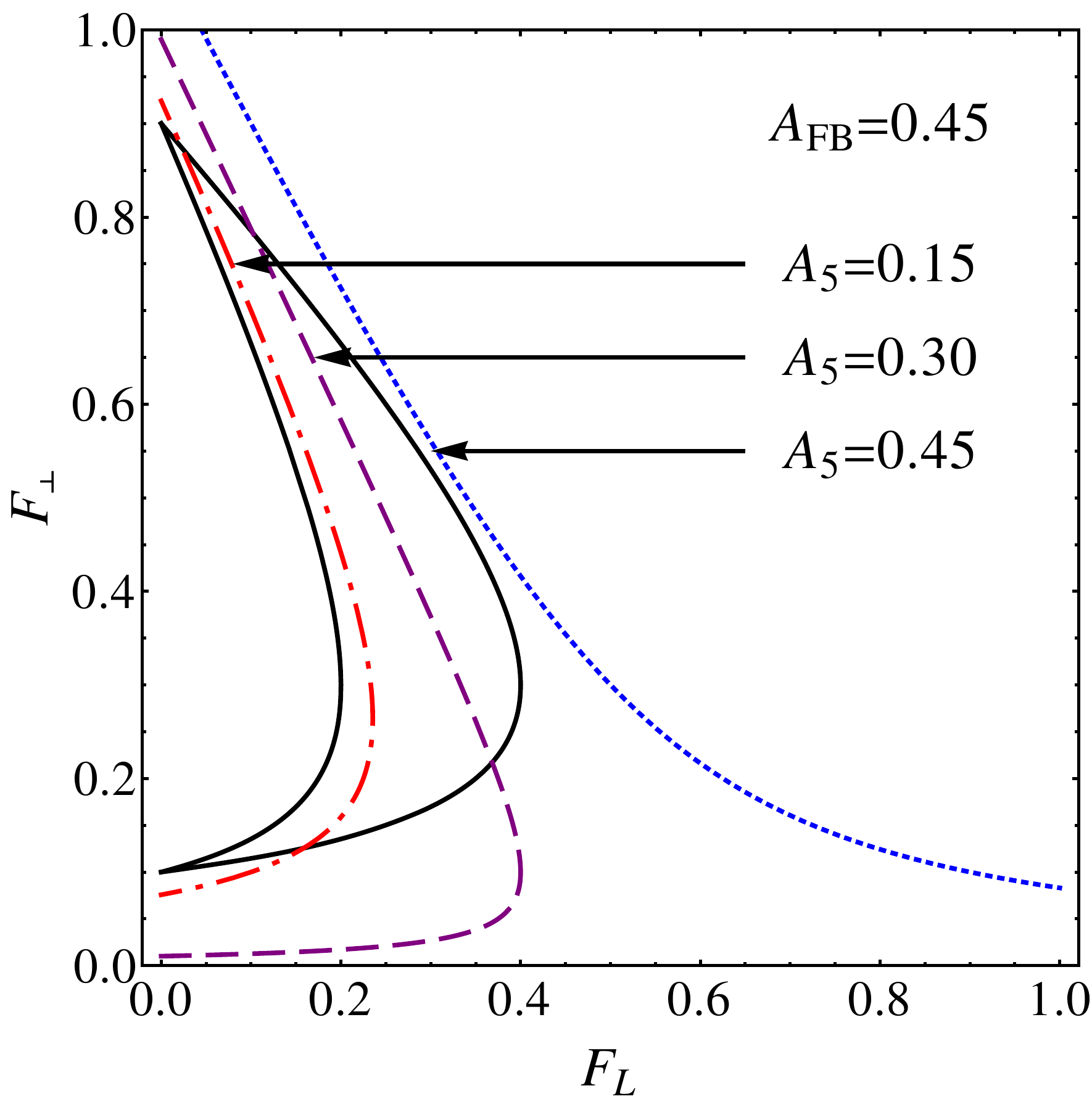}
\caption{The requirement that $Z_1, Z_2, Z_3$ must be real, for any
  consistent set of independent observables $\AFB$, $F_L$, $F_\perp$
  and $A_5$ constrains the allowed $F_L$-$F_\perp$ parameter space to
  lie only within the solid black lines. $A_4$ is given by
  Eq.~(\ref{eq:Obs-relation}).  Even with in the allowed
  $F_L$-$F_\perp$ domain only the region on the right is allowed
  depending on the values of $\AFB$ and $A_5$.  In the four figures we
  have sampled values of $\AFB$ and $A_5$ are as depicted. {\em There
    is no hadronic assumption made in obtaining the constraints
    depicted in these plots.}}
  \label{fig:bounds}
\end{figure*}

The arguments made above for $r_0+r_\perp$ can be repeated for
$r_{\!\wedge}+r_{\perp}$. One easily solves using Eqs.
~(\ref{eq2:Fperp}),~(\ref{eq2:A4}) and ~(\ref{eq2:A5AFB}):
\begin{widetext}
\begin{equation}
  r_{\!\wedge}+r_\perp= \pm\frac{\sqrt{\Gf}}{\sqrt{2}\mathcal{F}_\perp}
  \Big(\mathsf{P_3^2}(F_L+F_\|+\sqrt{2}\pi A_4)+F_\perp \pm \mathsf{P_3}
  Z_3\Big)^{\nicefrac{1}{2}}
\end{equation}
where, $\mathsf{P_3}$ has been defined in Eq.~(\ref{eq:P_3}) and we have defined 
\begin{equation}
Z_3 =\sqrt{4 (F_L+F_\|+\sqrt{2}\pi
  A_4)F_\perp-\frac{16}{9}(\AFB+\sqrt{2}A_5)^2}.
\end{equation}
Eq.~(\ref{eq2:A5AFB}) implies that $r_{\!\wedge}+r_\perp=0$ when
$\AFB+\sqrt{2}A_5=0$, hence,
\begin{equation}
\label{eq3:w+v}
r_{\!\wedge}+r_\perp\Big|_{\AFB+\sqrt{2}A_5=0}=\pm\frac{\sqrt{\Gf}}
{\sqrt{2}\mathcal{F}_\perp} \Big(\sqrt{F_\perp}
\pm \mathsf{P_3}\sqrt{F_L+F_\|+\sqrt{2}\pi A_4}\Big)=0
\end{equation}
\end{widetext}
Once again we choose the positive sign to fix the sign ambiguity since
$\mathsf{P_3}$ is negative. At the zero crossing points of $\AFB+\sqrt{2}A_5$
we hence have the equality,
\begin{equation}
  \label{eq:10}
\mathsf{P_3} =\dsp -\frac{\dsp \sqrt{F_\perp}}{\dsp\sqrt{F_L+F_\perp+\sqrt{2}\pi
    A_4}}\Bigg|_{\AFB+\sqrt{2}A_5=0}~.  
\end{equation}
Hence, the zero crossing of $\AFB+\sqrt{2}A_5$ enables the measurement
of form factor ratio $\mathsf{P_3}$ as well, in terms of
observables. Note however, that $\mathsf{P_3}$ is not independent and
related to $\mathsf{P_1}$ and $\mathsf{P_2}$ (see
Eq.~(\ref{eq:P_3}). The consequences of this relation will be
discussed later. The new solutions to $C_{10}$, $C_9$ and
$\widetilde{\mathcal{G}}_\parallel+\widetilde{\mathcal{G}}_0$:
\begin{widetext}
\begin{align}
\label{eq3:solC10}  
C_{10}=&\frac{\sqrt{\Gf}}{\sqrt{2}(\mathcal{F}_0+\mathcal{F}_\|)}
\frac{2}{3}\frac{A_{\text FB}+\sqrt{2}A_5}{\bigg[\pm\sqrt{\mathsf{P_3^2}
    (F_L+F_{\|}+\sqrt{2}\pi A_4)+F_\perp+ \mathsf{P_3} Z_3}\bigg]}~.\\
\label{eq3:solC9}  
C_9=& \frac{\sqrt{\Gf}}{\sqrt{2}(\mathcal{F}_0+\mathcal{F}_\|)}
\frac{\Big((F_L+F_\|+\sqrt{2}\pi
  A_4)\mathsf{P_3}\mathsf{P_3'}-F_\perp\Big)-\frac{1}{2}(\mathsf{P_3}-\mathsf{P_3'})Z_3}  {\bigg[\pm\sqrt{\mathsf{P_3^2} 
    (F_L+F_{\|}+\sqrt{2}\pi A_4)+F_\perp + \mathsf{P_3} 
    Z_3}\bigg]},\\
  \label{eq3:solC7}
\widetilde{\mathcal{G}}_\parallel+\widetilde{\mathcal{G}}_0=&\frac{\sqrt{\Gf}}{\sqrt{2}}
  \frac{\Big(\mathsf{P_3^2}(F_L+F_{\|}+\sqrt{2}\pi A_4)-F_\perp\Big)}
  {\bigg[\pm(\mathsf{P_3}-\mathsf{P_3'})\sqrt{\mathsf{P_3^2}(F_L+F_{\|}+\sqrt{2}\pi A_4)+F_\perp+
      \mathsf{P_3} Z_3}\bigg]}.
\end{align}
While these solutions may look more complicated they can also be
obtained from Eqs.~(\ref{eq:C10}), (\ref{eq:C9}) and (\ref{eq:C7}) by
the replacements $F_\|\to F_L+F_\|+\sqrt{2}\pi A_4$, $\AFB\to
\AFB+\sqrt{2}A_5$, $\mathcal{F}_\| \to \mathcal{F}_\|+\mathcal{F}_0$,
$\widetilde{\mathcal{G}}_\| \to \widetilde{\mathcal{G}}_\|+\widetilde{\mathcal{G}}_0$, which also imply
$r_\|\to r_{\!\wedge}$, $\mathsf{P_1}\to \mathsf{P_3}$ and $\mathsf{P_1'}\to \mathsf{P_3'}$.

Once again straightforward extrimization with respect to all the non
observables in Eq~(\ref{eq3:solC10}) results in the following bounds
on the form factor ratio $\mathsf{P_3}$,
\begin{equation}
   \label{eq1:P3sq}
   \mathsf{P_3^2}\lessgtr \frac{4(F_L+F_\|+\sqrt{2}\pi A_4)
     F_\perp-\tfrac{16}{9}(\AFB+\sqrt{2}A_5)^2} 
   {(F_L+F_\|+\sqrt{2}\pi A_4)^2}
   \quad\forall\quad (F_L+F_\|+\sqrt{2}\pi A_4) F_\perp \lessgtr
   \frac{2}{7}\Big(\frac{4(\AFB+\sqrt{2}A_5)}{3}\big)^2 \nn
\end{equation}
%
These bounds are a very good test of our understanding of the form factors.
Similar relations that can be derived from
Eqs.~(\ref{eq3:solC10}) -- (\ref{eq3:solC7}):
\begin{align}
  \label{eq3:C9by10}
  &\frac{C_9}{C_{10}}=\frac{2 ((F_L+F_\|+\sqrt{2}\pi A_4)\mathsf{P_3}
    \mathsf{P_3'}-F_\perp) -(\mathsf{P_3}-\mathsf{P_3'})
    Z_3}{\frac{4}{3}(\AFB+\sqrt{2}A_5) (\mathsf{P_3}-\mathsf{P_3'})}~,\\
  \label{eq3:C7by10}
  & \frac{\widetilde{\mathcal G}_\|+\widetilde{\mathcal G}_0}{C_{10}}=
 \frac{3}{2}({\cal F_\|+\cal
      F}_0)\frac{(\mathsf{P_3^2} (F_L+F_\|+\sqrt{2}\pi A_4)-F_\perp)}
 {(\AFB+\sqrt{2}A_5) (\mathsf{P_3}-\mathsf{P_3'})}~,
\end{align} 

Eqs.~(\ref{eq3:C9by10}) can be inverted to
obtain expressions for  $\AFB+\sqrt{2} A_5$ akin to the
expression for $\AFB$ obtained in Eq.~(\ref{eq:A_FB}). %
One easily derives:
\begin{align}
  \label{eq:A_FBA_5}
&\AFB+\sqrt{2}A_5 =\frac{3\big(RX_3-\sqrt{Y_3(\mathsf{P_3}-\mathsf{P_3'})^2(1+R^2)-X_3^2}\,\big)}
 {4(\mathsf{P_3}-\mathsf{P_3'})(1+R^2)}  
\end{align}
where $X_2=2(F_L\mathsf{P_3}\mathsf{P_3'}-F_\perp)$, $Y_2=4F_LF_\perp$,
$X_3=2((F_L+F_\|+\sqrt{2}\pi A_4)\mathsf{P_3}\mathsf{P_3'}-F_\perp)$ and
$Y_3=4(F_L+F_\|+\sqrt{2}\pi A_4)F_\perp$.

From Eqs.~\eqref{eq3:C9by10},~\eqref{eq3:C7by10} we can obtain yet another 
important relation, which is of the same kind as we obtained earlier in Eqs.
~(\ref{eq:7}) and (\ref{eq2:7a})
\begin{equation}
  \label{eq2:7b}
    (\frac{2}{3}\frac{C_7}{C_{10}}\mathsf{P_3^{'\!'}}
    -\frac{4}{3}\frac{C_9}{C_{10}}\mathsf{P_3} )(\AFB+\sqrt{2}A_5) 
 =\Big[(\mathsf{P_3^2}(F_L+F_\|+\sqrt{2}\pi A_4)+F_\perp+ \mathsf{P_3}Z_3\Big] >0
\end{equation}
\end{widetext}
where $\mathsf{P_3^{'\!'}}=(\mathcal G_0+\mathcal{G}_\|)/(\mathcal
F_0+\mathcal{F}_\|)\, (\mathsf{P_3}+\mathsf{P_3'})>0$. This is easily
verified to be true at leading order for the entire $q^2$ domain.  We
have shown earlier by doing a power expansion in $\AFB$ and $A_5$ ,
that respectively $(\mathsf{P_1^2}F_\|+F_\perp+\mathsf{P_1} Z_1)$ and
$(\mathsf{P_2^2}F_L+F_\perp+\mathsf{P_2}Z_2)$ are always
positive. It is easy to see that similar arguments can be made for the
positivity of $(\mathsf{P_3^2} (F_L+F_{\|}+ 2\sqrt{2}
\pi A_4) +F_\perp+ \mathsf{P_3}Z_3)$ by considering
expansions in $\AFB+\sqrt{2}A_5$.  These equations are equally useful
to determine the sign of $C_7$ as discussed earlier, however, the form
factors involved are not completely free from HQET form factor.

Eq.~(\ref{eq2:solC10})--(\ref{eq2:solC7}) and
Eq~(\ref{eq3:solC10})--(\ref{eq3:solC7}) have been expressed in terms
of from factor ratios $\mathsf{P_2}$, $\mathsf{P_2'}$, $\mathsf{P_3}$, $\mathsf{P_3'}$, which are not
completely free from the hadronic form factors, both at large 
and at low recoil. The form factor ratios $\mathsf{P_1}$ and $\mathsf{P_1'}$ on the 
other hand is completely free from the Isgur-Wise form factors
$\xi_\|$ and $\xi_\perp$ in the limit of heavy quark and large
recoil of the vector meson. We can express the form factor ratios
$\mathsf{P_2}$, $\mathsf{P_2'}$, $\mathsf{P_3}$, $\mathsf{P_3'}$ in terms of $\mathsf{P_1}$ and $\mathsf{P_1'}$.
Equating the relations obtained for $C_9/C_{10}$ and $C_7/C_{10}$ in 
Eqs.~(\ref{eq:C9by10}), (\ref{eq:C7by10}) with those in 
Eqs.~(\ref{eq2:C9by10}), (\ref{eq2:C7by10}) we obtain relations only
between form $\mathsf{P_1}$, $\mathsf{P_1'}$, $\mathsf{P_2}$, $\mathsf{P_2'}$ and observables. The two
equations so obtained can be used to solve for $\mathsf{P_2}$ and $\mathsf{P_2'}$ in
terms of $\mathsf{P_1}$ and $\mathsf{P_1'}$. 
%
\begin{eqnarray}
  \label{eq:P2}
\mathsf{P_2}&=&\frac{2\mathsf{P_1}\AFB F_\perp}{\sqrt{2}A_5(2F_\perp+Z_1\mathsf{P_1})-Z_2\mathsf{P_1}\AFB×}\\
\label{eq:P2p}
 \mathsf{P_2'}&=&\frac{\sqrt{2}A_5\Big(F_\perp-F_\|\mathsf{P_1^2}
   \Big)\mathsf{P_2^2}\mathsf{P_1'}} 
{\AFB T_2(\mathsf{P_1}-\mathsf{P_1'})+\sqrt{2}A_5\Big(F_\perp-F_\|
\mathsf{P_1^2} \Big)\mathsf{P_2} \mathsf{P_1'}}~~~~
\end{eqnarray}
where
\begin{equation}
  \label{eq:T2}
T_2=  \mathsf{P_1}(F_\perp-F_L \mathsf{P_2^2})
\end{equation}

We emphasize that while $\mathsf{P_2}$ and $\mathsf{P_2'}$ on the left
hand side depend on the Isgur-Wise wave functions $\xi_\|$ and
$\xi_\perp$, $\mathsf{P_1}$ and $\mathsf{P_1'}$ are independent of
them. These two equation can be used to obtain information about the
wave functions. Eq.~(\ref{eq:P2}) is also very important in the sense
that the domain of observables is itself constrained by the terms
under the radical signs must be positive to obtain real
$\mathsf{P_2}$. Similar relations for $\mathsf{P_3}$ and
$\mathsf{P_3'}$ in terms of $\mathsf{P_1}$ and $\mathsf{P_1'}$ can be
obtained by using Eq.~(\ref{eq:C9by10}), (\ref{eq:C7by10}) and
Eq.~(\ref{eq3:C9by10}), (\ref{eq3:C7by10}) to get:
%
 \begin{eqnarray}
  \label{eq:P3}
\mathsf{P_3}&=&\frac{2\mathsf{P_1}\AFB F_\perp}{(\AFB+\sqrt{2}A_5)(2F_\perp+Z_1\mathsf{P_1})-Z_3\mathsf{P_1}\AFB×},~~\\
  \label{eq:P3p}
\mathsf{P_3'}&=&\displaystyle\frac{(\AFB+\sqrt{2}A_5)(F_\perp-F_\|\mathsf{P_1^2})\mathsf{P_3^2}\mathsf{P_1'}}
{\AFB T_3(\mathsf{P_1}-\mathsf{P_1'})+\sqrt{2}A_5(F_\perp-F_\|\mathsf{P_1^2})\mathsf{P_3^2}\mathsf{P_1'}},~~~~
\end{eqnarray}
where 
\begin{eqnarray}
\label{eq:T3}
T_3&=&\mathsf{P_1}
\Bigg[F_\perp(1+\mathsf{P_3^2})-\mathsf{P_3^2}(1+\sqrt{2}\pi A_4)\Bigg]
\end{eqnarray}

\begin{figure}[!htbp]
 \begin{center}
  \includegraphics[width=0.35\textwidth]{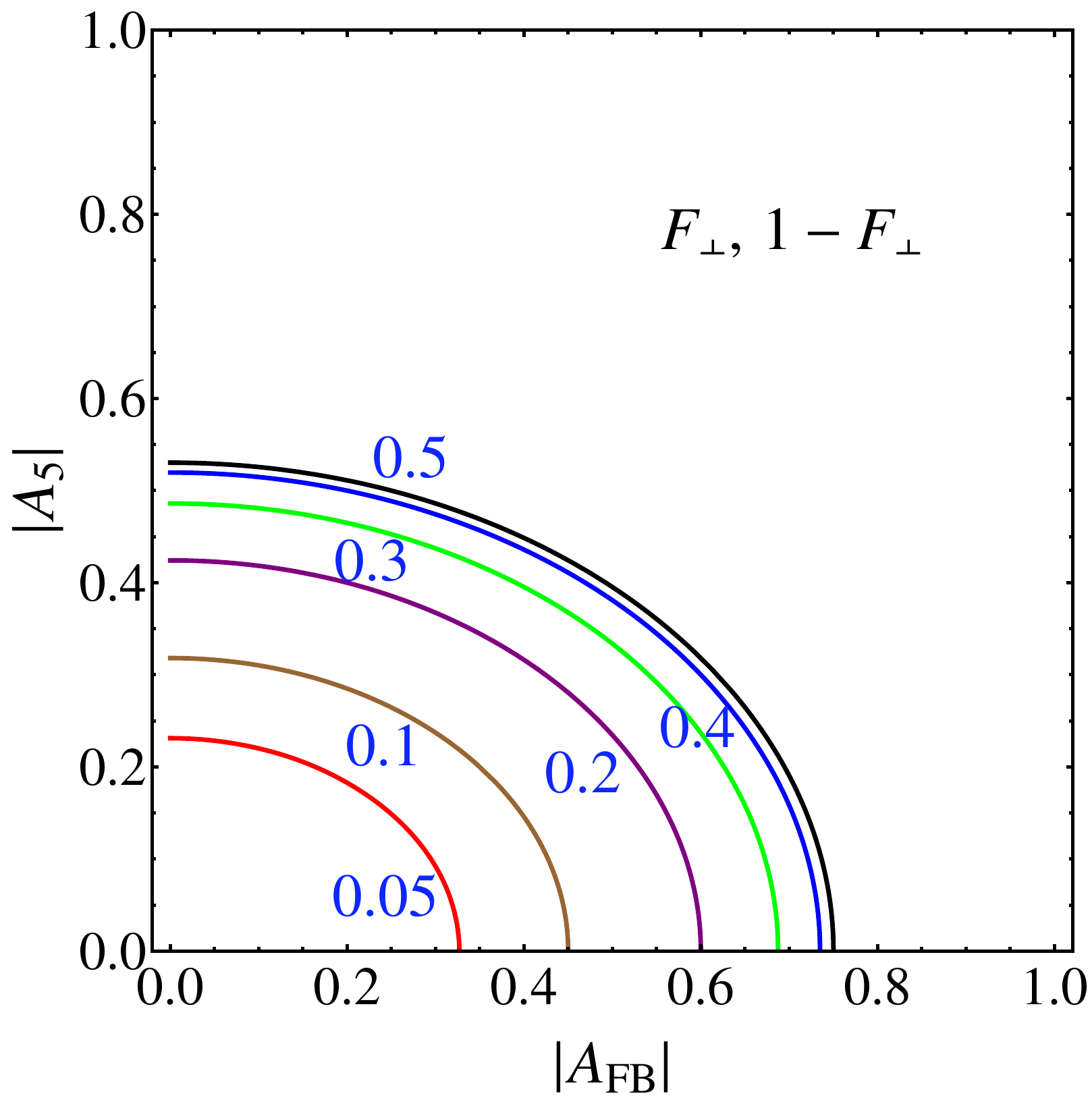}
   \caption{The constraint on $\AFB$, $A_5$ and $F_\perp$ arrived at by
   Eq.~(\ref{eq:realZ4}). The depicted values correspond to both
   $F_\perp$ and $1-F_\perp$.}
   \label{fig:A5vsqsq}
  \end{center}
\end{figure}
As emphasized earlier the Wilson coefficients are real constants
except for the non resonant regions.  This implies that just like
$Z_1$, both $Z_2$ and $Z_3$ are always real if resonant regions and CP
violation are excluded:
\begin{align}
  \label{eq:realZ2}
  4 F_LF_\perp &\geq \frac{16}{9} (\sqrt{2}A_5)^2~.\\
  \label{eq:realZ3}
  4 (F_L+F_{\|}+\sqrt{2}\pi A_4)F_\perp&\ge\frac{16}{9}
  (A_{\text FB}+\sqrt{2}A_5)^2
\end{align}
The combination of bounds in Eqs.~(\ref{eq:realZ1}) and (\ref{eq:realZ2})
results in yet another interesting bound among observables alone but 
involving only $\AFB^2$, $A_5^2$ and $F_\perp$:
\begin{align}
  \label{eq:realZ4}
  4 (1-F_\perp) F_\perp\geq \frac{16}{9}(\AFB^2+2 A_5^2).
\end{align}
In Fig.~\ref{fig:A5vsqsq} we depict the constraint on $\AFB$, $A_5$
and $F_\perp$ arrived at by Eq.~(\ref{eq:realZ4}). We emphasize that
like the bounds derived in Eqs.~(\ref{eq:realZ1}), (\ref{eq:realZ2})
and (\ref{eq:realZ3}) this bound is also completely free from any
hadronic uncertainty.

\begin{figure*}[thb]
  \centering
\includegraphics[width=0.4\textwidth]{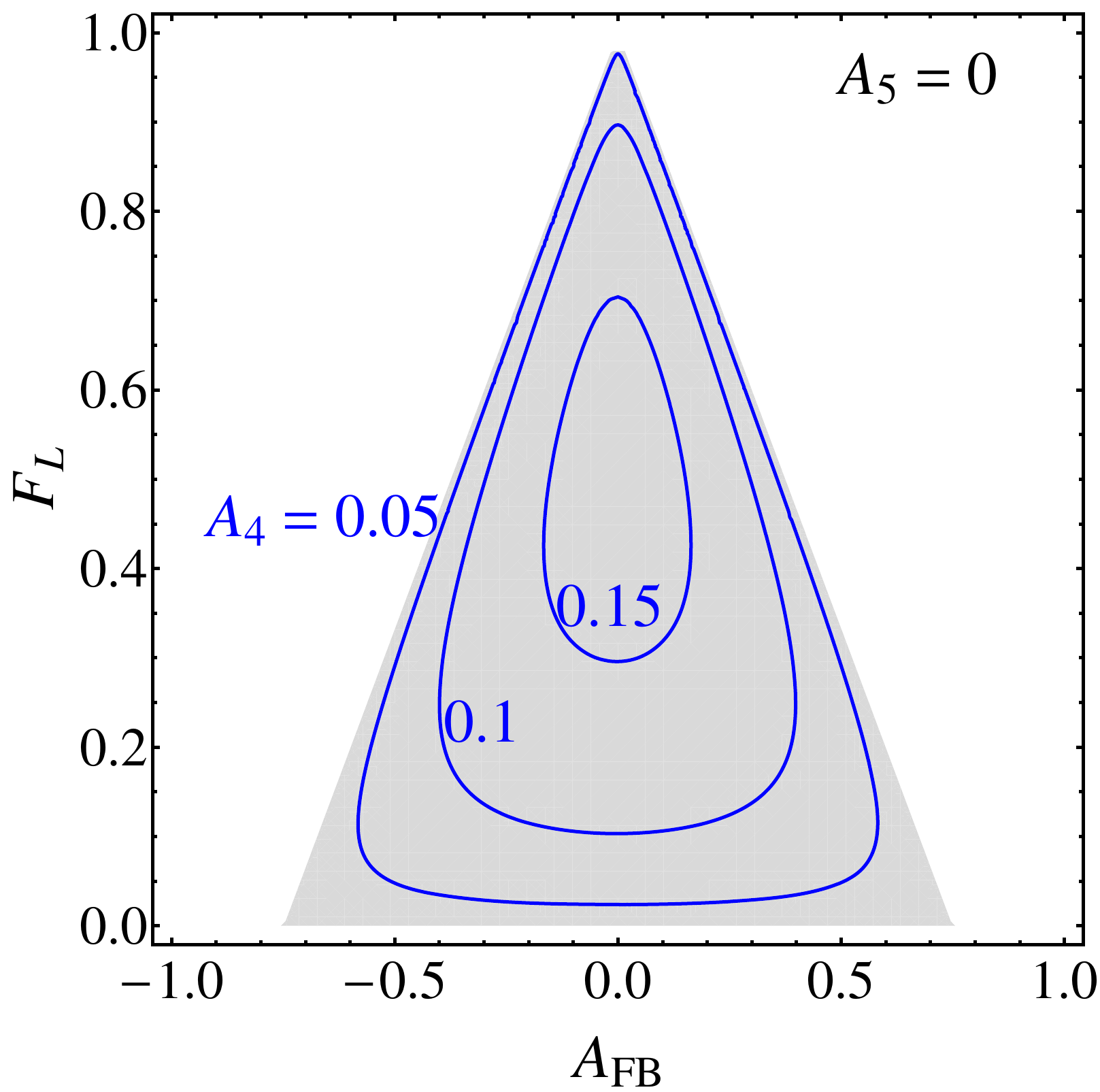}
\includegraphics[width=0.4\textwidth]{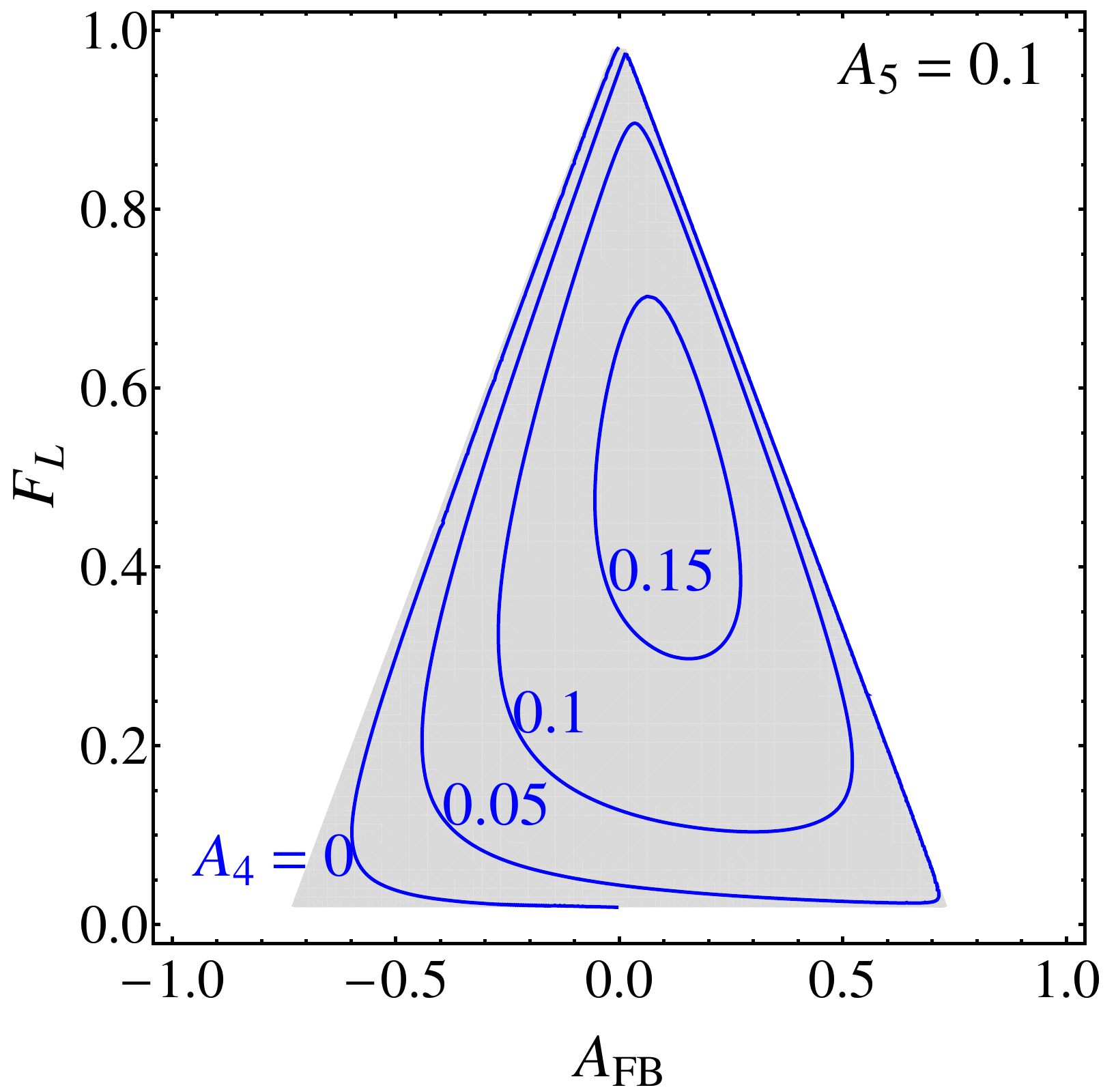}
\includegraphics[width=0.4\textwidth]{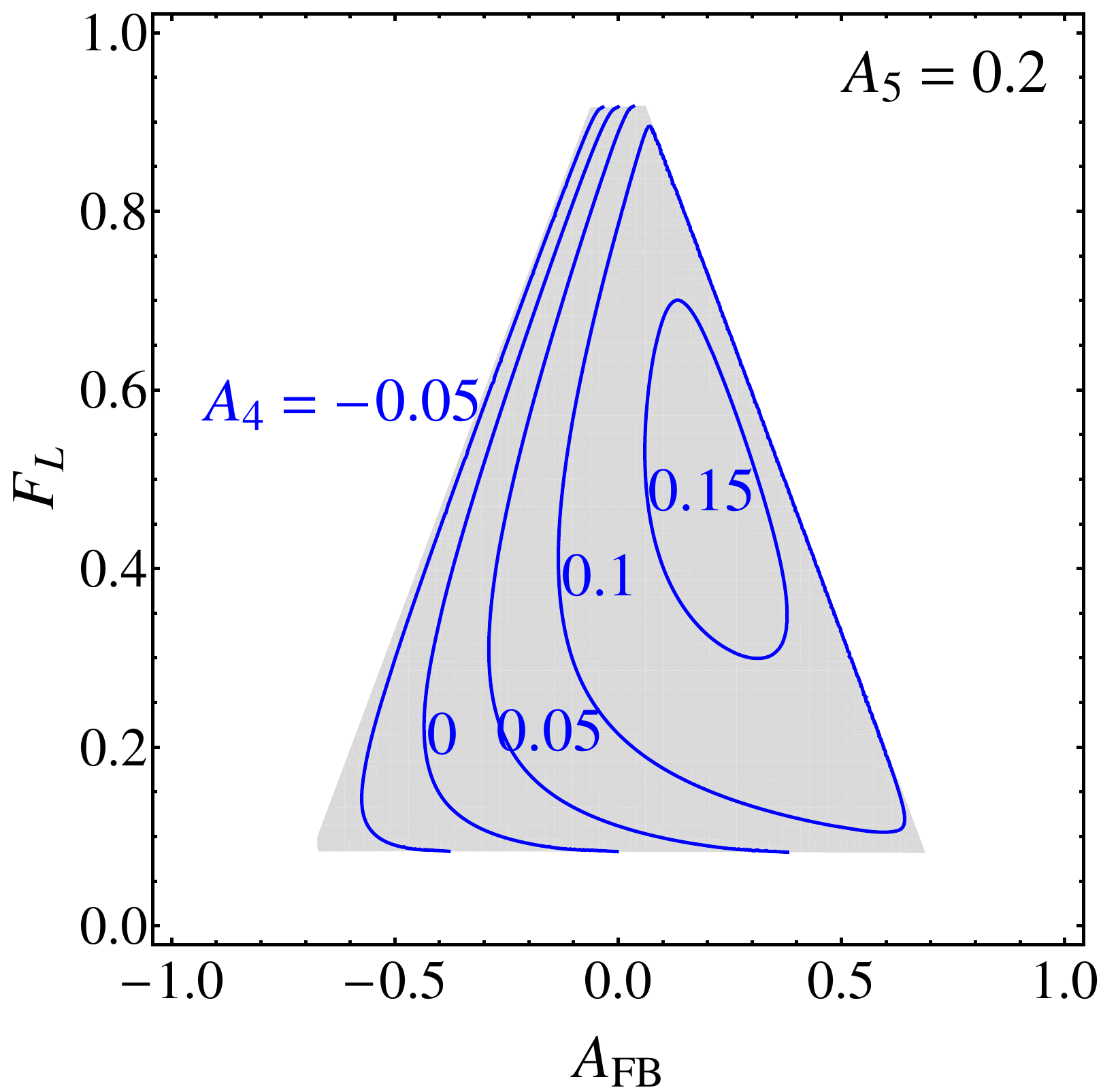}
\includegraphics[width=0.4\textwidth]{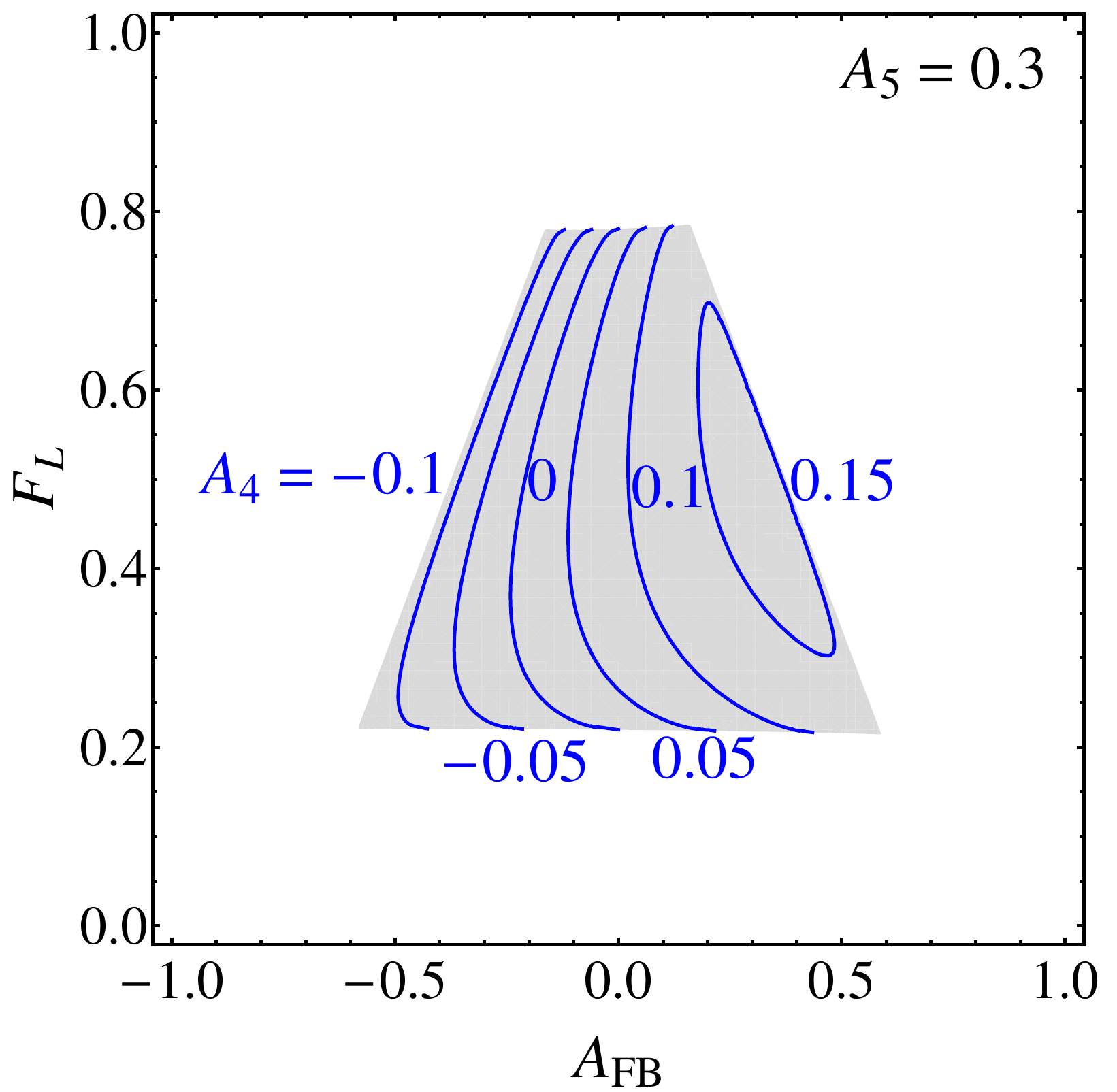}
\caption{The allowed region in the $F_L$-$F_\perp$ parameter space,
  shaded as gray, for $R=-1$ and different values of $A_5$. The values
  of $\mathsf{P_1}$ and $\mathsf{P_1'}$ are averaged over $1~\gev^2\le
  q^2\le 6~\gev^2$.  The blue lines correspond to the value of $A_4$
  that is estimated using Eq.~(\ref{eq:Obs-relation}).}
  \label{fig:teardrop}
\end{figure*}
\begin{figure}[thb]
  \centering
\includegraphics[width=0.4\textwidth]{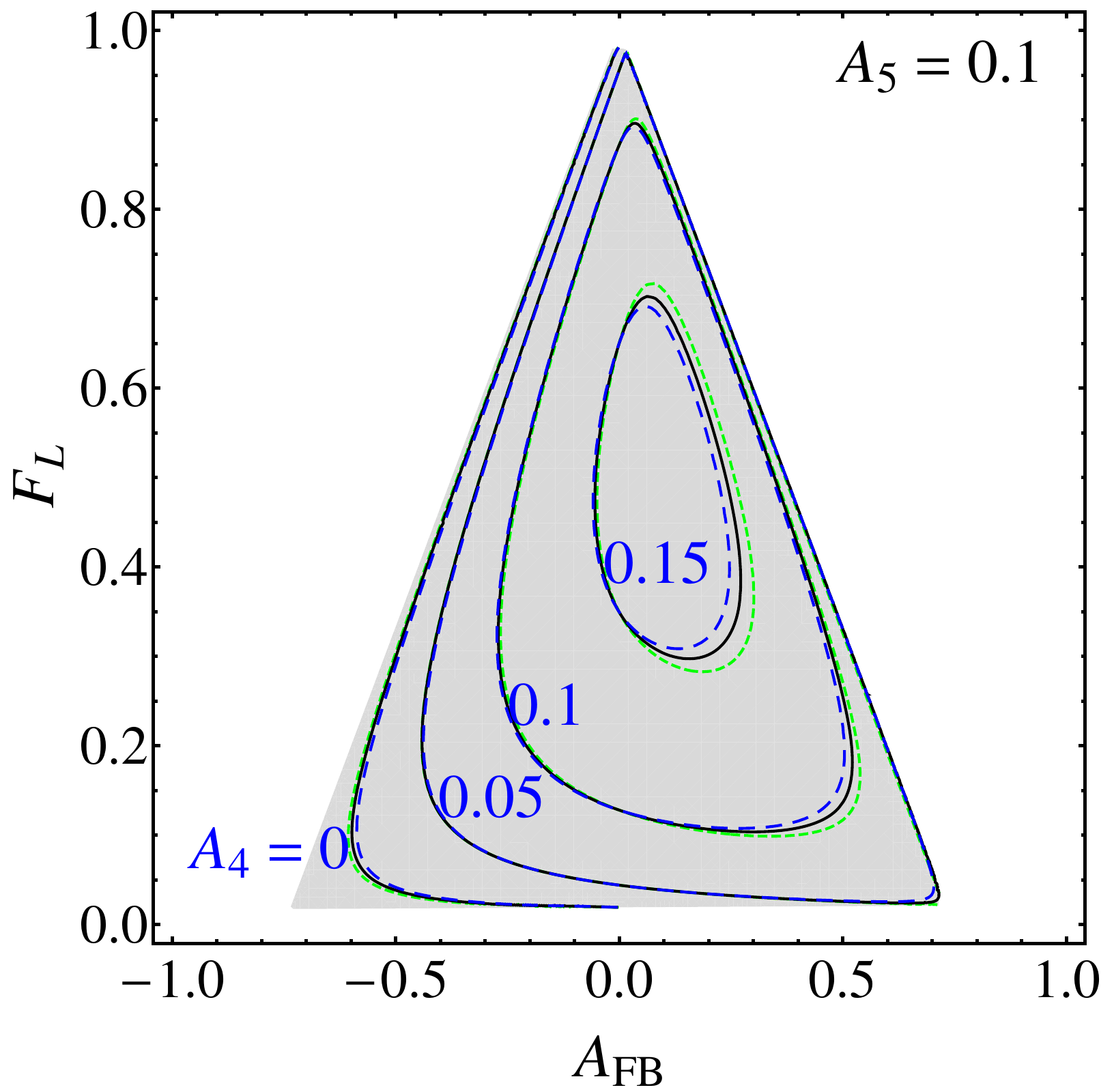}
\caption{The same as Fig.~\ref{fig:teardrop}, but studying the variation in
  $R$. The small dashed (green) curves are for the case $R=10$ while the big
  dashed (blue) curve correspond to $R=-10$. The solid black curves
  are for Standard model value of $R=-1$. Note the insensitivity to the
value of $R$ for the large recoil region $1~\gev^2\le
  q^2\le 6~\gev^2$.}
  \label{fig:teardrop-R}
\end{figure}
In Eq.~(\ref{eq:P_3}) we showed that $\mathsf{P_3}$ is not independent but
related to $\mathsf{P_1}$ and $\mathsf{P_2}$. $\mathsf{P_3}$ and $\mathsf{P_2}$ are themselves expressed
in terms of observables and $\mathsf{P_1}$ in Eqs.~(\ref{eq:P3}) and
(\ref{eq:P2}) respectively. This constraint results in an interesting relation
that depends on observables alone:
\begin{equation}
  \label{eq:Z-constraint}
  Z_3=Z_1+Z_2.
\end{equation}
We use this relation to solve for $A_4$ leading to
\begin{widetext}
\begin{equation}
  \label{eq:Obs-relation}
  A_4=
\frac{8 A_5 \AFB}{9 \pi F_\perp}+\sqrt{2}\,\frac{\sqrt{F_LF_\perp
       -\frac{8}{9} A_5^2}\sqrt{F_\|F_\perp -\frac{4}{9}\AFB^2}}{\pi F_\perp}
\end{equation}
\end{widetext}
Since $F_\perp$ is already predicted in Eq.~(\ref{eq:AFB-FP}) in terms
of the already measured observables $F_L$ and $\AFB$ and $\mathsf{P_1}$, $\mathsf{P_1'}$
and $R$, we can estimate $A_4$ in terms of $A_5$. The correlations
predicted by Eq.~(\ref{eq:Obs-relation}) would have to hold unless NP
contributes. In Fig.~\ref{fig:teardrop} we present the correlation
between the observables. It may be noted that Eq.~(\ref{eq:Obs-relation}) is a
relation involving only observables without any assumptions of
hadronic form factors, hence its violation must be an unambiguous
signal of NP.

Let us summarize the approach that has led to these solutions. We have
six observables, the decay width of $B\to \kstar \ell^+\ell^-$, $\Gf$,
the helicity fractions $F_L$ and $F_\perp$ and the angular asymmetries
$\AFB$, $A_4$ and $A_5$. These six observables are expressed in terms
of eight theoretical parameters in the most general approach. The
parameters being the six effective form factors $\mathcal{F}_0$,
$\mathcal{F}_\|$, $\mathcal{F}_\perp$, $\widetilde{\mathcal{G}}_0$,
$\widetilde{\mathcal{G}}_\|$ and $\widetilde{\mathcal{G}}_\perp$ and
the two Wilson coefficients $C_9$ and $C_{10}$. Three of the
observables $\Gf$, $F_L$ and $\AFB$ have already been measured by
several experiments.  We assume three further inputs-- the ratio
$R=C_9/C_{10}$ as it is theoretically reliably estimated in SM and the
ratios $\mathsf{P_1}$ and $\mathsf{P_1'}$ of form factors as defined
in section.  $\mathsf{P_1}$ and $\mathsf{P_1'}$ are accurately
predicted theoretically in the heavy quark limit to be free from
higher order corrections and the known universal form factors $\xi_\|$
and $\xi_\perp$.  These inputs allow us to estimate $F_\perp$. We find
that making assumption of one further observable $A_5$ we are able to
predict the only remaining observable $A_4$, completely free from
hadronic parameters or estimate of $R$. Clearly only five of the
observables are independent in SM and $\mathcal{F}_\|$ remains
unsolved given all the observables possible. It has also been realized
earlier~\cite{Egede:2010zc} following a different approach that there
exist symmetries in the angular distribution which reduce the number
of independent observable. {\em We emphasize that in our approach,
  $C_9/C_{10}$ and all the expressions independent of Wilson
  coefficients are ``clean" in the large recoil limit.}

\section{The low recoil limit}
\label{sec:low-recoil-detail}

In Sec~\ref{sec:low-recoil} we found that (see
Eqs.~\eqref{eq:form-factors-equal} and \eqref{eq:r-equal}) in the
low-recoil limit the form-factors satisfied the conditions
\begin{equation*}
\frac{\mathcal{G}_\|}{\mathcal{F}_\|}=
\frac{\mathcal{G}_\perp}{\mathcal{F}_\perp}=
\frac{\mathcal{G}_0}{\mathcal{F}_0}=\hat{\kappa},
\end{equation*}
which implies that 
\begin{equation*}
  r_\| = r_\perp = r_0 = r_\wedge\equiv r.
\end{equation*}
This reduces the number of independent relations (see
Eqs. \eqref{eq1:FL}--\eqref{eq1:AFB}) and the low recoil limit thus
needs to be treated more carefully. In this limit the Wilson coefficients
$C_7$ and $C_9$ cannot be solved following the approach in
Appendix~\ref{sec:appendix-1} as is obvious from Eq.~\eqref{eq:21}.
We will however be able to solve for $r$ and in turn for $C_7$ and
$C_9$ if $\hat{\kappa}$ is assumed or equivalently with the additional
input of $\mathcal{G}_\|$, since $\mathcal{F}_\|$ is in any case a
required input. This results in one additional constraint
relations between observables. In this section we derive a new
relations among observables that will test the validity of the
assumption on the form factors in the low recoil limit. We will also
elaborate on various other constraints in this limit.

We begin by considering Eq.~\eqref{eq:def1}, \eqref{eq:def2} and
\eqref{eq:def3} in the low recoil limit. Clearly since $r^2+C_{10}^2$
is independent of helicity, Eqs.~\eqref{eq:def1} and \eqref{eq:def2}
reduce to the same equation, hence, we have 
\begin{subequations}
  \label{eq:def-soft}
\begin{align}
  r^2+C_{10}^2=& \frac{\Gf F_\|}{2\mathcal{F}_\|^2}= \frac{\Gf
    F_\perp}{2\mathcal{F}_\perp^2}\equiv\frac{\hat{F}\Gf}{2}  \\ 
  4 r C_{10}=&\frac{2 \AFB \Gf}{3
    \mathcal{F}_\|\mathcal{F}_\perp}\equiv \frac{4 \AFB}{3
    \sqrt{F_\|F_\perp}}\frac{\hat{F}\Gf}{2},
\end{align}  
\end{subequations}
where 
\begin{equation}
  \label{eq:fhat}
  \hat{F}\equiv\frac{F_\|}{\mathcal{F}_\|^2}= \frac{F_\perp}{\mathcal{F}_\perp^2}.
\end{equation}
Eq.~\eqref{eq:new-P-relation} then implies that 
\begin{equation}
  \label{eq:new-relation}
\mathsf{P_1^2}=\mathsf{P_1'}^2=\frac{F_\perp}{F_\|}=
\frac{\mathcal{F}_\perp^2}{\mathcal{F}_\|^2}.   
\end{equation}
It is obvious from Eq.~\eqref{eq:def-soft} that we can solve for $r^2$
and $C_{10}^2$: 
\begin{align}
  \label{eq:18}
r^2
    =&\frac{\hat{F}\Gf}{4} \Big(1+\frac{Z_1}{\sqrt{2 F_\|F_\perp}}\Big)\\
C_{10}^2 
     =&\frac{\hat{F}\Gf}{4} \Big(1-\frac{Z_1}{\sqrt{2 F_\|F_\perp}}\Big)
\end{align}
The sign of $r/C_{10}$ is fixed such that 
\begin{equation}
  \label{eq:rbyC10}
  \frac{r}{C_{10}}=\frac{3}{4}\,\frac{2\sqrt{F_\perp F_\|}+Z_1}{\AFB},
\end{equation}
in order to satisfy the limit derived by appropriate combination of
Eqs.~\eqref{eq:C7} and \eqref{eq:C9}.

\begin{figure*}[!hbt]
  \centering
\includegraphics[width=0.4\textwidth]{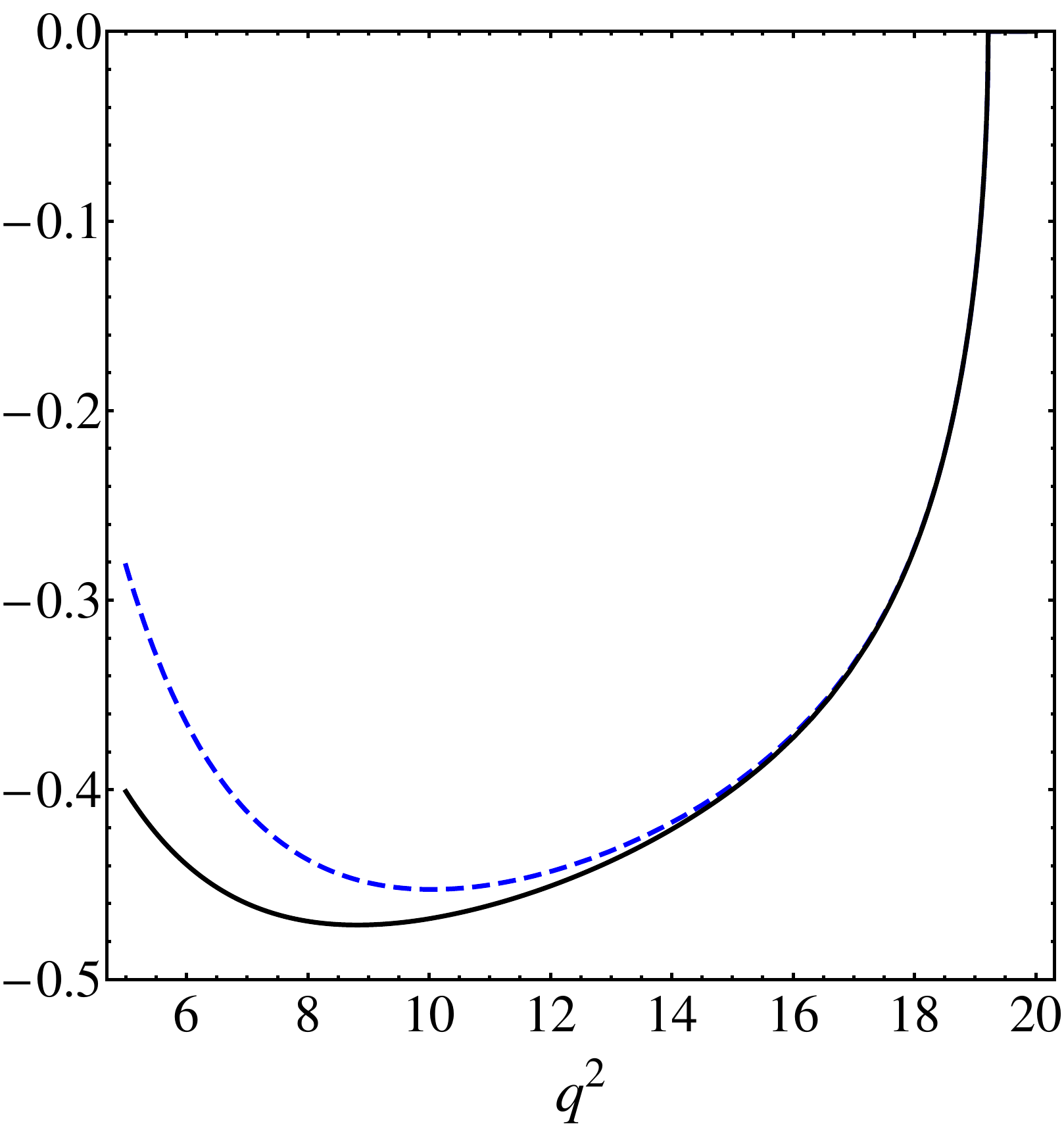}
\includegraphics[width=0.4\textwidth]{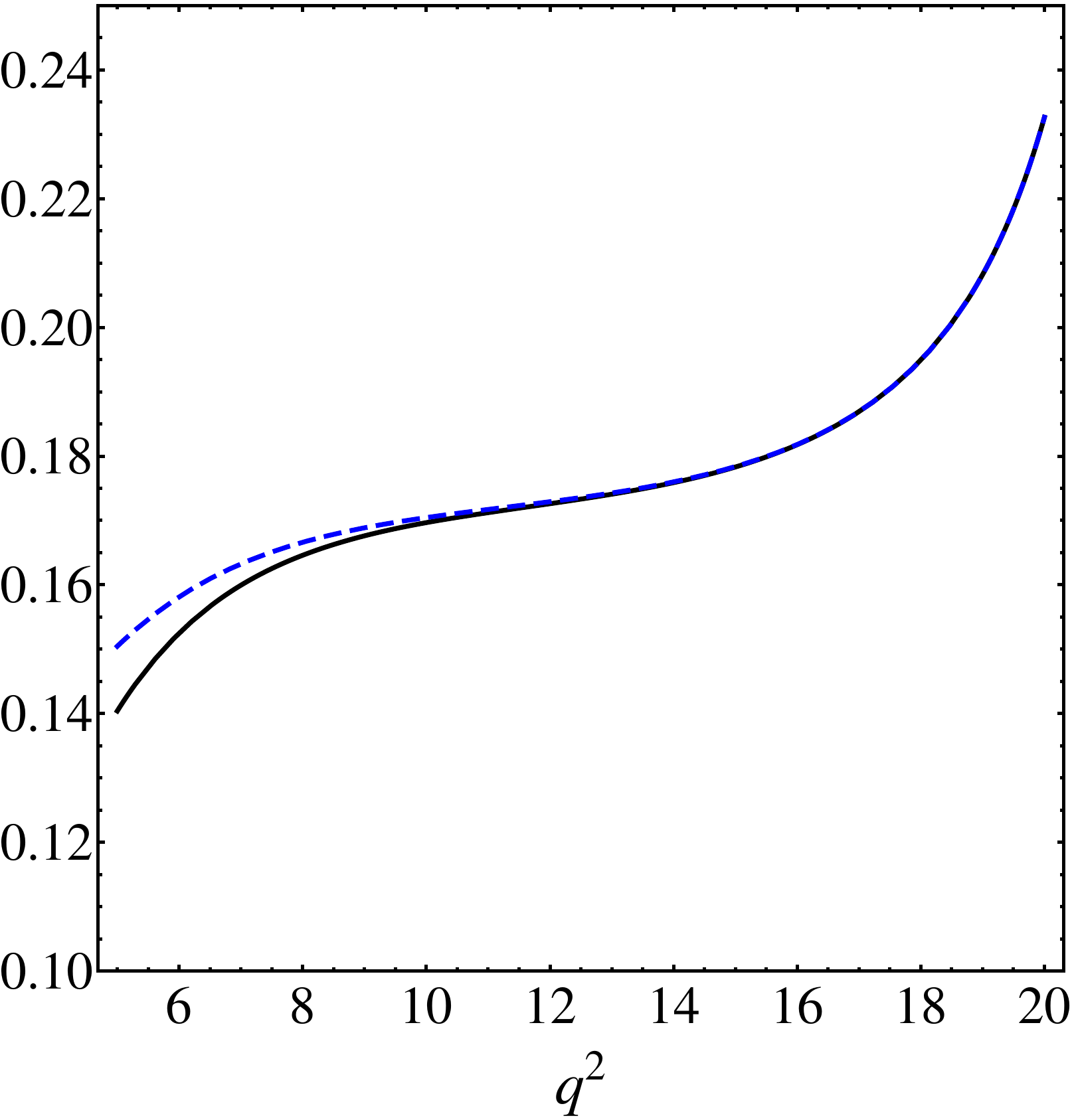}
\caption{In the figure to the left the left hand side (solid curve)
  and right hand side (dashed blue curve) of
  Eq.~\eqref{eq:low-relation1} are plotted. The figure on the right is
  the corresponding figure for Eq.~\eqref{eq:low-relation2}. These
  figures demonstrate the domain of validity in $q^2$ for the low
  recoil approximation and the region where new physics can be
  tested. The values are estimated using the form factors given in
  Table~\ref{table:nonlin}.}
  \label{fig:low recoil}
\end{figure*}
In the low recoil limit ``$r$'' is same not just for $\|$ and $\perp$
helicities but for all three helicities. This requires in analogy with
Eq.~\eqref{eq:new-relation} that,
\begin{subequations}
  \label{eq:P_i}
\begin{align}
\mathsf{P_1^2}=\mathsf{P_1'}^2=&\frac{F_\perp}{F_\|}=\frac{\mathcal{F}_\perp^2}
{\mathcal{F}_\|^2},\\  
\mathsf{P_2^2}=\mathsf{P_2'}^2=&\frac{F_\perp}{F_L}=\frac{\mathcal{F}_\perp^2}
{\mathcal{F}_0^2},\\
\mathsf{P_3^2}=\mathsf{P_3'}^2=&\frac{F_\perp}{(F_L+F_\|)}=
\frac{\mathcal{F}_\perp^2}{(\mathcal{F}_0^2+\mathcal{F}_\|^2 ) }.  
\end{align}
\end{subequations}
One can hence, measure $\mathsf{P_1}$, $\mathsf{P_2}$ and
$\mathsf{P_3}$ in the low recoil region in terms of the ratio of
helicity fractions.  Hence, the value $C_{10}^2 \mathcal{F}_\|^2$ can
be expressed in terms of observables alone. In the large recoil case
$C_{10}^2 \mathcal{F}_\|^2$ depended on $\mathsf{P_1}$ and
$\mathsf{P_2}$. The form factor $\mathsf{P_1}=\mathsf{P_1'}$ can be
measured, enabling a possibility of verifying the estimates of
presented in Table~\ref{table:nonlin}.  To derive a relation between
observables that is valid at low recoil and tests the validity of the
approximation we note Eq.~\eqref{eq:def-soft} leads
to the generalized relation
\begin{align}
  \label{eq:r-generalized}
  \frac{ r^2+C_{10}^2}{2 r C_{10}}=&
  \frac{2}{3}\frac{\AFB}{\sqrt{F_\|F_\perp}}\nn\\ 
  =&\frac{2}{3}\frac{\sqrt{2}A_5}{\sqrt{F_LF_\perp}}\nn\\
  =&\frac{2}{3}\frac{(\AFB+\sqrt{2}A_5)}
  {\sqrt{(1-F_\perp+\sqrt{2}\pi A_4)F_\perp}}
\end{align}
The equalities on the left side of the above equation yields two
interesting relations
\begin{align}
  \label{eq:low-relation1}
\sqrt{2}A_5=&{\AFB}\frac{\sqrt{F_L}}{\sqrt{F_\|}}\\
  \label{eq:low-relation2}
A_4=&\frac{\sqrt{2}}{\pi}\sqrt{F_LF_\|}
\end{align}
It is easily seen by direct substitution of
Eq.~\eqref{eq:low-relation1} in Eq.~\eqref{eq:Obs-relation} that it
reduces to Eq.~\eqref{eq:low-relation2}, hence it is not
independent. It is emphasized that a reasonable validity of the low
recoil approximation requires large $q^2$ and not the exact equality
of form factors as derived Eq.~\eqref{eq:P_i}. Even though the
values of the form factors depicted in Table~\ref{table:nonlin} are
not exactly equal, the low recoil approximation works well as can be
seen from Fig.~\ref{fig:low recoil} where we plot the left hand and
right hand of Eqs.~\eqref{eq:low-relation1} and
\eqref{eq:low-relation2}. These figures demonstrate the domain of
validity of the low recoil approximation and the region where new
physics can be tested. The values of observables are estimated using
the form factors given in Table~\ref{table:nonlin}.

We emphasize that the relation derived in
Eqs.~\eqref{eq:low-relation1} and \eqref{eq:low-relation2} are
extremely important both in testing the validity of the low recoil
approximation and the presence of New Physics. The value of $A_5$
predicted by these relations tests the validity of the low recoil
approximation, whereas the value of $A_4$ verifies the validity of
SM. If both the relations are found to be valid it would prove both
the validity of the low recoil limit and the absence of New
Physics. On the other hand if both the relations fail we must conclude
that low recoil limit is not valid. The presence of New Physics could
still be tested by the validity Eq.~\eqref{eq:Obs-relation} even in
this large $q^2$ domain. The remaining meaningful possibility is that
Eq.~\eqref{eq:low-relation1} holds and \eqref{eq:low-relation2} is
violated. This would imply validity of low recoil limit but signal the
presence of New Physics.  It is interesting to note that one should
expect from Eqs.~\eqref{eq:low-relation1} and \eqref{eq:low-relation2}
a very tiny product of asymmetries $A_4$ and $A_5$.
\begin{equation}
  \label{eq:low-tight}
A_4A_5=\frac{\AFB F_L}{\pi}
\end{equation}
since the right hand side $\AFB$ and $F_L$ have already been
measured.  {\em We emphasize that even in the low recoil limit,
  $C_9/C_{10}$ and all the expressions independent of Wilson
  coefficients are independent of the universal form factors $\xi_\|$ and
  $\xi_\perp$.}

\section{Conclusions}
\label{sec:conclusions}

In this paper we have derived several important new results.  After a
brief introduction, we discuss the differential decay distribution of
$B\to \kstar \ell^+\ell^-$ and introduced the observables $\Gf$,
$F_L$, $F_\perp$, $\AFB$, $A_4$ and $A_5$. While the partial decay
rate $\Gamma_f$ can be measured by angular integration, the other
observables require a study of angular distributions. We showed how
uni-angular distributions in the azimuthal angle $\phi$ can be used to
measure the helicity fraction $F_\perp$. $F_L$ and $\AFB$ have already
been measured by studying the uni-angular distribution in
$\thl$. $A_4$ and $A_5$ can only be measured by a complete angular
analysis involving $\thl$ and $\phi$ requiring higher
statistics. After setting up our notation and defining the observables
in terms of form factors, we expressed the amplitude in the most
general form within the Standard Model as $\mathcal{A}_\lambda^{
  L,R}=C_{L,R} \mathcal{F}_\lambda -\widetilde{\mathcal{G}}_\lambda$,
where $\lambda=\{0,\perp,\| \}$ is the helicity of the $\kstar$,
$C_{L,R}=C_9^\eff\mp C_{10}$ and $L,R$ defines the chirality of the
$\ell^-$.  The form factors $\mathcal{F}_\lambda$ and
$\widetilde{\mathcal{G}}_\lambda$ are expressed in terms of
conventional $B\to \kstar$ form factors $V$, $A_{1,2}$ and
$T_{1,2,3}$. To be exact $\widetilde{\mathcal{G}}_\lambda\equiv C_7
\mathcal{G}_\lambda+\cdots$ with the dots representing the higher
order and non factorizable contributions and only at leading order
$\mathcal{G}_\lambda$'s are related to $T_{1,2,3}$. It may be noted that
even at leading order $C_7$ and $\mathcal{G}_\lambda$ cannot be
separated and $C_7$ can only be defined at leading order on assuming
$\mathcal{G}_\lambda$. The six observables are thus defined in terms
eight parameters, the six form factors $\mathcal{F}_\lambda$,
$\widetilde{\mathcal{G}}_\lambda$ and two Wilson Coefficients
$C_{9,10}$. Hence only six theoretical parameters can be eliminated in
terms of observables and a minimum of two reliable theoretical inputs
are needed, to resolve between new physics and hadronic contributions.
This is made possible by the significant advances in our understanding
of form-factors that permit us to make truly these reliable
inputs. One of our achievements are derivations of ``clean relations''
that permit the verifications of these hadronic inputs.

The $B\to \kstar$ form factors are estimated using heavy quark
effective theory and the treatment varies on the recoil energy of the
$\kstar$. At large recoil the ratio of the form factors
$\mathsf{P_1}=\mathcal{F}_\perp/\mathcal{F}_\|$ and $\mathsf{P_1'}
=\widetilde{\mathcal{G}}_\perp/ \widetilde{\mathcal{G}}_\|$ are
reliably evaluated at ${\cal O}(1/m_b)$ to be free from universal wave
functions and are unaltered by non-factorizable contributions and
higher order corrections in $\alpha_s$. In the large recoil limit we
therefore choose $\mathsf{P_1}$ and $\mathsf{P_1'}$ as the two inputs
in addition to observables.  In the low recoil limit the relations
$\mathsf{P_1}=\mathsf{P_1'}$ between the form factors serves as an
additional input.

We summarize briefly a few significant new results.  The simple
analytic derivation and solutions to the Wilson coefficients in terms
of the observables and ``clean'' form factors was achieved by defining
new variables $r_\lambda=\widetilde{\mathcal{G}}_\lambda/
\mathcal{F}_\lambda -C_9$. These enable solutions to $C_9$ and $C_{10}$ in
terms of observables, $\mathsf{P_1}$, $\mathsf{P_1'}$ and the form
factor $\mathcal{F}_\|$ to be
\begin{align*}
 C_9&=\dsp\frac{\sqrt{\Gf}}{\sqrt{2}\mathcal{F}_\|}\, \frac{(F_\|\mathsf{P_1}
    \mathsf{P_1'}-F_\perp) -\tfrac{1}{2}(\mathsf{P_1}-\mathsf{P_1'})Z_1}{\Big[\pm(\mathsf{P_1}\!-\!\mathsf{P_1'})\sqrt{\mathsf{P_1^2}
    F_\|+F_\perp+ \mathsf{P_1} Z_1}\,\Big]},\\
 C_{10}&=\dsp\frac{\sqrt{\Gf}}{\sqrt{2}\mathcal{F}_\|}\frac{2}{3}\frac{
    \AFB}{\Big[\pm\sqrt{\mathsf{P_1^2} F_\|+F_\perp+ \mathsf{P_1}
      Z_1}\,\Big]}.
\end{align*}
where $Z_1$ is expressed in terms of observables in
Eq.~\eqref{eq:Z_1}. 
Two additional solutions for $C_9$ and $C_{10}$ can be obtained in terms of
different observables. These are obtained by the replacements 
\begin{itemize}
\item  $F_\|\to F_L$, $\AFB \to \sqrt{2}A_5$,
$\mathcal{F}_\| \to \mathcal{F}_0$, $\mathcal{G}_\| \to
\mathcal{G}_0$, which also imply that $r_\|\to r_0$, $\mathsf{P_1}\to
\mathsf{P_2}$ and $\mathsf{P_1'} \to \mathsf{P_2'}$. 
\item  $F_\|\to
F_L+F_\|+\sqrt{2}\pi A_4$, $\AFB\to \AFB+\sqrt{2}A_5$, $\mathcal{F}_\|
\to \mathcal{F}_\|+\mathcal{F}_0$, $\mathcal{G}_\| \to
\mathcal{G}_\|+\mathcal{G}_0$, which also imply $r_\|\to
r_{\!\wedge}$, $\mathsf{P_1}\to \mathsf{P_3}$ and $\mathsf{P_1'}\to
\mathsf{P_3'}$.
\end{itemize}

We found that the form factor ratios
$\mathsf{P_1}$, $\mathsf{P_2}$ and $\mathsf{P_3}$ can be directly measured in
terms of the ratio of helicity fractions at $q^2$ corresponding to the
zero crossings of asymmetries $\AFB$, $A_5$ and $\AFB+\sqrt{2}A_5$
respectively by the relations:
\begin{align*}
\mathsf{P_1}&=\dsp -\frac{\sqrt{F_\perp}}{\sqrt{F_\|}}\Bigg|_{\AFB=0}\qquad
\mathsf{P_2}=\dsp
-\frac{\dsp\sqrt{F_\perp}}{\dsp\sqrt{F_L}}\Bigg|_{A_5=0}\\
\mathsf{P_3}&=\dsp -\frac{\dsp \sqrt{F_\perp}}{\dsp\sqrt{F_L+F_\perp+\sqrt{2}\pi
    A_4}}\Bigg|_{\AFB+\sqrt{2}A_5=0}~ 
\end{align*}

Since we have neglected the tiny $CP$
violation in the standard model, we
find that the observables must satisfy the following inequalities
which are completely free from any hadronic uncertainties and 
hence clean. These relations are,
\begin{align*}
  4 F_\|F_\perp&\geq \frac{16}{9} \AFB^2 \\
 4 F_LF_\perp &\geq \frac{16}{9} (\sqrt{2}A_5)^2~.\\
  4 (1-F_\perp) F_\perp&\geq \frac{16}{9}(\AFB^2+2 A_5^2),\\
  4 (F_L+F_{\|}+\sqrt{2}\pi A_4)F_\perp&\ge\frac{16}{9}
  (A_{\text FB}+\sqrt{2}A_5)^2.
\end{align*}
In Fig.~\ref{fig:bounds} we have plotted the constraints on
$F_L-F_\perp$ that depends only on observables. The condition $4
F_\|F_\perp\geq 16/9 \AFB^2$ implies that if $|\AFB|$ is large $F_L$
must be small so that $4 F_\|F_\perp$ can be sufficiently large.  Our
approach is sensitive enough to already show tensions in the
data~\cite{Moriond-talk}.

 Clearly, expressions for $C_9$ and $C_{10}$ are not
``clean.'' However, the ratio $C_9/C_{10}$ is obtained as a ``clean
expression''. Assuming the theoretical estimate of $C_9/C_{10}$ which
is reliably evaluated at NNLL in Standard Model we ``cleanly''
predicted $F_\perp$ in Eq.~\eqref{eq:AFB-FP}.  The correlation between
$\AFB$, $F_L$ and $F_\perp$ have been plotted in
Figs.~\ref{fig:C10-FP},\ref{fig:C10-FP-R}, \ref{fig:wedgeAFB-HQET} and
\ref{fig:wedgeAFB-LCSR}. We showed that the valid domain of
$\AFB$ is constrained in terms of $F_L$ as follows:
\begin{equation*}
   \dsp\frac{-3 (1-F_L)}{4} T_{-}\leq A_{FB}\leq\frac{3 (1-F_L)}{4} T_{+},
\end{equation*}
where $T_\pm$ is given in terms of $\mathsf{P_1}$, $\mathsf{P_1'}$ and
$R$ in Eq.~\eqref{eq:AFB-FL}. It is interesting to note {\em that $F_L$ and
$F_\perp$ are constrained with Standard Model to lie in a very narrow
region, well approximated by a line} as shown in
Fig.~\ref{fig:wedgeAFB-HQET} and ~\ref{fig:wedgeAFB-LCSR}. The
effective photon vertex $\widetilde{\mathcal{G}}_\|$ and
$\widetilde{\mathcal{G}}_0$ can also be expressed as a ``clean
expression''.

The $C_9/C_{10}$ and $C_7/C_{10}$ ratios in Eqs.~(\ref{eq:C9by10}) 
ratio in (\ref{eq:C7by10}) were combined to obtain
\begin{equation*}
  \Big(\frac{2}{3} \frac{C_9}{C_{10}}\,\mathsf{P_1^{'\!'}}-\frac{4}{3}
  \frac{C_7}{C_{10}}\,\mathsf{P_1}\Big) \AFB =(\mathsf{P_1}^2
  F_{\|}+F_\perp+ \mathsf{P_1}Z)>0.
\end{equation*}
If the $\AFB$ zero crossing is confirmed~\cite{Moriond-talk} with
$\AFB>0$ at small $q^2$, then based on the signs of the from factors
it is unambiguously concluded that the signs of $C_7/C_{10}$ and
$C_9/C_{10}$ are in agreement with the Standard Model,
i.e. $C_7/C_{10}>0$ and $C_9/C_{10}>0$ as long as other constraints
like $Z_1^2>0$ hold. In Ref.~\cite{Moriond-talk} the zero crossing is
indeed seen. However, in the $2\gev^2 \le q^2\le 4.3\gev^2$ bin
$Z_1^2>0$ is only marginally satisfied. These conclusions are exact and
not altered by any hadronic uncertainties.

We have obtained three sets of $C_9/C_{10}$ and $C_7/C_{10}$ solutions
involving difference observables and form factor ratios. Since, the
form factor ratios $\mathsf{P}_1$ and $\mathsf{P}_1'$ are the ones
that are most reliably estimated in both large recoil and low
recoil limits, we obtain relations for $\mathsf{P_2}$, $\mathsf{P_2'}$
and $\mathsf{P_3}$, $\mathsf{P_3'}$ in terms of $\mathsf{P_1}$,
$\mathsf{P_1'}$ and observables. Equating the relations obtained for
$C_9/C_{10}$ and $C_7/C_{10}$ in Eqs.~(\ref{eq:C9by10}),
(\ref{eq:C7by10}) with those in Eqs.~(\ref{eq2:C9by10}),
(\ref{eq2:C7by10}) and Eqs.~(\ref{eq3:C9by10}), (\ref{eq3:C7by10}) we
get:
\begin{eqnarray*}
\mathsf{P_2}&=&\frac{2\mathsf{P_1}\AFB F_\perp}{\sqrt{2}A_5(2F_\perp+Z_1\mathsf{P_1})-Z_2\mathsf{P_1}\AFB×}\\
 \mathsf{P_2'}&=&\frac{\sqrt{2}A_5\Big(F_\perp-F_\|\mathsf{P_1^2}
   \Big)\mathsf{P_2^2}\mathsf{P_1'}} 
{\AFB T_2(\mathsf{P_1}-\mathsf{P_1'})+\sqrt{2}A_5\Big(F_\perp-F_\|
\mathsf{P_1^2} \Big)\mathsf{P_2} \mathsf{P_1'}}\\
\mathsf{P_3}&=&\frac{2\mathsf{P_1}\AFB F_\perp}{(\AFB+\sqrt{2}A_5)(2F_\perp+Z_1\mathsf{P_1})-Z_3\mathsf{P_1}\AFB×},\\
\mathsf{P_3'}&=&\displaystyle\frac{(\AFB+\sqrt{2}A_5)(F_\perp-F_\|\mathsf{P_1^2})\mathsf{P_3^2}\mathsf{P_1'}}
{\AFB
  T_3(\mathsf{P_1}-\mathsf{P_1'})+\sqrt{2}A_5(F_\perp-F_\|\mathsf{P_1^2})\mathsf{P_3^2}\mathsf{P_1'}},
\end{eqnarray*}
where $T_2= \mathsf{P_1}(F_\perp-F_L \mathsf{P_2^2})$ and
$T_3=\mathsf{P_1}\big[F_\perp(1+\mathsf{P_3^2})-\mathsf{P_3^2}(1+\sqrt{2}\pi
A_4)\big]$. Even though $\mathsf{P_2}$, $\mathsf{P_2'}$ and
$\mathsf{P_3}$, $\mathsf{P_3'}$ inherently depend on $\xi_\|$ and
$\xi_\perp$ we have expressed them in terms of ``clean relations''
above. {\em  Hence, in our approach, all the expressions for observables are
``clean,'' with only the Wilson coefficients $C_7$, $C_9$ and $C_{10}$
being expressed in terms of only one form factor $\mathcal{G}_\|$ or
$\mathcal{F}_\|$.} 

We have derived significant constraints between observables that can
be used to test for New Physics.  The constraint purely in terms of
observables arises since $\mathsf{P_2}$ and $\mathsf{P_3}$ are
expressed in terms of observables and $\mathsf{P_1}$ while
$\mathsf{P_3}$ itself is related in Eq.~\eqref{eq:P_3} to $\mathsf{P_1}$
and $\mathsf{P_2}$. We obtain the interesting
constraint~\eqref{eq:Obs-relation} among observables:
\begin{equation*}
    A_4=
\frac{8 A_5 \AFB}{9 \pi F_\perp}+\sqrt{2}\,\frac{\sqrt{F_LF_\perp
       -\frac{8}{9} A_5^2}\sqrt{F_\|F_\perp -\frac{4}{9}\AFB^2}}{\pi F_\perp}.
\end{equation*}
The observables $A_4$ and $A_5$ also impose constraints on the
parameter space. In Fig.~\ref{fig:bounds} we plot constraints
on the parameter space of $F_L$--$F_\perp$ that depend purely on
observables $\AFB$ and $A_5$ with $A_4$ being calculated in terms of
the above relation between observables. As can be seen the parameter
space is highly constrained in the Standard Model.

We introduced six observables of which three $\Gf$, $F_L$ and $\AFB$
have already been measured. We showed that $ F_\perp$ can expressed in
terms of  $\mathsf{P_1}$, $\mathsf{P_1'}$ and the ratio $C_9/C_{10}$.
If we further choose a value for $A_5$, $A_4$ can be obtained. In
Fig.~\ref{fig:teardrop} we depict the constraints in the $\AFB$--$F_L$
parameter space. These constraints and the constraints obtained in
Fig.~\ref{fig:C10-FP} fix completely the parameter space and predict
the values of yet unmeasured observables.


We pay special attention to the low recoil limit and derive two new relations  
\begin{align*}
  \sqrt{2}A_5=&{\AFB}\frac{\sqrt{F_L}}{\sqrt{F_\|}}\\
A_4=&\frac{\sqrt{2}}{\pi}\sqrt{F_LF_\|}
\end{align*}
in terms
of observables alone. These two relations allow us to test not only
the validity of the low recoil approximation but also the presence of
New Physics. The value of $A_5$ predicted by these relations tests the
validity of the low recoil approximation, whereas the value of $A_4$
verifies the validity of SM.  If both relations hold we verify that
the low recoil approximation is correct and that no new physics can
exist. If both relation fail we can conclude that the low recoil
approximation fails but {\em one can never-the-less still test for new
  physics } by Eq.~\eqref{eq:Obs-relation}, which is valid in
general. If $A_5$ is accurately predicted but $A_4$ does not have the
value given by these two relations one can conclude that there is new
physics and that the low recoil limit is accurate. 

In this paper we re-examined the new physics discovery potential of
the mode $B\to \kstar \ell^+\ell^-$. This modes has an advantage as a
multitude of observables can be measured via angular analysis. We
showed how the multitude of “related observables” obtained from $B\to
\kstar \ell^+\ell^-$ can provide many new “clean tests” of the
Standard Model and discriminate new physics contributions from
hadronic effects.  The hallmark of these tests is that most of them
are independent of the unknown form factors $\xi_\|$ and $\xi_\perp$
in heavy quark effective theory. In the large recoil limit (at ${\cal
  O}(1/m_b)$) these relations are valid to all orders in $\alpha_s$.
We derive a relation between observables that is free of form factors
and Wilson coefficients, the violation of which will be an unambiguous
signal of New Physics. We also obtained for the first time relations
between observables and form factors that are independent of Wilson
coefficients and enable verification of hadronic estimates. We show
how form factor ratios can be measured directly from helicity fraction
with out any assumptions what so ever. We find that the allowed
parameter space for observables is very tightly constrained in
Standard Model, thereby providing clean signals of New Physics. We
examine in detail both the large-recoil and low-recoil regions of the
$\kstar$ meson and probe special features valid in the two limits.
Another new relation involving only observables that would verify the
validity of the relations between form-factors assumed in the
low-recoil region was also derived.  The several relations and
constraints derived will provide unambiguous signals of New Physics if
it contributes to these decays. {\em We emphasize that in our approach,
  $C_9/C_{10}$ and all the expressions independent of Wilson
  coefficients are ``clean" in the large recoil limit and in the low
  recoil limit they are reliably calculated as they do not depend on
  the universal form factors $\xi_\|$ and $\xi_\perp$.}

\appendix

\section{Derivation of Wilson Coefficients} 
\label{sec:appendix-1}

Below we present the solution of $r_\|+r_\perp$. The solutions of $r_0+r_\perp$ and
$r_{\!\wedge}+r_\perp$ are identical. 

We start with the expression involving $r_\|$ and $r_\perp$ in terms of
observables as expressed in Eqs.~(\ref{eq2:Fparallel}), (\ref{eq2:FL})
and (\ref{eq2:AFB}):
\begin{eqnarray}
  \label{eq:def1}
r_\|^2+C_{10}^2 &=&\frac{F_\|\Gf}{2\mathcal{F}_\|^2}\\
\label{eq:def2}
r_\perp^2+C_{10}^2 &=&\frac{F_\perp\Gf}{2\mathcal{F}_\perp^2}\\
\label{eq:def3}
2C_{10}(r_\|+r_\perp) &=&\frac{2}{3}\frac{\AFB\Gf}{\mathcal{F}_\perp\mathcal{F}_\|}.
  \end{eqnarray}
We can write 
\begin{eqnarray}
  \label{eq:16}
\frac{F_\|F_\perp\Gf^2}{4\mathcal{F}_\|^2\mathcal{F}_\perp^2}&=&
(r_\|r_\perp-C_{10})^2+C_{10}^2(r_\|+r_\perp)^2\nn\\   
           &=&(r_\|r_\perp-C_{10})^2+\frac{\AFB^2\Gf^2}{9\mathcal{F}_\|^2
             \mathcal{F}_\perp^2}\nn 
\end{eqnarray}
hence
\begin{eqnarray}
\label{eq1:uv}
r_\|r_\perp-C_{10}^2 &=&\pm\frac{\Gf}{2\mathcal{F}_\|
             \mathcal{F}_\perp} \sqrt{F_\|F_\perp-\frac{4\AFB^2}{9}}.
\end{eqnarray}
Now we can express $C_{10}^2$ in terms of $r_\|^2$  using
Eq.~(\ref{eq:def1}) or in
terms of $r_\perp^2$ using Eq.~(\ref{eq:def2}), to re-express $u v-C_{10}^2$:
\begin{eqnarray}
  \label{eq2:uv}
 2r_\|r_\perp -2C_{10}^2&=& 2 r_\|r_\perp
 -\Big(\frac{F_\|\Gf}{2\mathcal{F}_\|^2}-r_\|^2\Big)-
 \Big(\frac{F_\perp\Gf}{2\mathcal{F}_\perp^2}-r_\perp^2\Big)\nn\\ 
&=&\Big[(r_\|+r_\perp)^2-\frac{F_\|\Gf}{2\mathcal{F}_\|^2}-
\frac{F_\perp\Gf}{2\mathcal{F}_\perp^2}\Big]  
\end{eqnarray}
Equating Eqs.~(\ref{eq1:uv}) and (\ref{eq2:uv}) we get 
\begin{align}
  \label{eq:19}
  r_\|+r_\perp&=\pm\Bigg[\frac{F_\|\Gf}{2\mathcal{F}_\|^2}
  +\frac{F_\perp\Gf}{2\mathcal{F}_\perp^2}\pm  
  \frac{\Gf}{2\mathcal{F}_\|\mathcal{F}_\perp}Z_1\Bigg]^{\nicefrac{1}{2}}\nn
  \\[2.ex]
&=\frac{\pm\sqrt{\Gf}}{\sqrt{2}\mathcal{F}_\perp}\Big[\mathsf{P_1^2}
F_\|+F_\perp\pm \mathsf{P_1}Z_1 \Big]^{\nicefrac{1}{2}}
\end{align}
where $Z_1=\sqrt{4F_\|F_\perp-\tfrac{16}{9}\AFB^2}$.
Now, Eqs.~(\ref{eq:def1}) and ~(\ref{eq:def2}) imply: 
\begin{equation}
  \label{eq:20}
r_\|^2-r_\perp^2=
\frac{F_\|\Gf}{2\mathcal{F}_\|^2}-
\frac{F_\perp\Gf}{2\mathcal{F}_\perp^2},
\end{equation}
which gives $r_\|-r_\perp$ to be,
\begin{equation}
  \label{eq:21}
r_\|-r_\perp= \dsp\frac{\pm\sqrt{\Gf}}{\sqrt{2}\mathcal{F}_\perp}\frac{\dsp
  \mathsf{P_1^2} F_\| -F_\perp}{\dsp\Big[\mathsf{P_1^2}
F_\|+F_\perp\pm \mathsf{P_1} Z_1\Big]^{\nicefrac{1}{2}}}
\end{equation}
$C_{10}$ is readily solved using Eq.~(\ref{eq:def3}) and the
expression for $r_\|+r_\perp$ obtained above. $C_7$ and $C_9$ are also
easily solved using Eq.~(\ref{eq:uvwz}) and the expressions for
$r_\|-r_\perp$. The solutions for $C_7$, $C_9$ and $C_{10}$ are
presented in Eqs.~(\ref{eq:C7}), (\ref{eq:C9}) and (\ref{eq:C10})
respectively.

\section{Form Factor Calculations} 
\label{sec:appendix-inputs}
In this appendix we discuss the calculations of form factors and the
form factor ratios. In our numerical analysis we have calculated the 
average value of the form factor $\mathcal{F}_\|$ and the two
form factor ratios $\mathsf{P_1}$ and $\mathsf{P_1'}$ in differnt $q^2$ regions.

As has already been discussed in Sec.(\ref{sec:large-recoil}),
at large recoil region the heavy quark symmetry applies and
the seven form factors $V, A_{1,2,3}, T_{1,2,3}$ are functions
of Isgur-Wise form factors $\xi_\|(q^2)$ and $\xi_\perp(q^2)$~\cite{Isgur-Wise}.
These two form factors are parameterized as~\cite{Beneke:2001at}
\begin{eqnarray}
\xi_\perp(q^2)&=&\xi_\perp(0)\Bigg(\frac{1}{1-q^2/m_B^2}\Bigg)^2\nn\\
\xi_\|(q^2)&=&\xi_\|(0)\Bigg(\frac{1}{1-q^2/m_B^2}\Bigg)^3\nn
\end{eqnarray}
where $\xi_\perp(0)=0.266\pm0.032$ and $\xi_\|(0)=0.118\pm0.008$
~\cite{Altmannshofer:2008dz}.
The two ratios $\mathsf{P_1}, \mathsf{P_1'}$ (see Eqs~\ref{eq:P_1-simple} 
and~\ref{eq:P_1prime-simple}) are independent of Isgur-Wise form factors, and
only $\mathcal{F}_\|$ (see Eq.~\ref{eq:form-factors}) is dependent on $\xi_\perp$.
In Table.~(\ref{table:nonlin-2}) we have calculated the 
the values of $\mathsf{P_1}$, $\mathsf{P_1'}$ and $\mathcal{F}_\|$ 
averaged over each $q^2$ bin
used by the recent experiments~\cite{Aaij:2011aa}.

\begin{table}[ht]
 \centering
\begin{tabular}{|c|c|c|c|c|c|c|}
\hline
\hline
$\gev^2$&0.10-2&2-4.3&4.3-8.68&10.09-12.86&1-6\\
\hline
$\mathsf{P_1}$ &-0.8924&-0.9286&-0.9034&-0.8337&-0.9259\\
\hline
$\mathsf{P_1'}$ &-0.9189&-0.9561&-0.9302&-0.8585&-0.9533\\
\hline
$\mathcal{F}_\|(10^{-12})$ &-5.7667&-11.330&-17.4311&-25.8917&-11.8692\\
\hline
\hline
\end{tabular}
\caption{The form factor ratios $\mathsf{P_1}, \mathsf{P_1'}$ and $\mathcal{F}_\|$ 
averaged over different $q^2$ bins at large recoil.}
\label{table:nonlin-2}
\end{table}

At low recoil the seven form factors $V, A_{1,2,3}, T_{1,2,3}$
are parameterized ~\cite{DAs} as:
\begin{eqnarray}
 V(q^2)&=&\frac{r_1}{1-q^2/m_R^2}+\frac{r_2}{1-q^2/m_{\text{fit}}^2}\nn\\
A_1(q^2)&=&\frac{r_2}{1-q^2/m_{\text{fit}}^2}\nn\\
A_2(q^2)&=&\frac{r_1}{1-q^2/m_{\text{fit}}^2}+\frac{r_2}{(1-q^2/m_{\text{fit}}^2)^2}\nn\\
\label{eq:ff-LCSR}
T_1(q^2)&=&\frac{r-1}{1-q^2/m_R62}+\frac{r_2}{1-q^2/m_{\text{fit}}^2)^2}\\
T_2(q^2)&=&\frac{r_2}{1-q^2/m_{\text{fit}}^2)^2}\nn\\
T_3(q^2)&=&\frac{m_B^2-m_{K^*}}{q^2}(\tilde{T}_3(q^2)-T_2(q^2))\nn
\end{eqnarray}
where $\tilde{T}_3$ has same parameterization as $A_1$. The parameters
$r_1, r_2, m_R^2, m_{\text{fit}}^2$ for each of the above form factors
has benn taken from ~\cite{DAs}. Following the above parameterization,
the ratios $\mathsf{P_1}, \mathsf{P_1'}$ and $\mathcal{F}_\|$ has been calculated in 
the low recoil region, averaged over each $q^2$ bins 
and has been shown in Table.(\ref{table:nonlin}).

\begin{table}[ht]
 \centering
\begin{tabular}{|c|c|c|}
\hline
\hline
$\gev^2$ &14.18-16&16-19\\
\hline
$\mathsf{P_1}$ &-0.6836&-0.4719\\
\hline
$\mathsf{P_1'}$ &-0.7093&-0.4952\\
\hline
$\mathcal{F}_\|(10^{-12})$ &-27.8735&-25.0050\\
\hline
\hline
\end{tabular}
\caption{The form factor ratios $\mathsf{P_1}, \mathsf{P_1'}$ and $\mathcal{F}_\|$ 
averaged over different $q^2$ bins at low recoil.}
\label{table:nonlin}
\end{table}

\end{document}